\documentclass[preprint,3p,times]{elsarticle}
\usepackage{subfigure} 
\usepackage{setspace}
\usepackage{color}
\usepackage{booktabs}           
\usepackage{multirow}  
\usepackage{xcolor}  

\definecolor{myblue}{RGB}{0,0,139}
\definecolor{myred}{RGB}{139,0,0}
\definecolor{mygreen}{RGB}{0,100,0}

\newcommand{\revA}[1]{#1}
\newcommand{\revB}[1]{#1}
\newcommand{\revAB}[1]{#1}
\usepackage{amssymb}
\usepackage{amsmath}
\journal{Journal of Computational Physics}

\begin{document}

\begin{frontmatter}



\title{Implicit discretization schemes for full-kinetic ion and drift-kinetic electron simulations}


\author[SWIP,Tsinghua]{Zilong Li} %% Author name
\author[Colorado]{Yang Chen}
\author[SWIP]{Haotian Chen\corref{cor1}}
\ead{chenhaotian@swip.ac.cn}
\cortext[cor1]{Corresponding author}
\author[IPP]{Lei Ye}
\author[Tsinghua]{Zhe Gao}
\author[SWIP]{Wei Chen}
\affiliation[SWIP]{organization={Southwestern Institute of Physics},%Department and Organization
            city={Chengdu},
            postcode={610041}, 
            country={China}}

\affiliation[Tsinghua]{organization={Department of Engineering Physics,Tsinghua University},%Department and Organization
            city={Beijing},
            postcode={100084}, 
            country={China}}

\affiliation[Colorado]{organization={Department of Physics, University of Colorado at Boulder},%Department and Organization
            city={Boulder},
            postcode={80309}, 
            country={USA}}

\affiliation[IPP]{organization={Institute of Plasma Physics, Chinese Academy of Science},%Department and Organization
            city={Hefei},
            postcode={230031}, 
            country={China}}
%% Abstract
\begin{abstract}
We present a new electromagnetic plasma simulation model with full-kinetic ions and drift-kinetic electrons.
This model (termed as FIDES) solves the electric field using the implicit perpendicular Ohm's law and a novel implicit parallel Ampere's law, where the latter requires an implicit scheme for the parallel electric field in advancing the electron weights.
To suppress unphysical high-frequency instabilities, ion weights are advanced using an implicit scheme for perpendicular electric fields.
Simulations of perpendicular and parallel waves validate the model's capability in handling high-frequency physics.
Low-frequency wave simulations demonstrate that the implicit parallel Ampere's law can mitigate the cancellation problem more effectively than the conventional schemes using the parallel Ohm's law.
To reduce the numerical damping from implicit time-stepping, we develop a second-order scheme for particle pushing.
Meanwhile, an integrated strategy combining the first- and second-order schemes is employed to suppress odd-even decoupling while maintaining the accuracy of the second-order formulation. 
\end{abstract}


% \begin{highlights}
% \end{highlights}

%% Keywords

\end{frontmatter}



\section{Introduction}

Despite its success in low-frequency physics, where the wave frequency $\omega$ is much less than the ion cyclotron frequency $\Omega_{ci}$, the gyrokinetic simulation is limited by the so-called nonlinear gyrokinetic ordering assumptions \cite{ChenH2024,ChenL1982}.
These ordering assumptions break down in regimes of growing practical importance.
For example, high-frequency $(\omega\sim\Omega_{ci})$ waves, which are essential for applications including wave heating and current drive, violate the condition $\omega/\Omega_{ci}\ll1$ and thus lie beyond the scope of standard gyrokinetics \cite{ChenH2024}.
In the plasma pedestal region where the equilibrium pressure gradient scale length $L_{p}$ can be comparable to the ion Larmor radius $\rho_{i}$, the scale separation assumption fails.
More importantly, as a perturbative theory, gyrokinetics cannot self-consistently perform full-f non-perturbative simulations.
As shown in \cite{ChenH2024}, the gyrokinetic framework becomes invalid when the perturbative amplitude is beyond a quantitative threshold.
Additionally, this perturbative nature constrains its formal accuracy. 
The widely employed gyrokinetic equation is accurate to $O(\lambda^{2})$ ($\lambda=\rho_{i}/L_{p}$) \cite{Lee1983,Lee1987,ChenY2003, ChenY2007}.
An accuracy imbalance arises in the quasi-neutrality equations for $k_{\perp}\rho_{i}\sim\lambda$, where the ion polarization density is computed at $k^{2}_{\perp}\rho_{i}^{2}\phi\sim O(\lambda^{2}\phi)$ but the perturbed ion density remains accurate to $O(\lambda)$ \cite{Parra2008}.

A fully kinetic description of ions may overcome the limitations in gyrokinetic simulations and is now within reach for modern supercomputers.
But resolving electron gyromotion remains prohibitively expensive.
In this paper, we propose a novel full-kinetic ion and drift-kinetic electron simulation (FIDES) model.
This scheme is promising since, in most scenarios of interest, the ordering assumptions $(\omega/\Omega_{ce}\ll 1,k_{\perp}\rho_{e}\ll1)$ of the drift kinetic equation are well satisfied, enabling a balance between computational efficiency and the preservation of essential physics.
More importantly, the drift kinetic equation is valid for fluctuations with amplitudes comparable to the equilibrium level \cite{ChenH2024}.
Therefore, the resulting full-kinetic ion drift-kinetic electron simulation model is inherently suitable for full-f non-perturbative simulations.
\revAB{In practice, however, self-consistently performing non-perturbative simulations poses significant numerical challenges.
As an initial step in the development of the FIDES model, the algorithm presented in this paper employs the $\delta f$ method.
This choice is motivated by its advantage in reducing particle noise and by the extensive experience accumulated within this framework.
When the perturbation becomes sufficiently large that the marker distribution departs substantially from the background Maxwellian, however, the $\delta f$ scheme not only loses its noise-reduction benefit but also induces severe convergence difficulties in the iterative field solver of FIDES.
Consequently, the present algorithm is not designed to handle non-perturbative scenarios in which the perturbed distribution function becomes comparable to the equilibrium distribution.
To demonstrate the core concepts, we implement the $\delta f$ method in a slab geometry for the present study.}
The model adopts the implicit perpendicular Ohm's law for perpendicular electric fields $\mathbf{E}_{\perp}$, but, unlike existing models \cite{ChenY2009,ChengJ2013,ChenH2021,ChenH2023}, employs a novel implicit parallel Ampere's law for the parallel electric field $E_{\|}$, in which the lower-order electron velocity moment can mitigate the cancellation problem.
This formulation requires an implicit $E_{\|}$ scheme for electron weight pushing.
Ion weights are advanced by an implicit $\mathbf{E}_{\perp}$ scheme to suppress unphysical high-frequency instabilities.
Numerical tests demonstrate that the FIDES model can correctly capture high-frequency wave physics and address the cancellation problem in low-frequency wave simulations.
To improve the accuracy of wave dynamics, we further implement a second-order time-stepping scheme and employ an integrated strategy combining first- and second-order schemes to suppress associated odd-even decoupling while preserving numerical accuracy.

% \cite{ChenY2009,ChengJ2013,ChenH2021,ChenH2023,LinY2005,BaoJ2016,yu2022}

% Similar researches have been conducted in this direction.
% Many existing models are electrostatic \cite{Benjamin2016,ZhaoD2016,Raeth2024}, which restricts their applicability to high-frequency electromagnetic waves.
% The electromagnetic models in GeFi and GTC employ the quasi-neutrality equation as the field equation \cite{LinY2005,BaoJ2016,yu2022}; however, this approach requires an explicit expression for the electron polarization density and encounters an accuracy imbalance for $k_{\perp}\rho_{e}\ll 1$ \cite{Parra2008}. 
% In contrast, the electromagnetic model developed by Chen and Parker casts the Ampere's law into the generalized Ohm's law, thereby avoiding direct use of the quasi-neutrality equation and enabling direct solution for electric field \cite{ChenY2009,ChengJ2013}.
% Nevertheless, a numerical cancellation problem arises in the parallel Ohm's law when electrons behave adiabatically for $k_{\|}v_{te}\gg \omega$ \cite{ChenY2009,ChenH2021,ChenH2023}.
% In this regime, the leading-order terms $E_{1\|}$ and $-\nabla_{\|}\delta p_{e\|}$ nearly cancel, so that the numerical error in $\delta p_{e\|}$ can lead to a severe loss of accuracy.   

The rest of the paper is structured as follows. Section 2 presents the FIDES models and the simulation examples are discussed in section 3.
Section 4 introduces the second-order scheme and the conclusions are given in section 5.

\section{Numerical model}
\subsection{\revAB{Notation and Normalization}}

\revAB{In this subsection, we define the notation used throughout the paper for both field and particle quantities. A summary is provided in Table~\ref{tab:notation} for quick reference.}

\subsubsection{\revAB{Field quantities}}

\revAB{Field quantities are functions defined on the spatial grid. They may carry superscripts, subscripts, or both. For example, \(E_{1\parallel}^{n}(\mathbf{x}_{g})\) denotes the perturbed electric field parallel to the background magnetic field, evaluated at the grid point \(\mathbf{x}_{g}\) and at time step \(t^{n}=n\Delta t\).}

\revAB{The superscripts and subscripts used for electromagnetic fields are defined as follows:}
\revAB{
\begin{itemize}
    \item \textbf{Superscripts:}
    \begin{itemize}
        \item \(n\): time step index, with \(t^{n}=n\Delta t\).
        \item \(k\): iteration index within a time step (e.g., the \(k\)-th iteration of the field solver).
        \item \(*\): intermediate quantity in the implicit discretization scheme.
        \item \(\text{Imp}\): implicit quantity in the implicit discretization scheme.
    \end{itemize}
    \item \textbf{Subscripts:}
    \begin{itemize}
        \item \(0\): equilibrium (background) quantity.
        \item \(1\): perturbed (fluctuating) quantity.
        \item \(\parallel\) or \(\perp\): direction parallel or perpendicular to the background magnetic field.
        \item \(x, y, z\): Cartesian components when applicable.
    \end{itemize}
\end{itemize}}

\revAB{When we refer to quantities in Fourier space, we denote them with $\tilde{\left(\cdot\right)}$ or $F\left\{\cdot\right\}$ for spatial Fourier transformation, and $\hat{\left(\cdot\right)}$ for time Fourier transformation, respectively. The wavenumber and frequency are indicated as arguments. For instance, \(\tilde{E}_{1z}(k_{z})\) is the spatial Fourier amplitude of the perturbed electric field in the z direction at wavenumber \(k_{z}\). }

\subsubsection{\revAB{Particle quantities}}

\revAB{Particle quantities are defined for each species and each marker particle. They carry additional subscripts to indicate the species and the particle index. For example, \(\mathbf{v}_{ij\perp}^{n+1}\) represents the perpendicular velocity of the \(j\)-th ion at time step \(t^{n+1}\).}

\revAB{The notation for particle quantities follows:}
\revAB{\begin{itemize}
    \item \textbf{Species subscripts:}
    \begin{itemize}
        \item \(i\): ion.
        \item \(e\): electron.
    \end{itemize}
    \item \textbf{Particle index:}
    \begin{itemize}
        \item \(j\): index of the marker particle.
    \end{itemize}
    \item \textbf{Other subscripts and superscripts:} The same conventions for direction (\(\parallel\), \(\perp\)) and time (\(n\), \(k\), \(*\)) as defined for field quantities apply consistently to particle quantities.
\end{itemize}}

\begin{table}[htbp]
    \centering
    \caption{Summary of notation used in this paper.}
    \label{tab:notation}
    \begin{tabular}{@{}lll@{}}
        \toprule
        \textbf{Symbol} & \textbf{Meaning} & \textbf{Example} \\
        \midrule
        \multicolumn{3}{l}{\textbf{Superscripts}} \\
        \(n\)      & Time step index, \(t^n = n\Delta t\)                & \(E_{1\parallel}^{n}\) \\
        \(k\)      & Iteration index within a time step                 & \(E_{1\parallel}^{k}\) \\
        \(*\)      & Intermediate quantity in implicit scheme           & \(\mathbf{J}_{i\perp}^{*}\) \\
        \(\text{Imp}\)      & Implicit quantity in implicit scheme           & \(\mathbf{J}_{i}^{\text{Imp}}\) \\
        \midrule
        \multicolumn{3}{l}{\textbf{Subscripts}} \\
        \(0\)      & Equilibrium (background) quantity                  & \(B_0\) \\
        \(1\)      & Perturbed (fluctuating) quantity                   & \(E_{1\parallel}\) \\
        \(\parallel\) & Parallel to background magnetic field            & \(E_{1\parallel}\) \\
        \(\perp\)    & Perpendicular to background magnetic field       & \(\mathbf{v}_{ij\perp}\) \\
        \(x, y, z\) & Cartesian components                              & \(E_{1x}\) \\
        \(g\)       & Physical quantity at the grid point               & \(\mathbf{x}_{g}\) \\
        \(i\)      & Ion                                                & \({v}_{ij\|}\) \\
        \(e\)      & Electron                                           & \({v}_{ej\|}\) \\
        \(j\)      & Marker particle index                              & \({v}_{ij\|}\) \\
        \midrule
        \multicolumn{3}{l}{\textbf{Other symbols}} \\
        \(\tilde{(\cdot)}, F\left\{\cdot\right\}\) & Space Fourier transformed quantity                        & \(\tilde{E}_{1z}^{n}(k_{z})\) \\
        \(\hat{(\cdot)}\) & Time Fourier transformed quantity                        & \(\hat{E}_{1z}(k_{z},\omega)\) \\
        \bottomrule
    \end{tabular} 
\end{table}

\subsubsection{\revAB{Normalization}}

\revAB{We adopt the following normalization conventions in this paper.
Velocities $v_{i}$ and $v_{e}$ are normalized to $c_{s}=\sqrt{T_{e}/m_{i}}$.
The magnetic moment $\mu_{e}$ is normalized to $T_{e}/B_{0}$.
Time $\Delta t$ is scaled by the inverse ion cyclotron frequency $\Omega_{ci}^{-1}=m_{i}/eB_{0}$, and lengths by the Larmor radius $\rho_{s}=c_{s}/\Omega_{ci}$.
The magnetic field $B$ is normalized to the background field $B_{0}$, the electric field $E$ to $T_{e}/e\rho_{s}$, the current density $J$ to $en_{e}c_{s}$, and the pressure $p$ to $n_{e}T_{e}$.}

\revAB{For the sake of clarity and without loss of generality, in this paper, we present our formulations in a shearless slab geometry.
In this configuration, a uniform background magnetic field aligns with the z direction, while all equilibrium non-uniformities (e.g., in density $n_{e}$) are along the x direction.
The core idea of the FIDES algorithm is independent of this specific choice and can be readily extended to more complex geometries.}

In the following sections, we present both an implicit and an explicit discretization scheme for FIDES. 

\subsection{Implicit discretization scheme}
\label{subsec:model I}
The FIDES model employs a full-kinetic description for ions, governed by the Vlasov equation
\begin{equation}
\frac{\partial f_i}{\partial t}+\mathbf{v}_{i} \cdot \nabla f_i+\frac{q_{i}}{m_{i}}\left(\mathbf{E}+\mathbf{v}_{i}\times\mathbf{B}\right) \cdot\frac{\partial f_i}{\partial \mathbf{v}_{i}}=0,
\end{equation}
while the electrons are described by the drift kinetic equation
\begin{equation}
\frac{\partial f_e}{\partial t}+\mathbf{v}_G \cdot \nabla f_e+\dot{\varepsilon}_{e} \frac{\partial f_e}{\partial \varepsilon_{e}}=0.
\end{equation}
\revA{The guiding center velocity is}
\begin{equation}
\mathbf{v}_G=v_{e\|}\left({\mathbf{b}}+\frac{\mathbf{B}_{1\perp}}{B_{0}}\right)+\mathbf{v}_D+\mathbf{v}_E,
\label{eq:guiding center velocity}
\end{equation}
\revA{with $\mathbf{b}=\mathbf{B}_{0}/B_{0}$. 
Here $\mathbf{v}_D$ contains the grad-B and curvature drift, and $\mathbf{v}_E$ is the $\mathbf{E}\times\mathbf{B}$ drift,}
\begin{equation}
  \begin{aligned}
    &\mathbf{v}_{D}=-\frac{m_{e}v_{e\|}^{2}}{eB_{0}}\mathbf{b}\times\left(\mathbf{b}\cdot\nabla\right)\mathbf{b}-\frac{\mu_{e}}{eB_{0}}\mathbf{b}\times\nabla B_{0},\\
    &\mathbf{v}_{E}=\frac{\mathbf{E}_{1}\times\mathbf{b}}{B_{0}},
  \end{aligned}
\end{equation}
\revA{where the subscript 1 denotes perturbed quantities. The perturbed magnetic field is assumed to be much smaller than the equilibrium magnetic field.
Therefore, the amplitude of the total magnetic field is $|\mathbf{B}_{0}+\mathbf{B}_{1}|\approx B_{0}+B_{1\|}$ and the direction of the total magnetic field is $\left(\mathbf{B}_{0}+\mathbf{B}_{1}\right)/|\mathbf{B}_{0}+\mathbf{B}_{1}|\approx \mathbf{b}+\mathbf{B}_{1\perp}/B_{0}$ in Eq.~(\ref{eq:guiding center velocity}).}
The time derivative of the electron kinetic energy $\varepsilon_{e}=m_{e}v^{2}/2$ is 
\begin{equation}
\dot{\varepsilon}_{e}=-e \mathbf{v}_G \cdot \mathbf{E}_1-\mu_{e}\mathbf{b}\cdot \nabla\times\mathbf{E}_{1},
\end{equation}
where the magnetic moment $\mu_{e}=m_{e}v_{\perp}^{2}/2B_{0}$ is an adiabatic invariant in the drift kinetic model.

\revB{To reduce the particle noise in simulations, we employ the $\delta f$ method for both species in the present work \cite{Parker1993}.
However, $\delta f$ scheme suffers from severe numerical problems when the perturbed distribution function is comparable to the equilibrium distribution function.
Simulating large perturbation problems lies beyond the scope of the present work.}

\revB{Regarding the field equations, the implicit discretization scheme is built upon a reduced form of the Maxwell equations}
\begin{equation}
\begin{aligned}
  \frac{\partial \mathbf{B}}{\partial t}=&-\nabla \times \mathbf{E}, \\
  \nabla \times \mathbf{B}=&\mu_0 \mathbf{J}.
  \label{eq:Maxwell equation}
\end{aligned}
\end{equation}
\revB{where the displacement current term in Ampere's law is dropped.}
\revAB{This omission is motivated by the need to numerically eliminate plasma oscillations at the plasma frequency $\omega_{pe}=\sqrt{n_{e}e^{2}/m_{e}\varepsilon_{0}}$ (with  $\varepsilon_{0}$ the vacuum permittivity), which can be comparable to the electron cyclotron frequency.
The plasma oscillations are not relevant to the physics of interest and, if retained,  would violate the ordering assumptions of the electron drift kinetic equation.}

\revB{
An important consequence of this reduced form of Ampere's law is that it inherently ensures the quasi-neutrality condition.
Taking the divergence of Ampere's law and noting that the divergence of a curl is zero, we obtain $\nabla\cdot\mathbf{J}=0$.
Meanwhile, the charge conservation equation holds as $\partial (q_{i}n_{i}-en_{e})/\partial t+\nabla\cdot{\mathbf{J}}=0$, which can be derived from the first-order moment of the ion Vlasov equation and electron drift kinetic equation.
Combining $\nabla\cdot\mathbf{J}=0$ with the charge conservation equation yields  $\partial (q_{i}n_{i}-en_{e})/\partial t=0$.
Thus, within this formulation, quasi-neutrality is consistently maintained.
}
% However, unlike in gyrokinetic simulations, the quasi-neutrality equation is not explicitly employed as a field equation here. 
% This is because, within our fully kinetic ion model, the ion polarization density lacks a simple explicit expression in terms of the electric potential.

\revB{The original Maxwell equations (\ref{eq:Maxwell equation}) are ill-posed for explicit particle-pushing schemes. 
This is attributed to the fact that particle noise breaks the $\nabla\cdot\mathbf{J}=0$ condition required by Ampere's law, resulting in inaccurate electromagnetic fields.
A common solution is to transform the Ampere's law into the generalized Ohm's law \cite{ChenY2009,ChengJ2013,ChenH2021,ChenH2023}.
Here, in the context of particle-in-cell (PIC) magnetic confinement fusion (MCF) simulations, the generalized Ohm's law is used to solve for the electric field using the current computed from particles.
This usage differs from the classical Ohm's law, which serves as a constitutive relation expressing the current in terms of the electric field.
Depending on the direction relative to the magnetic field, the generalized Ohm's law  can be split into a parallel component and a perpendicular component.
In previous work \cite{ChenY2009,ChengJ2013,ChenH2021,ChenH2023}, both components have been used as field equations.
However, a numerical cancellation problem arises in the parallel Ohm's law when electrons behave adiabatically for $k_{\|}v_{te}\gg \omega$ \cite{ChenY2009,ChenH2021,ChenH2023}.
In this regime, the leading-order terms $E_{1\|}$ and $-\nabla_{\|} p_{1,e\|}$ nearly cancel, and governing dynamics appears at the next order. 
Consequently, numerical noise in $p_{1,e\|}$ can lead to a severe loss of accuracy.}

\revB{To address the numerical cancellation problem inherent in the parallel Ohm's law, we propose using the Ampere's law directly to solve for the electric field.
Since $E_{\|}$ does not appear explicitly in Ampere's law but enters implicitly through the electron response, an implicit treatment of the $E_{\|}$ contribution to the electron response has to be used.
This leads to the formulation of an implicit parallel Ampere's law, which employs a lower-order electron velocity moment and, as shown in Section 3, effectively mitigates the cancellation problem.
In our algorithm, we adopt this approach by replacing the parallel Ohm's law with the implicit parallel Ampere's law, while retaining the perpendicular Ohm's law.
The electric field is then obtained by solving these two equations simultaneously as a coupled system.}

To develop an implicit scheme for the parallel Ampere's law, we advance electrons using an implicit $E_{\|}$ scheme
\begin{equation}
  \begin{aligned}
  \frac{w_e^*-w_e^n}{\Delta t} & =-\left\{\frac{\partial \ln n_e}{\partial x}+\left[\frac{m_{e}\left(v_{e\|}^{n}\right)^{2}}{2m_{i}}+\mu_{e} B_{0}-\frac{3}{2}\right] \frac{\partial \ln T_e}{\partial x}\right\}\left({E_{1 y}^n}+v_{e \|}^n {B_{1 x}^n}\right)-\mu_{e} \mathbf{b}\cdot\nabla\times\mathbf{E}_{1}^{n}, \\
  \frac{w_e^{n+1}-w_e^*}{\Delta t} & =-v_{e \|}^{n+1} E_{1 \|}^{n+1},
  \label{eq:electron weight pushing}
  \end{aligned}
\end{equation}
\revB{which have been normalized according to the conventions in section~2.1.3. Here, the intermediate electron weight $w_{e}^{*}$ is first advanced from $w_{e}^{n}$ via the first equation. 
Once the field equations are solved, the electron weight $w_{e}^{n+1}$ at time step $t^{n+1}$ is obtained by advancing $w_{e}^{*}$ using the updated parallel electric field $E_{1\|}^{n+1}$.
Since the magnetic moment $\mu_{e}$ is a constant of motion, its superscript $n$ has been omitted in Eq.~(\ref{eq:electron weight pushing}).
}
Under the implicit $E_{\|}$ scheme, the electron parallel current \revAB{$J_{e\|}^{n+1}$} contains two parts
\begin{equation}
\begin{aligned}
& J_{e \|}^{n+1}\left(\mathbf{x}_g\right)=J_{e \|}^*\left(\mathbf{x}_g\right)+\frac{1}{N_p} \sum_j \Delta t\left(v_{e j \|}^{n+1}\right)^2 E_{1 \|}^{n+1}\left(\mathbf{x}_{e j}^{n+1}\right) S\left(\mathbf{x}_g-\mathbf{x}_{e j}^{n+1}\right), \\
& J_{e \|}^*\left(\mathbf{x}_g\right)=-\frac{1}{N_p} \sum_j w_{e j}^* v_{e j \|}^{n+1} S\left(\mathbf{x}_g-\mathbf{x}_{e j}^{n+1}\right),
\label{eq:electron current}
\end{aligned}
\end{equation}
where \revAB{$J_{e\|}^{*}$} and the second term of \revAB{$J_{e\|}^{n+1}$} are, respectively, the \revAB{intermediate} and implicit parts in the electron parallel current.
In Eq.~(\ref{eq:electron current}), $S\left(\mathbf{x}_g-\mathbf{x}_{e j}^{n+1}\right)$ is the shape function, where $\mathbf{x}_g$ and $\mathbf{x}_{e j}^{n+1}$ are the positions of the grid point and the electron, respectively.
\revAB{In the FIDES algorithm,  the particle positions and velocities at $t^{n+1}$ are computed before accumulating particle moments.
In linear simulations, the particle trajectories are advanced along unperturbed orbits, independently of the perturbed fields.
In nonlinear simulations, the particle positions and velocities are updated iteratively using the fields from previous iteration, so that $v_{ej\|}^{n+1}$ and $\mathbf{x}_{ej}^{n+1}$ are available when the particle moments are computed.}

\revAB{Substituting the electron parallel current (\ref{eq:electron current}) into the parallel Ampere's law of Eq.~(\ref{eq:Maxwell equation}), we can get}
\begin{equation}
  \mathbf{b}\cdot\nabla\times\mathbf{B}^{n+1}_{1}=\beta_{e}\left(J_{e\|}^{*}+J_{i\|}^{n+1}\right)+\frac{\beta_e}{N_p} \sum_j \Delta t\left(v_{e j \|}^{n+1}\right)^2 E_{1 \|}^{n+1}\left(\mathbf{x}_{e j}^{n+1}\right) S\left(\mathbf{x}_g-\mathbf{x}_{e j}^{n+1}\right),
\end{equation}
\revAB{where $\beta_{e}=\mu_{0}n_{e}T_{e}/B_{0}^{2}$ and $N_{p}$ is the particle number in one grid. Next, substituting  $\mathbf{B}_{1}^{n+1}$ with $\mathbf{B}_{1}^{n}-\Delta t\nabla\times\mathbf{E}_{1}^{n+1}$ and collecting all terms that depend on the electric field at time step $t^{n+1}$ on the left-hand side, we obtain the implicit parallel Ampere's law, which serves as one of the equations for the electric field,}
\begin{equation}
  \begin{aligned}
  \Delta t \mathbf{b} \cdot \nabla \times \nabla \times \mathbf{E}_1^{n+1}+\frac{\beta_e}{N_p} \sum_j \Delta t\left(v_{e j \|}^{n+1}\right)^2 E_{1 \|}^{n+1}\left(\mathbf{x}_{e j}^{n+1}\right) S\left(\mathbf{x}_g-\mathbf{x}_{e j}^{n+1}\right)=\mathbf{b} \cdot \nabla \times \mathbf{B}_1^n-\beta_e\left(J_{e \|}^*+J_{i \|}^{n+1}\right).
  \end{aligned}
\label{eq:implicit parallel Ampere's law}
\end{equation}
\revAB{After computing the ion and electron currents, and given $\mathbf{B}_{1}^{n}$, this equation is solved together with the perpendicular Ohm's law (introduced later) to obtain the electric field at time step $t^{n+1}$.
Note that regardless of whether the field equations are solved using a finite difference method or a spectral method, the left-hand side of Eq.~(\ref{eq:implicit parallel Ampere's law}) contains a discrete summation term over electrons, $\left(\beta_{e}/N_{p}\right)\sum_j\Delta t\left(v_{e j \|}^{n+1}\right)^2 E_{1 \|}^{n+1}\left(\mathbf{x}_{e j}^{n+1}\right) S\left(\mathbf{x}_g-\mathbf{x}_{e j}^{n+1}\right)$.
This term renders the discretized coefficient matrix of field equations complex and time-step-dependent, precluding the use of cost-saving techniques such as performing a single LU decomposition and storing the matrix for use in later time steps.
Therefore, in FIDES, we propose to solve the implicit parallel Ampere's law and perpendicular Ohm's law in an iterative manner.
In particular, we first find an approximate expression for $\left(\beta_{e}/N_{p}\right)\sum_j\Delta t\left(v_{e j \|}^{n+1}\right)^2 E_{1 \|}^{n+1}\left(\mathbf{x}_{e j}^{n+1}\right) S\left(\mathbf{x}_g-\mathbf{x}_{e j}^{n+1}\right)$,
}
\begin{equation}
  \begin{aligned}
    &\frac{\beta_{e}}{N_{p}}\sum_j \Delta t\left(v_{e j \|}^{n+1}\right)^2 E_{1 \|}^{n+1}\left(\mathbf{x}_{e j}^{n+1}\right) S\left(\mathbf{x}_g-\mathbf{x}_{e j}^{n+1}\right)=\frac{\beta_{e}\Delta t}{N_{p}}\int_{-\infty}^{+\infty} dv_{\|} \left[v_{\|}^{2}\sum_{j}E_{1\|}^{n+1}\left(\mathbf{x}_{ej}^{n+1}\right)S\left(\mathbf{x}_{g}-\mathbf{x}_{ej}^{n+1}\right)\delta \left(v_{\|}-v_{ej\|}^{n+1}\right)\right]\\
    &\approx \frac{\beta_{e}\Delta t\Delta V}{N_{p}}\int_{-\infty}^{+\infty} dv_{\|} \left[v_{\|}^{2}\sum_{j}E_{1\|}^{n+1}\left(\mathbf{x}_{ej}^{n+1}\right)\delta\left(\mathbf{x}_{g}-\mathbf{x}_{ej}^{n+1}\right)\delta \left(v_{\|}-v_{ej\|}^{n+1}\right)\right]\\
    &= \frac{\beta_{e}\Delta t\Delta V}{N_{p}}\int_{-\infty}^{+\infty} dv_{\|} \left[v_{\|}^{2}E_{1\|}^{n+1}\left(\mathbf{x}_{g}\right)\sum_{j}\delta\left(\mathbf{x}_{g}-\mathbf{x}_{ej}^{n+1}\right)\delta \left(v_{\|}-v_{ej\|}^{n+1}\right)\right].
  \end{aligned}
  \label{eq:iteration term 1}
\end{equation}
\revAB{Here, we have approximated the shape function as $S\left(\mathbf{x}_{g}-\mathbf{x}_{ej}^{n+1}\right)\approx \Delta V \delta \left(\mathbf{x}_{g}-\mathbf{x}_{ej}^{n+1}\right)$, where $\Delta V$ is the volume of a grid cell.
This approximation becomes increasingly accurate as the grid size decreases.
In the limit of a sufficiently large number of marker particles, we can approximate $\left(\Delta V/N_{p}\right)\sum_{j}\delta\left(\mathbf{x}_{g}-\mathbf{x}_{ej}^{n+1}\right)\delta \left(v_{\|}-v_{ej\|}^{n+1}\right)$ as the normalized marker distribution $exp[-m_{e}v_{\|}^{2}/\left(2m_{i}\right)]/(2\pi m_{i}/m_{e})^{0.5}$ in Eq.~(\ref{eq:iteration term 1}), }
\begin{equation}
  \begin{aligned}
    &\frac{\beta_{e}\Delta t\Delta V}{N_{p}}\int_{-\infty}^{+\infty} dv_{\|} \left[v_{\|}^{2}E_{1\|}^{n+1}\left(\mathbf{x}_{g}\right)\sum_{j}\delta\left(\mathbf{x}_{g}-\mathbf{x}_{ej}^{n+1}\right)\delta \left(v_{\|}-v_{ej\|}^{n+1}\right)\right]\approx \beta_{e}\Delta t E_{1\|}^{n+1}\left(\mathbf{x}_{g}\right)\int_{-\infty}^{\infty} dv_{\|}\left[\frac{v_{\|}^{2}}{(2\pi m_{i}/m_{e})^{0.5}}exp\left(-\frac{m_{e}v_{\|}^{2}}{2m_{i}}\right)\right]\\
    &=\Delta t\beta_{e}\frac{m_{i}}{m_{e}}E_{1\|}^{n+1}\left(\mathbf{x}_{g}\right).
  \end{aligned}
  \label{eq:iteration term 2}
\end{equation}
\revAB{Substituting Eq.~(\ref{eq:iteration term 2}) into Eq.~(\ref{eq:implicit parallel Ampere's law}), treating the difference between $\left(\beta_{e}/N_{p}\right)\sum_j\Delta t\left(v_{e j \|}^{n+1}\right)^2 E_{1 \|}^{n+1}\left(\mathbf{x}_{e j}^{n+1}\right) S\left(\mathbf{x}_g-\mathbf{x}_{e j}^{n+1}\right)$ and $\Delta t\beta_{e}\left({m_{i}}/{m_{e}}\right)E_{1\|}^{n+1}\left(\mathbf{x}_{g}\right)$ as a small perturbation and moving them to the right-hand side, we can get the iterative form of the implicit parallel Ampere's law,}
\begin{equation}
  \begin{aligned}
  \Delta t \mathbf{b} \cdot \nabla \times \nabla \times \mathbf{E}_1^{k+1}+\Delta t \beta_e \frac{m_i}{m_e} E_{1 \|}^{k+1}=&\mathbf{b} \cdot \nabla \times \mathbf{B}_1^n-\beta_e\left(J_{e \|}^*+J_{i \|}^{n+1}\right)\\
  &+\beta_e \Delta t\left[\frac{m_i}{m_e} E_{1 \|}^k-\frac{1}{N_p} \sum_j\left(v_{e j \|}^{n+1}\right)^2 E_{1 \|}^k\left(\mathbf{x}_{e j}^{n+1}\right) S\left(\mathbf{x}_g-\mathbf{x}_{e j}^{n+1}\right)\right],
  \label{eq:iterative implicit ampere law}
  \end{aligned}
\end{equation}
\revAB{where we have updated the superscript notation to distinguish between time step and iteration: the electric field at time step $n+1$ is now denoted with iteration indices $k$ and $k+1$ within the iterative loop.
Specifically, for the initial guess $(k=0)$, we set $\mathbf{E}_{1}^{k}=\mathbf{E}_{1}^{n}$.
In FIDES, the iteration loop terminates when the relative change between successive iterates falls below a prescribed threshold, $\|\mathbf{E}_{1d}^{k+1}-\mathbf{E}_{1d}^{k}\|_{2}/\|\mathbf{E}_{1d}^{k}\|_{2}<\text{tol}$, $(d=x,~y,~z)$, where $||\cdot||_{2}$ denotes the global 2-norm (i.e., the Euclidean norm of the vector formed by collecting the electric field values at all grid points).
The converged solution is then assigned as $\mathbf{E}_{1}^{n+1}=\mathbf{E}_{1}^{k+1}$.
When the particle number and grid number are sufficient, convergence is typically achieved within a few steps; a convergence test can be seen in Fig.~\ref{fig: iterative_solver}.
It is important to note that for quantities which remain unchanged during the iteration loop, such as $\mathbf{B}_{1}^{n}$ in Eq.~(\ref{eq:iterative implicit ampere law}), the superscript continues to denote the time step n, not the iteration index.
}

\revAB{Upon convergence, the iterative form of the implicit parallel Ampere's law, Eq.~(\ref{eq:iterative implicit ampere law}), reverts to its original formulation, Eq.~(\ref{eq:implicit parallel Ampere's law}). 
Comparing the two equations reveals that the coefficient matrix in Eq.~(\ref{eq:iterative implicit ampere law}) is considerably simpler and, importantly, independent of the time step. This allows us to perform a single LU decomposition and store the matrix, significantly reducing the computational cost of solving the field equations.}

For typical parameters in fusion plasmas, the presence of high-frequency waves with perpendicular electric fields (e.g., compressional Alfvén, ion Bernstein, and extraordinary waves) forces an extremely small timestep $\Omega_{ci}\Delta t<0.01$.
To eliminate this strict limitation, we push ions in an implicit $\mathbf{E}_{\perp}$ scheme \cite{ChenY2009,ChengJ2013}
\begin{equation}
\begin{aligned}
  & \frac{w_i^*-w_i^n}{\Delta t}=\frac{T_e}{T_i} v_{i \|}^n E_{1 \|}^n-\left(E_{1 y}^n+v_{iz}^{n} B_{1 x}^n-v_{ix}^{n} B_{1 z}^n\right)\left\{\frac{\partial \ln n_i}{\partial x}+\left[\frac{T_e \left(v_i^n\right)^2}{2 T_i}-\frac{3}{2}\right] \frac{\partial \ln T_i}{\partial x}\right\}, \\
  & \frac{w_i^{n+1}-w_i^*}{\Delta t}=\frac{T_e}{T_i} \mathbf{v}_{i \perp}^{n+1} \cdot \mathbf{E}_{1 \perp}^{n+1}.
  \label{eq:ion weight pushing}
\end{aligned}
\end{equation}
Here, the ion equilibrium distribution is assumed as a local Maxwellian distribution and we can get the ion perpendicular current
\begin{equation}
\begin{aligned}
& \mathbf{J}_{i \perp}^{n+1}\left(\mathbf{x}_g\right)=\mathbf{J}_{i \perp}^*\left(\mathbf{x}_g\right)+\frac{\Delta t}{N_p} \frac{T_e}{T_i} \sum_j \mathbf{v}_{ij \perp}^{n+1} \mathbf{v}_{ij \perp}^{n+1} \cdot \mathbf{E}_{1 \perp}^{n+1}\left(\mathbf{x}_{ij}^{n+1}\right) S\left(\mathbf{x}_g-\mathbf{x}_{ij}^{n+1}\right), \\
& \mathbf{J}_{i \perp}^*\left(\mathbf{x}_g\right)=\frac{1}{N_p} \sum_j w_{ij}^* \mathbf{v}_{ij \perp}^{n+1} S\left(\mathbf{x}_g-\mathbf{x}_{ij}^{n+1}\right).
\label{eq:ion perpendicular current}
\end{aligned}
\end{equation}

Given that the perturbed electron distribution is independent of gyrophase in the drift kinetic model, it follows that the perpendicular electron current is alternatively determined from the electron perpendicular momentum equation
\begin{equation}
\mathbf{J}_{e \perp}^{n+1}=-\mathbf{E}_1^{n+1} \times \mathbf{b}+\mathbf{b} \times \nabla p_{1,e\perp}^{n+1},
\label{eq:electron perpendicular current}
\end{equation}
\revA{where $p_{1,e\perp}^{n+1}$ is the perturbed electron perpendicular pressure.}

\revAB{Substituting Eq.~(\ref{eq:ion perpendicular current}) and (\ref{eq:electron perpendicular current}) into the perpendicular Ampere's law of Eq.~(\ref{eq:Maxwell equation}), we can get}
\begin{equation}
  \begin{aligned}
    \nabla\times\mathbf{B}_{1}^{n+1}-\mathbf{b}\cdot\nabla\times\mathbf{B}_{1}^{n+1}=\beta_{e}\mathbf{J}_{i\perp}^{*}+\frac{\beta_{e}\Delta t}{N_p} \frac{T_e}{T_i} \sum_j \mathbf{v}_{ij \perp}^{n+1} \mathbf{v}_{ij \perp}^{n+1} \cdot \mathbf{E}_{1 \perp}^{n+1}\left(\mathbf{x}_{ij}^{n+1}\right) S\left(\mathbf{x}_g-\mathbf{x}_{ij}^{n+1}\right)-\beta_{e}\mathbf{E}_1^{n+1} \times \mathbf{b}+\beta_{e}\mathbf{b} \times \nabla p_{1,e\perp}^{n+1}.
  \end{aligned}
\end{equation}
\revAB{Then, taking the cross product of this equation with $\mathbf{b}$, substituting the magnetic field $\mathbf{B}_{1}^{n+1}$ using Faraday's law, and moving all terms involving $\mathbf{E}_{1}^{n+1}$ to the left-hand side, we obtain the implicit perpendicular Ohm's law.}
\begin{equation}
  \begin{aligned}
  &\beta_{e}\mathbf{E}_{1 \perp}^{n+1}+\frac{\beta_{e}\Delta t}{N_p} \frac{T_e}{T_i} \sum_j \left(\mathbf{v}_{ij \perp}^{n+1} \times \mathbf{b}\right)\left[\mathbf{v}_{ij \perp}^{n+1} \cdot \mathbf{E}_{1 \perp}^{n+1}\left(\mathbf{x}_{ij}^{n+1}\right)\right] S\left(\mathbf{x}_g-\mathbf{x}_{ij}^{n+1}\right)-{\Delta t} \mathbf{b} \times\left(\nabla \times \nabla \times \mathbf{E}_1^{n+1}\right)\\
  &=-\mathbf{b} \times\left(\nabla \times \mathbf{B}_1^n\right)-\beta_{e}\nabla_{\perp} p_{1,e\perp }^{n+1}-\beta_{e}\mathbf{J}_{i \perp}^* \times \mathbf{b}.
  \label{eq:implicit perpendicular Ohm's law}
  \end{aligned}
\end{equation}
\revAB{For reasons similar to those discussed for the implicit parallel Ampere's law, Eq.~(\ref{eq:implicit perpendicular Ohm's law}) is also solved iteratively in FIDES.
Following the same approach of Eq.~(\ref{eq:iteration term 1}), we introduce the approximation $S\left(\mathbf{x}_{g}-\mathbf{x}_{ij}^{n+1}\right)\approx \Delta V \delta \left(\mathbf{x}_{g}-\mathbf{x}_{ij}^{n+1}\right)$.
Under the assumption of a sufficiently large number of marker particles,  $\left(\Delta V/N_{p}\right)\sum_{j}\delta\left(\mathbf{x}_{g}-\mathbf{x}_{ij}^{n+1}\right)\delta \left(\mathbf{v}-\mathbf{v}_{ij}^{n+1}\right)$ is then approximated by the normalized ion marker distribution $exp\left[-(T_{e}v^{2})/(2T_{i})\right]/(2\pi T_{i}/T_{e})^{1.5}$, yielding}
\begin{equation}
  \begin{aligned}
    &\frac{\beta_{e}\Delta t}{N_p} \frac{T_e}{T_i} \sum_j \left(\mathbf{v}_{ij \perp}^{n+1} \times \mathbf{b}\right)\left[\mathbf{v}_{ij \perp}^{n+1} \cdot \mathbf{E}_{1 \perp}^{n+1}\left(\mathbf{x}_{ij}^{n+1}\right)\right] S\left(\mathbf{x}_g-\mathbf{x}_{ij}^{n+1}\right)\\
    &\approx \frac{\beta_{e}\Delta t\Delta V}{N_{p}}\frac{T_e}{T_i} \int d\mathbf{v}\left\{\left(\mathbf{v}\times\mathbf{b}\right)\left[\mathbf{v}_{\perp}\cdot\mathbf{E}_{1\perp}^{n+1}\left({\mathbf{x}_{g}}\right)\right]\sum_{j}\delta\left(\mathbf{x}_{g}-\mathbf{x}_{ij}^{n+1}\right)\delta\left(\mathbf{v}-\mathbf{v}_{ij}^{n}\right)\right\}\\
    &\approx \beta_{e}\Delta t\frac{T_{e}}{T_{i}}\int d\mathbf{v}\left\{\left(\mathbf{v}\times\mathbf{b}\right)\left[\mathbf{v}_{\perp}\cdot\mathbf{E}_{1\perp}^{n+1}\left({\mathbf{x}_{g}}\right)\right]\frac{1}{\left(2\pi T_{i}/T_{e}\right)^{1.5}}exp\left(-\frac{T_{e}v^{2}}{2T_{i}}\right)\right\}\\
    &=\beta_{e}\Delta t\mathbf{E}_{1\perp}^{n+1}\left(\mathbf{x}_{g}\right)\times\mathbf{b}.
    \end{aligned}
    \label{eq:iteration term 3}
\end{equation}

\revAB{It should be mentioned that the perturbed electron perpendicular pressure $p_{1,e\perp}^{n+1}$ also contains an implicit contribution, arising from the implicit $E_{\|}$ scheme used to advance electron weights (see Eq.~(\ref{eq:electron weight pushing})),}
\begin{equation}
  \begin{aligned}
    p_{1,e\perp}^{n+1}\left(\mathbf{x}_{g}\right)=\frac{1}{N_{p}}\sum_{j}w_{ej}^{*}\mu_{ej}B_{0}S\left(\mathbf{x}_{g}-\mathbf{x}_{ej}^{n+1}\right)-\frac{\Delta t}{N_{p}}\sum_{j}\mu_{ej}B_{0}v_{ej\|}^{n+1}E_{1\|}^{n+1}\left(\mathbf{x}_{ej}^{n+1}\right) S\left(\mathbf{x}_{g}-\mathbf{x}_{ej}^{n+1}\right).
  \end{aligned}
  \label{eq:perpendicular electron pressure}
\end{equation}
\revAB{Following a derivation analogous to that of Eq.~(\ref{eq:iteration term 3}), the second term (the implicit part) on the right-hand side of Eq.~(\ref{eq:perpendicular electron pressure}) can be approximated by zero. 
This approximation does not affect the iterative form of Eq.~(\ref{eq:implicit perpendicular Ohm's law}).
Therefore, we have not expanded the expression for $p_{1,e\perp}^{n+1}$ in Eq.~(\ref{eq:implicit perpendicular Ohm's law}).
However, in the actual FIDES implementation, the implicit part of $p_{1,e\perp}^{n+1}$ is retained and computed iteratively. 
This is because the particle weights are updated after each field equation iteration,  and the particle moments including $p_{1,e\perp}^{n+1}$ are subsequently recalculated as a whole.}

\revAB{Substituting Eq.~(\ref{eq:iteration term 3}) into Eq.~(\ref{eq:implicit perpendicular Ohm's law}) and moving the difference between the discrete summation term and its approximation term to the right-hand side, we obtain the iterative form of the implicit perpendicular Ohm's law, }
\begin{equation}
  \begin{aligned}
  &\beta_{e}\mathbf{E}_{1 \perp}^{k+1}+\beta_{e}\Delta t\mathbf{E}_{1\perp}^{k+1}\times\mathbf{b}-{\Delta t} \mathbf{b} \times\left(\nabla \times \nabla \times \mathbf{E}_1^{k+1}\right)\\
  &=-\mathbf{b} \times\left(\nabla \times \mathbf{B}_1^n\right)-\beta_{e}\nabla_{\perp} p_{1,e\perp }^{n+1}-\beta_{e}\mathbf{J}_{i \perp}^* \times \mathbf{b}-\beta_{e}\Delta t\mathbf{b}\times\left\{\mathbf{E}_{1\perp}^{k}-\frac{1}{N_p} \frac{T_e}{T_i} \sum_j \mathbf{v}_{ij \perp}^{n+1} \left[\mathbf{v}_{ij \perp}^{n+1} \cdot \mathbf{E}_{1 \perp}^{k}\left(\mathbf{x}_{ij}^{n+1}\right)\right] S\left(\mathbf{x}_g-\mathbf{x}_{ij}^{n+1}\right)\right\},
  \label{eq:iterative implicit perpendicular Ohm's law}
  \end{aligned}
\end{equation}
\revAB{where the superscript of the the electric field has been changed from time step $n+1$ to iteration indices $k$ and $k+1$.
The iteration process of Eq.~(\ref{eq:iterative implicit perpendicular Ohm's law}) is similar to that of Eq.~(\ref{eq:iterative implicit ampere law}).
}

\revB{In FIDES, the implicit parallel Ampere's law (\ref{eq:iterative implicit ampere law}) and the implicit perpendicular Ohm's law (\ref{eq:iterative implicit perpendicular Ohm's law}) are solved simultaneously as a coupled system to obtain the electric field.
The numerical method employed depends on the complexity of the equilibrium profile.
For problems with a uniform equilibrium, a spectral method is used. 
For cases where equilibrium non-uniformities are restricted to the x direction, we apply a hybrid approach that combines a finite difference discretization in x with a Fourier decomposition in the y and z directions.
For more complex problems, the dimension of the coefficient matrix becomes large. 
In such cases, cost-efficient methods are required, such as precomputing and storing an LU decomposition or employing a Krylov subspace method.
In this paper, for the purpose of demonstration, we present the spectral method for solving Eq.~(\ref{eq:iterative implicit ampere law}) and Eq.~(\ref{eq:iterative implicit perpendicular Ohm's law}).
After Fourier transformation in x, y and z directions, the field equations become
}
\begin{equation}
  \left[\begin{array}{lll}
  D_{x x} & D_{x y} & D_{x z} \\
  D_{y x} & D_{y y} & D_{y z} \\
  D_{z x} & D_{z y} & D_{z z}
  \end{array}\right]\left[\begin{array}{l}
  \tilde{E}_{1 x}^{k+1}\left(\mathbf{k}\right) \\
  \tilde{E}_{1 y}^{k+1}\left(\mathbf{k}\right)\\
  \tilde{E}_{1 z}^{k+1}\left(\mathbf{k}\right)
  \end{array}\right]=\left[\begin{array}{l}
  \tilde{r}_x\left(\mathbf{k}\right) \\
  \tilde{r}_y\left(\mathbf{k}\right)\\
  \tilde{r}_z\left(\mathbf{k}\right)
  \end{array}\right],
  \label{eq:field equations fourier form}
\end{equation}
\revB{where elements of the coefficient matrix are}
\begin{equation}
  \begin{aligned}
  &D_{xx}=\beta_{e}-\Delta tk_{x}k_{y},\\
  &D_{xy}=\Delta t\left(\beta_{e}+{k_{x}^{2}+k_{z}^{2}}\right),\\
  &D_{xz}=-\Delta tk_{y}k_{z},\\
  &D_{yx}=-\Delta t\left(\beta_{e}+{k_{y}^{2}+k_{z}^{2}}\right),\\
  &D_{yy}=\beta_{e}+\Delta tk_{x}k_{y},\\
  &D_{yz}=\Delta tk_{x}k_{z},\\
  &D_{zx}=-\Delta tk_{x}k_{z},\\
  &D_{zy}=-\Delta tk_{y}k_{z},\\
  &D_{zz}=\Delta t\beta_{e}\frac{m_{i}}{m_{e}}+\Delta t\left(k_{x}^{2}+k_{y}^{2}\right).
  \label{eq:coeffients of Ohm's law}
  \end{aligned}
\end{equation}
\revB{On the right-hand side of Eq.~(\ref{eq:field equations fourier form}), $\tilde{r}_{x}\left(\mathbf{k}\right), \tilde{r}_{y}\left(\mathbf{k}\right)$ and $\tilde{r}_{z}\left(\mathbf{k}\right)$ can be expressed as }
\begin{equation}
  \begin{aligned}
    \tilde{r}_{x}\left(\mathbf{k}\right)=&-ik_{x}\tilde{B}_{1z}^{n}\left(\mathbf{k}\right)+ik_{z}\tilde{B}_{1x}^{n}\left(\mathbf{k}\right)-ik_{x}\beta_{e}\tilde{p}_{1,e\perp}^{n+1}\left(\mathbf{k}\right)-\beta_{e}\tilde{J}_{iy}^{*}\left(\mathbf{k}\right)+\beta_{e}\Delta t \tilde{E}_{1y}^{k}\left(\mathbf{k}\right)\\
    &-F\left\{\frac{\beta_{e}\Delta t}{N_{p}}\frac{T_{e}}{T_{i}}\sum_{j}v_{ijy}^{n+1}\left[\mathbf{v}_{ij\perp}^{n+1}\cdot\mathbf{E}_{1\perp}^{k}\left(\mathbf{x}_{ij}^{n+1}\right)\right]S\left(\mathbf{x}_{g}-\mathbf{x}_{ij}^{n+1}\right)\right\},\\
    \tilde{r}_{y}\left(\mathbf{k}\right)=&-ik_{y}\tilde{B}_{1z}^{n}\left(\mathbf{k}\right)+ik_{z}\tilde{B}_{1y}^{n}\left(\mathbf{k}\right)-ik_{y}\beta_{e}\tilde{p}_{1,e\perp}^{n+1}\left(\mathbf{k}\right)+\beta_{e}\tilde{J}_{ix}^{*}\left(\mathbf{k}\right)-\beta_{e}\Delta t\tilde{E}_{1x}^{n}\left(\mathbf{k}\right)\\
    &+F\left\{\frac{\beta_{e}\Delta t}{N_{p}}\frac{T_{e}}{T_{i}}\sum_{j}v_{ijx}^{n+1}\left[\mathbf{v}_{ij\perp}^{n+1}\cdot\mathbf{E}_{1\perp}^{k}\left(\mathbf{x}_{ij}^{n+1}\right)\right]S\left(\mathbf{x}_{g}-\mathbf{x}_{ij}^{n+1}\right)\right\},\\
    \tilde{r}_{z}\left(\mathbf{k}\right)=&ik_{x}\tilde{B}_{1y}^{n}\left(\mathbf{k}\right)-ik_{y}\tilde{B}_{1x}^{n}\left(\mathbf{k}\right)-\beta_{e}\tilde{J}_{e\|}^{*}\left(\mathbf{k}\right)-\beta_{e}\tilde{J}_{i\|}^{n+1}\left(\mathbf{k}\right)+\beta_{e}\Delta t\frac{m_{i}}{m_{e}}\tilde{E}_{1\|}^{n}\left(\mathbf{k}\right)\\
    &-F\left\{\frac{\beta_{e}\Delta t}{N_{p}}\sum_j\left(v_{e j \|}^{n+1}\right)^2 E_{1 \|}^k\left(\mathbf{x}_{e j}^{n+1}\right) S\left(\mathbf{x}_g-\mathbf{x}_{e j}^{n+1}\right)\right\},
  \end{aligned}
  \label{eq:rhs of field equations}
\end{equation}
\revB{where $\tilde{\left(\cdot\right)}$ and $F\left\{\cdot\right\}$ denote Fourier transformed quantities and $\mathbf{k}=k_{x}\hat{\mathbf{x}}+k_{y}\hat{\mathbf{y}}+k_{z}\hat{\mathbf{z}}$ is the wavenumber.
At each field iteration, the updated electric field $\tilde{\mathbf{E}}_{1}^{k+1}\left(\mathbf{k}\right)$ is obtained by solving Eq.~(\ref{eq:field equations fourier form}).
This Fourier-space solution is then filtered and transformed back to real space via an inverse Fourier transform, yielding $\mathbf{E}_{1}^{k+1}\left(\mathbf{x}_{g}\right)$ at the grid points.
}

After calculating the electric field, the Faraday's law is used to update the magnetic field,
\begin{equation}
\mathbf{B}_1^{n+1}=\mathbf{B}_1^n-\left(\nabla \times \mathbf{E}_1^{n+1}\right) \Delta t.
\label{eq:faraday's law}
\end{equation}

In summary, the field evolution in the implicit discretization scheme is governed by two sets of equations. The electric field is solved using the implicit parallel Ampere's law (\ref{eq:implicit parallel Ampere's law}) and the implicit perpendicular Ohm's law (\ref{eq:implicit perpendicular Ohm's law}), while the magnetic field is advanced via Faraday's law (\ref{eq:faraday's law}). 
For particle advancement, the electron and ion weights are updated using the implicit $E_{\|}$ scheme (\ref{eq:electron weight pushing}) and the implicit $\mathbf{E}_{\perp}$ scheme (\ref{eq:ion weight pushing}), respectively.

\begin{figure}[h]
  \centering
  \includegraphics[width=0.5\textwidth]{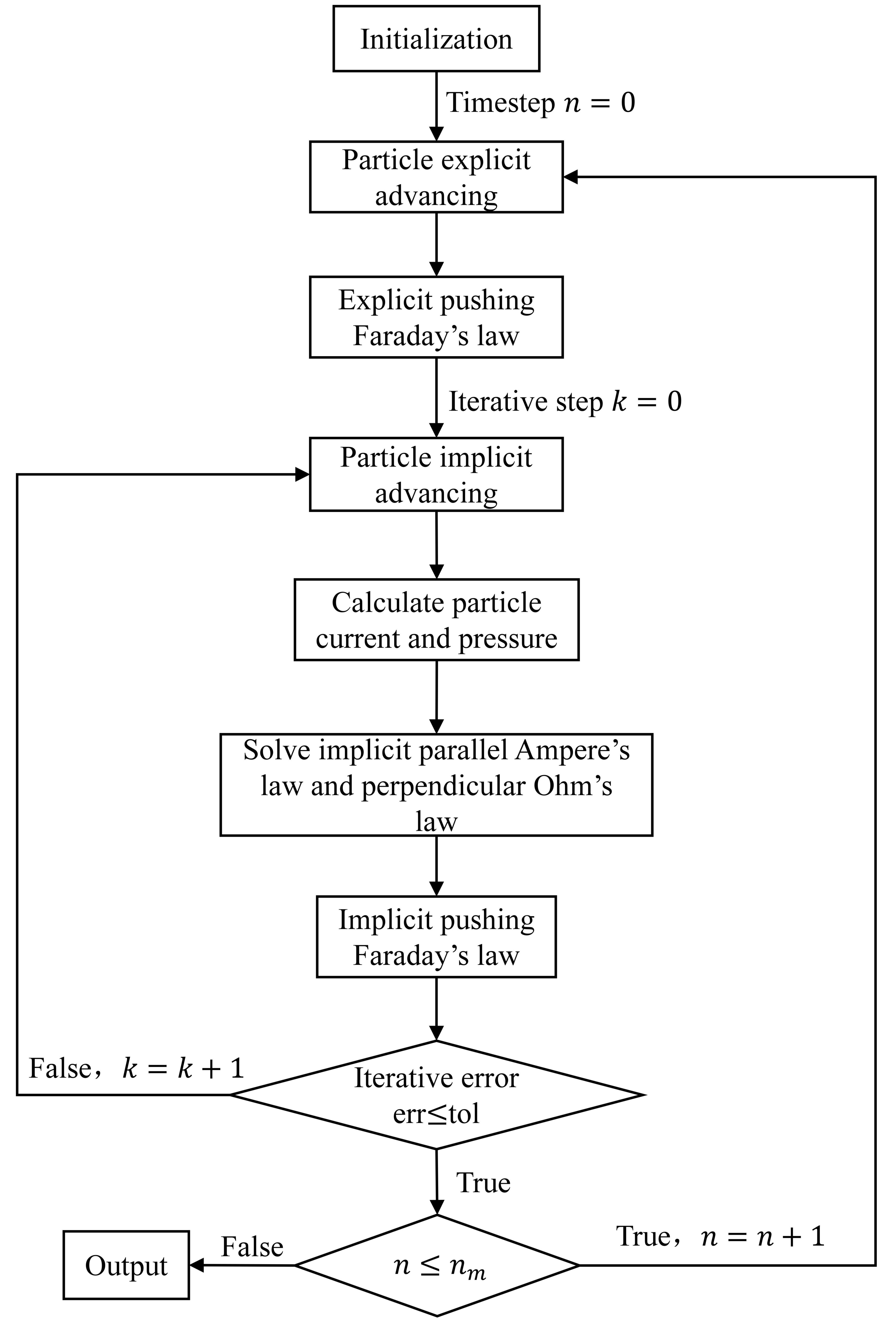}
  \caption{\label{fig: flowchart} The flowchart of FIDES algorithm.}
\end{figure}

\revAB{With the governing equations established, we can now specify the numerical algorithm of FIDES. The main steps in one time step are summarized below, and a corresponding flowchart is provided in Fig.~\ref{fig: flowchart},}

\revAB{1. Particle explicit advancing. }

\revAB{
In this step, the explicit parts of particle positions, velocities, and weights are computed. For linear simulations, only the equilibrium field contributes to the particle equations of motion; the perturbed field does not enter at this stage.
Since the equilibrium field is known from the initialization, particle positions and velocities at time step $t^{n+1}$ can be obtained directly.
The linear ion and electron motion equations in FIDES are given in Eqs.~(\ref{eq:linear ion motion}) and (\ref{eq:linear electron motion}), respectively,
}
\begin{equation}
  \begin{aligned}
  &\frac{\mathbf{x}^{n+1}_{i}-\mathbf{x}^{n}_{i}}{\Delta t}=\frac{\mathbf{v}_{i}^{n+1}+\mathbf{v}_{i}^{n}}{2},\\
  &\frac{\mathbf{v}_{i}^{n+1}-\mathbf{v}_{i}^{n}}{\Delta t}=\left(\frac{\mathbf{v}_{i}^{n+1}+\mathbf{v}_{i}^{n}}{2}\right)\times\mathbf{B}_{0}.
  \end{aligned}
  \label{eq:linear ion motion}
\end{equation}

\begin{equation}
  \begin{aligned}
    &\frac{\mathbf{x}^{n+1}_{e}-\mathbf{x}^{n}_{e}}{\Delta t}=\frac{v_{e\|}^{n+1}+v_{e\|}^{n}}{2}\hat{\mathbf{z}},\\
    &\frac{v_{e\|}^{n+1}-v_{e\|}^{n}}{\Delta t}=0.
  \end{aligned}
  \label{eq:linear electron motion}
\end{equation}

\revAB{For nonlinear simulations, the perturbed field affects particle motion and is divided in one explicit and one implicit steps.
In explicit advancing, the ion and electron motion equations are first pushed by the electromagnetic field at time step $t^{n}$, as shown by Eqs.~(\ref{eq:nonlinear ion motion 1}) and (\ref{eq:nonlinear electron motion 1}),}
\begin{equation}
  \begin{aligned}
  &\frac{\mathbf{x}^{*}_{i}-\mathbf{x}^{n}_{i}}{\Delta t/2}=\mathbf{v}_{i}^{n},\\
  &\frac{\mathbf{v}_{i}^{*}-\mathbf{v}_{i}^{n}}{\Delta t/2}=\mathbf{E}_{1}^{n}\left(\mathbf{x}_{i}^{n}\right)+\mathbf{v}_{i}^{n}\times\left[{\mathbf{B}_{0}+\mathbf{B}_{1}^{n}\left(\mathbf{x}_{i}^{n}\right)}\right],
  \end{aligned}
  \label{eq:nonlinear ion motion 1}
\end{equation}

\begin{equation}
  \begin{aligned}
  &\frac{\mathbf{x}^{*}_{e}-\mathbf{x}^{n}_{e}}{\Delta t/2}=v_{e\|}^{n}\hat{\mathbf{z}}+v_{e\|}^{n}\frac{\mathbf{B}_{1\perp}^{n}\left(\mathbf{x}_{e}^{n}\right)}{B_{0}}+\frac{\mathbf{E}_{1}^{n}\left(\mathbf{x}_{e}^{n}\right)\times\mathbf{b}}{B_{0}},\\
  &\frac{{v}_{e\|}^{*}-{v}_{e\|}^{n}}{\Delta t/2}=-\frac{m_{i}}{m_{e}}E_{1\|}^{n}\left(\mathbf{x}_{e}^{n}\right),
  \end{aligned}
  \label{eq:nonlinear electron motion 1}
\end{equation}
\revAB{where $\mathbf{x}^{*}_{i}$, $\mathbf{v}_{i}^{*}$, $\mathbf{x}_{e}^{*}$, and $v_{e\|}^{*}$ are intermediate particle positions and velocities.
The electron magnetic moment $\mu_{e}$ is a constant of motion and its equation is not shown here.}

\revAB{In particle explicit pushing, the electron and ion weights are advanced by the explicit equations of Eqs.~(\ref{eq:electron weight pushing}) and (\ref{eq:ion weight pushing}), which are rewritten here for clarity, }
\begin{equation}
  \begin{aligned}
  \frac{w_i^*-w_i^n}{\Delta t}&=\frac{T_e}{T_i} v_{i \|}^n E_{1 \|}^n\left(\mathbf{x}_{i}^{n}\right)-\left[E_{1 y}^n\left(\mathbf{x}_{i}^{n}\right)+v_{iz}^{n} B_{1 x}^n\left(\mathbf{x}_{i}^{n}\right)-v_{ix}^{n} B_{1 z}^n\left(\mathbf{x}_{i}^{n}\right)\right]\left\{\frac{\partial \ln n_i}{\partial x}+\left[\frac{T_e \left(v_i^n\right)^2}{2 T_i}-\frac{3}{2}\right] \frac{\partial \ln T_i}{\partial x}\right\},\\
  \frac{w_e^*-w_e^n}{\Delta t} & =-\left\{\frac{\partial \ln n_e}{\partial x}+\left[\frac{m_{e}\left(v_{e\|}^{n}\right)^{2}}{2m_{i}}+\mu_{e} B_{0}-\frac{3}{2}\right] \frac{\partial \ln T_e}{\partial x}\right\}\left[{E_{1 y}^n}\left(\mathbf{x}_{e}^{n}\right)+v_{e \|}^n {B_{1 x}^n}\left(\mathbf{x}_{e}^{n}\right)\right]-\mu_{e} \mathbf{b}\cdot\nabla\times\mathbf{E}_{1}^{n}\left(\mathbf{x}_{e}^{n}\right). \\
  \end{aligned}
  \label{eq: weight explicit pushing}
\end{equation}

\revAB{2. Explicit pushing Faraday's law}

\revAB{If Faraday's law is advanced using the second-order scheme discussed in section 4, the magnetic field is first updated to an intermediate value $\mathbf{B}_{1}^{*}$ via}
\begin{equation}
  \begin{aligned}
    \mathbf{B}^{*}_{1}=\mathbf{B}_{1}^{n}-\left(\Delta t/2\right)\nabla\times\mathbf{E}_{1}^{n},
  \end{aligned}
  \label{eq:explicit faraday's law}
\end{equation}
\revAB{where $\mathbf{B}_{1}^{*}$ denotes the intermediate perturbed magnetic field.
If Faraday's law is advanced with the first-order implicit scheme of Eq.~(\ref{eq:faraday's law}), this step is omitted, and we simply set $\mathbf{B}_{1}^{*}=\mathbf{B}_{1}^{n}$.
}

\revAB{3. Iteration loop}

\revAB{In the actual FIDES implementation, particle weights, nonlinear particle motion, and the perturbed electromagnetic fields are all updated iteratively.
The iterative perturbed electric field $\mathbf{E}_{1}^{k}$ plays a central role in the iteration loop, as all other iterative quantities depend on it.
For the initial guess, we set $\mathbf{E}_{1}^{k}=\mathbf{E}_{1}^{n}$ and $\mathbf{B}_{1}^{k}=\mathbf{B}_{1}^{*}$.  
For nonlinear simulations, we additionally set
$\mathbf{x}_{i}^{k}=\mathbf{x}_{i}^{*}$, $\mathbf{v}_{i}^{k}=\mathbf{v}_{i}^{*}$, $\mathbf{x}_{e}^{k}=\mathbf{x}_{e}^{*}$, $v_{e\|}^{k}=v_{e\|}^{*}$ for $k=0$.
}

\revAB{(1) Particle implicit advancing}

\revAB{For linear simulations, particle positions and velocities at time step $t^{n+1}$ have already been obtained by Eq.~(\ref{eq:linear ion motion}) and Eq.~(\ref{eq:linear electron motion}). While for nonlinear simulations, $\mathbf{x}_{i}^{k+1}$, $\mathbf{v}_{i}^{k+1}$, $\mathbf{x}_{e}^{k+1}$ and $v_{e\|}^{k+1}$ are modified by the perturbed electromagnetic fields $\mathbf{E}_{1}^{k}$ and $\mathbf{B}_{1}^{k}$,}
\begin{equation}
  \begin{aligned}
  &\frac{\mathbf{x}^{k+1}_{i}-\mathbf{x}^{*}_{i}}{\Delta t/2}=\mathbf{v}_{i}^{k},\\
  &\frac{\mathbf{v}_{i}^{k+1}-\mathbf{v}_{i}^{*}}{\Delta t/2}=\mathbf{E}_{1}^{k}\left(\mathbf{x}_{i}^{k}\right)+\mathbf{v}_{i}^{k}\times\left[{\mathbf{B}_{0}+\mathbf{B}_{1}^{k}\left(\mathbf{x}_{i}^{k}\right)}\right],
  \end{aligned}
  \label{eq:nonlinear ion motion 2}
\end{equation}

\begin{equation}
  \begin{aligned}
  &\frac{\mathbf{x}^{k+1}_{e}-\mathbf{x}^{*}_{e}}{\Delta t/2}=v_{e\|}^{k}\hat{\mathbf{z}}+v_{e\|}^{k}\frac{\mathbf{B}_{1\perp}^{k}\left(\mathbf{x}_{e}^{k}\right)}{B_{0}}+\frac{\mathbf{E}_{1}^{k}\left(\mathbf{x}_{e}^{k}\right)\times\mathbf{b}}{B_{0}},\\
  &\frac{{v}_{e\|}^{k+1}-{v}_{e\|}^{*}}{\Delta t/2}=-\frac{m_{i}}{m_{e}}E_{1\|}^{k}\left(\mathbf{x}_{e}^{k}\right).
  \end{aligned}
  \label{eq:nonlinear electron motion 2}
\end{equation}
\revAB{When the iteration converges, Eq.~(\ref{eq:nonlinear ion motion 1}) and (\ref{eq:nonlinear ion motion 2}),  Eq.~(\ref{eq:nonlinear electron motion 1}) and (\ref{eq:nonlinear electron motion 2}) constitute the second-order semi-implicit pushing schemes for ion and electron motion, respectively.}

\revAB{More importantly, for both linear and nonlinear simulations, the electron and ion weights are updated by the implicit equations of Eqs.~(\ref{eq:electron weight pushing}) and (\ref{eq:ion weight pushing}), which are rewritten here for clarity,}
\begin{equation}
  \begin{aligned}
    &\frac{w_i^{k+1}-w_i^*}{\Delta t}=\frac{T_e}{T_i} \mathbf{v}_{i \perp}^{k+1} \cdot \mathbf{E}_{1 \perp}^{k}\left(\mathbf{x}_{i}^{k+1}\right),\\
    &\frac{w_e^{k+1}-w_e^*}{\Delta t}=-v_{e \|}^{k+1} E_{1 \|}^{k}\left(\mathbf{x}_{e}^{k+1}\right),\\
  \end{aligned}
  \label{eq:implicit weight pushing}
\end{equation}
\revAB{with $\mathbf{x}_{i}^{k+1}=\mathbf{x}_{i}^{n+1}$, $\mathbf{v}_{i}^{k+1}=\mathbf{v}_{i}^{n+1}$, $\mathbf{x}_{e}^{k+1}=\mathbf{x}_{e}^{n+1}$ and $v_{e\|}^{k+1}=v_{e\|}^{n+1}$ for linear simulations.}

\revAB{(2) Calculate particle current and pressure}

\revAB{In the actual FIDES code, the particle moments are computed directly as a whole, rather than being split into explicit and implicit parts, }
\begin{equation}
  \begin{aligned}
    &\mathbf{J}_{i}^{k+1}\left(\mathbf{x}_{g}\right)=\frac{1}{N_{p}}\sum_{j}\mathbf{v}_{ij}^{k+1}w_{ij}^{k+1}S\left(\mathbf{x}_{g}-\mathbf{x}_{ij}^{k+1}\right),\\
    &J_{e\|}^{k+1}\left(\mathbf{x}_{g}\right)=-\frac{1}{N_{p}}\sum_{j}v_{ej\|}^{k+1}w_{ej}^{k+1}S\left(\mathbf{x}_{g}-\mathbf{x}_{ej}^{k+1}\right),\\
    &p_{1,e\perp}^{k+1}\left(\mathbf{x}_{g}\right)=\frac{1}{N_{p}}\sum_{j}\mu_{ej}B_{0}w_{ej}^{k+1}S\left(\mathbf{x}_{g}-\mathbf{x}_{ej}^{k+1}\right),
    \end{aligned}
    \label{eq:particle moment}
\end{equation}
\revAB{where the superscripts of particle moments have been changed from time step $n+1$ to iteration index $k+1$. At first glance, Eq.~(\ref{eq:particle moment}) may appear different from the earlier expressions in Eqs.~(\ref{eq:electron current}), (\ref{eq:ion perpendicular current}) and (\ref{eq:perpendicular electron pressure}).
In fact, they are equivalent for solving field equations (\ref{eq:iterative implicit ampere law}) and (\ref{eq:iterative implicit perpendicular Ohm's law}).
The earlier equations are introduced to theoretically illustrate the treatment of the implicit terms by separating them from the explicit parts.
In the code, however, it is more convenient and structurally clearer to compute the particle moments as unified quantities without such a split.
This distinction reflects the difference between the theoretical exposition and the numerical implementation.
}

\revAB{(3) Solve implicit parallel Ampere's law and perpendicular Ohm's law}

\revAB{Given $\mathbf{E}_{1}^{k}$, $\mathbf{B}_{1}^{n}$, and particle moments $\mathbf{J}_{i}^{k+1}$, $J_{e\|}^{k+1}$, and $p_{1,e\perp}^{k+1}$ computed from Eq.~(\ref{eq:particle moment}), the updated electric field $\mathbf{E}_{1}^{k+1}$ is obtained by solving the coupled system consisting of the implicit parallel Ampere's law and perpendicular Ohm's law,}
\begin{equation}
  \begin{aligned}
  \Delta t \mathbf{b} \cdot \nabla \times \nabla \times \mathbf{E}_1^{k+1}+\Delta t \beta_e \frac{m_i}{m_e} E_{1 \|}^{k+1}=&\mathbf{b} \cdot \nabla \times \mathbf{B}_1^n-\beta_e\left(J_{e \|}^{k+1}+J_{i \|}^{k+1}\right)+\beta_e \Delta t\frac{m_i}{m_e} E_{1 \|}^k,
  \label{eq:numerical implicit ampere law}
  \end{aligned}
\end{equation}

\begin{equation}
  \begin{aligned}
  &\beta_{e}\mathbf{E}_{1 \perp}^{k+1}+\beta_{e}\Delta t\mathbf{E}_{1\perp}^{k+1}\times\mathbf{b}-{\Delta t} \mathbf{b} \times\left(\nabla \times \nabla \times \mathbf{E}_1^{k+1}\right)\\
  &=-\mathbf{b} \times\left(\nabla \times \mathbf{B}_1^n\right)-\beta_{e}\nabla_{\perp} p_{1,e\perp }^{k+1}-\beta_{e}\mathbf{J}_{i \perp}^{k+1} \times \mathbf{b}+\beta_{e}\Delta t\mathbf{E}_{1\perp}^{k}\times\mathbf{b},
  \label{eq:numerical implicit perpendicular Ohm's law}
  \end{aligned}
\end{equation}
\revAB{which correspond to the numerical versions in FIDES of Eqs.~(\ref{eq:iterative implicit ampere law}) and (\ref{eq:iterative implicit perpendicular Ohm's law}), respectively.
The iteration logic and solution procedure of above equations have been discussed in details by Eqs.~(\ref{eq:implicit parallel Ampere's law})-(\ref{eq:iterative implicit ampere law}) and (\ref{eq:implicit perpendicular Ohm's law})-(\ref{eq:rhs of field equations}).}

\revAB{(4) Implicit pushing Faraday's law}

\revAB{For the first-order implicit scheme, the updated magnetic field $\mathbf{B}_{1}^{k+1}$ is obtained by}
\begin{equation}
  \mathbf{B}_{1}^{k+1}=\mathbf{B}_{1}^{n}-\Delta t\nabla \times\mathbf{E}_{1}^{k+1}.
  \label{eq:implicit faraday's law 1}
\end{equation}

\revAB{For the second-order semi-implicit scheme (see section 4), this step advances the magnetic field in Eq.~(\ref{eq:explicit faraday's law}), }
\begin{equation}
  \mathbf{B}_{1}^{k+1}=\mathbf{B}_{1}^{*}-\left(\Delta t/2\right)\nabla \times\mathbf{E}_{1}^{k+1}.
  \label{eq:implicit faraday's law 2}
\end{equation}

\revAB{The iteration convergence criterion in FIDES is based on the relative change of the electric field between successive iterates. 
Specifically, convergence is declared when $\|\mathbf{E}_{1d}^{k+1}-\mathbf{E}_{1d}^{k}\|_{2}/\|\mathbf{E}_{1d}^{k}\|_{2}<\text{tol}$, $(d=x,~y,~z)$. The tolerance is typically set to $\text{tol}=10^{-4}$.
Once the electric field $\mathbf{E}_{1}^{k+1}$ has converged, the perturbed magnetic field $\mathbf{B}_{1}^{k+1}$ is also converged, as can be seen in Eq.~(\ref{eq:implicit faraday's law 1}) or (\ref{eq:implicit faraday's law 2}).
Furthermore, the particle positions, velocities, and weights are driven to convergence accordingly, as they depend on the converged fields through Eqs.~(\ref{eq:nonlinear ion motion 2}), (\ref{eq:nonlinear electron motion 2}), and (\ref{eq:implicit weight pushing}).
}

\subsection{Explicit discretization scheme}

\revB{The conventional models cast Ampere's law into the generalized Ohm's law, which serves as the field equation for the electric field \cite{ChenY2009,ChengJ2013, ChenH2021,ChenH2023}.
While all of these works employ the generalized Ohm's law, they differ in the underlying particle models and implementation details.
Specifically, the simulation model established by Chen and Parker \cite{ChenY2009} combines Vlasov ions with drift kinetic electrons, advancing ion weights via an implicit $\mathbf{E}_{\perp}$ scheme and electron weights via an explicit scheme.
Cheng et al. \cite{ChengJ2013} adopt a full-kinetic ion model but use a fluid electron model (which oversimplifies electron kinetics) while employing a second-order semi-implicit scheme for ion pushing.
The GK-E$\&$B model \cite{ChenH2021,ChenH2023} is developed based on gyrokinetic models for both ions and electrons.
Among these, the method presented in \cite{ChenY2009} is the baseline method that we aim to improve.}

\revB{To establish a basis for comparison, we first summarize the main numerical procedure of the baseline method \cite{ChenY2009}.
A key distinction between the baseline method and our new algorithm lies in the treatment of electron weights. 
In the baseline method, the electron weight is advanced using an explicit scheme,}
\begin{equation}
  \begin{aligned}
  \frac{w_e^{n+1}-w_e^n}{\Delta t} =-\left\{\frac{\partial \ln n_e}{\partial x}+\left[\frac{m_{e}\left(v_{e\|}^{n}\right)^{2}}{2m_{i}}+\mu_{e} B_{0}-\frac{3}{2}\right] \frac{\partial \ln T_e}{\partial x}\right\}\left({E_{1 y}^n}+v_{e \|}^n {B_{1 x}^n}\right)-\mu_{e} \mathbf{b}\cdot\nabla\times\mathbf{E}_{1}^{n}-v_{e \|}^{n} E_{1 \|}^{n},
  \label{eq:explicit electron weight pushing}
  \end{aligned}
\end{equation}
\revB{whereas our method employs an implicit $E_{\|}$ scheme Eq.~(\ref{eq:electron weight pushing}).
This seemingly minor difference has profound implications for the overall algorithm structure.
In the baseline method, the electron weight is fully updated during particle explicit advancing, which makes the direct use of the parallel Ampere's law as a field equation ill-posed.
Instead, the parallel Ohm's law is used to close the system.}
\revAB{In the shearless slab geometry, the parallel Ohm's law takes the form,}
\begin{equation}
\begin{aligned}
&\left(1+\frac{m_e}{m_i}\right) E_{1 \|}^{n+1}+\frac{m_e}{m_i} \frac{1}{\beta_e} \mathbf{b} \cdot \nabla \times \nabla \times \mathbf{E}_1^{n+1}-\Delta t\left(\nabla \times \mathbf{E}_1^{n+1}\right) \cdot \left(\nabla p_{0,e\perp}-\frac{m_{e}}{m_{i}}\nabla p_{0,i\perp}\right)\\
&=-\nabla_{\|} p_{1,e\|}^{n+1}+\frac{m_e}{m_i} \nabla_{\|} p_{1,i\|}^{n+1}-\mathbf{B}_1^n \cdot \left(\nabla p_{0,e\perp}-\frac{m_{e}}{m_{i}}\nabla p_{0,i\perp}\right),
\end{aligned}
\label{eq:parallel Ohm's law}
\end{equation}
\revAB{where $p_{1,e\|}^{n+1}$ and $p_{1,i\|}^{n+1}$ are perturbed electron and ion parallel pressure, respectively,}
\begin{equation}
  \begin{aligned}
    &p_{1,e\|}^{n+1}\left(\mathbf{x}_{g}\right)=\frac{1}{N_{p}}\sum_{j}\frac{m_{e}}{m_{i}}v_{ej\|}^{2}w_{ej}^{n+1}S\left(\mathbf{x}_{g}-\mathbf{x}_{ej}^{n+1}\right),\\
    &p_{1,i\|}^{n+1}\left(\mathbf{x}_{g}\right)=\frac{1}{N_{p}}\sum_{j}v_{ij\|}^{2}w_{ij}^{n+1}S\left(\mathbf{x}_{g}-\mathbf{x}_{ij}^{n+1}\right).
  \end{aligned}
\end{equation}

\revB{To clearly distinguish the two schemes, we refer to the baseline method as the explicit discretization scheme and our new algorithm as the implicit discretization scheme, reflecting the distinct way the electron response to $E_{\|}$ is handled.}
\revAB{The treatment of ion pushing (\ref{eq:ion weight pushing}), the perpendicular Ohm's law (\ref{eq:implicit perpendicular Ohm's law}), and the magnetic field update equation (\ref{eq:faraday's law}) are all identical for the two schemes.}

\subsection{\revB{Origin of the cancellation problem}}

\revB{In the parallel Ohm's law (\ref{eq:parallel Ohm's law}), the ion contribution is of order $m_{e}/m_{i}$ relative to the electron contribution.
If ion terms are neglected entirely, the resulting simplified form is}
\begin{equation}
\begin{aligned}
  E_{1 \|}^{n+1}+\frac{m_e}{m_i} \frac{1}{\beta_e} \mathbf{b} \cdot \nabla \times \nabla \times \mathbf{E}_1^{n+1}-\Delta t\left(\nabla \times \mathbf{E}_1^{n+1}\right) \cdot \nabla p_{0,e\perp}=-\nabla_{\|} p_{1,e\|}^{n+1}-\mathbf{B}_1^n \cdot \nabla p_{0,e\perp},
\end{aligned}
\label{eq:parallel Ohm's law without ion terms}
\end{equation}
\revB{which fails to capture key physical processes such as the ion acoustic wave (IAW) and the ion temperature gradient instability (ITG), where ion parallel dynamics plays an essential role \cite{ChenY2009}.}

\revB{
When the ion terms are retained, numerical simulations face a different difficulty: the perturbed electron pressure $p_{1,e\|}$ contains numerical noise, which can overwhelm the small ion contribution, making it difficult to accurately compute the ion response.
This problem becomes particularly severe in the regime $k_{\|}v_{te} \gg \omega$, where electrons behave adiabatically \cite{ChenY2009, ChenH2021, ChenH2023}.
In this limit, the perturbed electron distribution can be written as $f_{e1} = e\phi f_{e0}/T_e+(f_{e1})_{na}$, where $\phi$ is the electric potential, and $(f_{e1})_{na}$ is the nonadiabatic part, which is much smaller than the adiabatic part $e\phi f_{e0}/T_e$.
}

\revB{
Theoretically, in the electrostatic limit, the adiabatic part of electron pressure $-\nabla_{\|} \left(p_{1,e\|}^{n+1}\right)_{a}=-\nabla_{\|}\phi$ cancels with the term $E_{1\|}^{n+1}$ in Eq.~(\ref{eq:parallel Ohm's law}). 
After this leading-order cancellation, the governing dynamics appear at the next order, i.e., terms of $O({m_{e}/m_{i}})$ and $O(k_{\perp}^{2}\rho_{s}^{2}m_e/(m_i\beta_{e}))$, leading to the following reduced equation,}
\begin{equation}
\begin{aligned}
&\frac{m_e}{m_i}E_{1 \|}^{n+1}+\frac{m_e}{m_i} \frac{1}{\beta_e} \mathbf{b} \cdot \nabla \times \nabla \times \mathbf{E}_1^{n+1}-\Delta t\left(\nabla \times \mathbf{E}_1^{n+1}\right) \cdot \left(\nabla p_{0,e\perp}-\frac{m_{e}}{m_{i}}\nabla p_{0,i\perp}\right)\\
&=-\nabla_{\|} \left(p_{1,e\|}^{n+1}\right)_{na}+\frac{m_e}{m_i} \nabla_{\|} p_{1,i\|}^{n+1}-\mathbf{B}_1^n \cdot \left(\nabla p_{0,e\perp}-\frac{m_{e}}{m_{i}}\nabla p_{0,i\perp}\right),
\end{aligned}
\label{eq:parallel Ohm's law after cancellation}
\end{equation}
\revB{where $\left(p_{1,e\|}^{n+1}\right)_{na}$ is the nonadiabatic part of the electron pressure. The accurate dispersion relation for $k_{\|}v_{te} \gg \omega$ can be derived from Eq.~(\ref{eq:parallel Ohm's law after cancellation})}.

\begin{figure*}[t]
  \centering
  \subfigure[$E_{1\|}$ and $-\nabla_{\|}p_{1,e\|}$]
  {
    \begin{minipage}[b]{.45\textwidth}
      \centering
      \includegraphics[width=\textwidth]{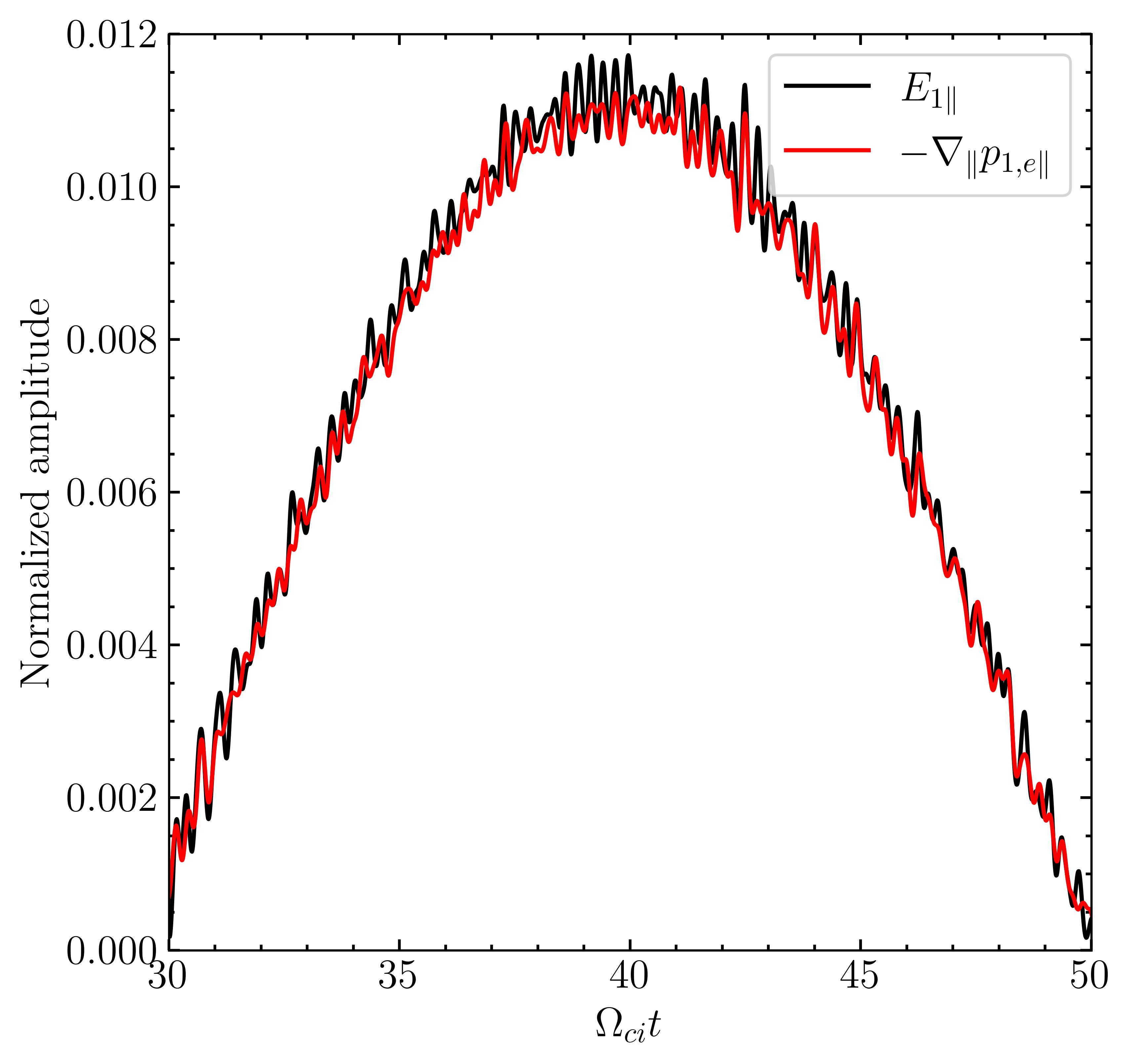}
    \end{minipage}
  }
  \subfigure[$E_{1\|}+\nabla_{\|}p_{1,e\|}$ and $\left({m_{e}}/{m_{i}}\right)\nabla_{\|}p_{1,i\|}$]
  {
    \begin{minipage}[b]{.45\textwidth}
      \centering
      \includegraphics[width=\textwidth]{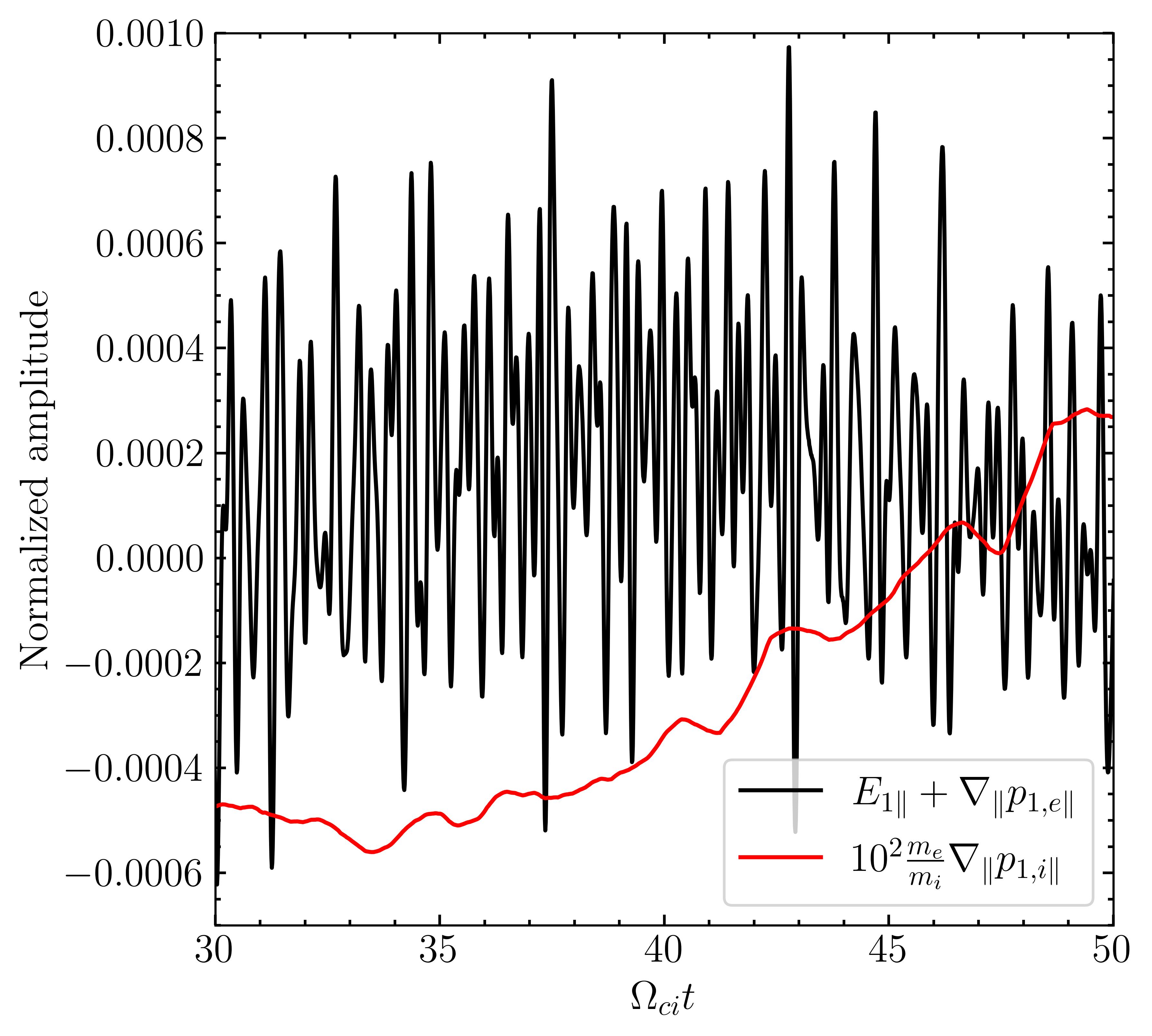}
    \end{minipage}
  }
  \caption{\label{fig:diagnostic plot} Diagnosis of the cancellation problem in the parallel Ohm's law, using the ion acoustic wave (IAW) simulation. (a) Time evolution of the two large terms, $\mathbf{E}_{1\|}$ and $-\nabla_{\|} p_{1,e\|}$, at a representative grid point. (b) Time evolution of the difference between $\mathbf{E}_{1\|}$ and $-\nabla_{\|} p_{1,e\|}$, and the physical ion contribution $\left({m_{e}}/{m_{i}}\right)\nabla_{\|}p_{1,i\|}$. 
  In Fig. (b), to avoid the ion signal being too weak, we amplify it by a factor of $10^{2}$, so the quantity actually plotted is $10^{2}\left(m_{e}/m_{i}\right)\nabla_{\|}p_{1,i\|}$.
  The simulation uses a grid of $n_{x}=n_{y}=2, n_{z}=32$ with $N_{p}=32$ particles per grid cell, a timestep $\Omega_{ci}\Delta t=0.01$, mass ratio $m_{i}/m_{e}=1836$, and the wave parameters $\beta_e=0.01,T_{i}/T_{e}=0.25,k_{x}\rho_{s}=k_{y}\rho_{s}=0,k_{z}\rho_{s}=0.1$.
  }
\end{figure*}

\revB{In practice, however, the numerically computed $-\nabla_{\|} p_{1, e\|}^{n+1}$ does not exactly cancel $E_{1\|}^{n+1}$ due to numerical noise.
This noise originates from the second-order velocity moment of the electron weights and the spatial derivative.
Furthermore, the adiabatic and nonadiabatic parts of the electron pressure cannot be distinguished in the simulation.
Take the IAW simulation as an example, as shown in Fig.~\ref{fig:diagnostic plot} (a), the two large terms in the parallel Ohm's law, $E_{1\|}^{n+1}$ and $-\nabla_{\|} p_{1,e\|}^{n+1}$, closely follow each other but suffer from high-frequency numerical noise.
Their difference, plotted in Fig.~\ref{fig:diagnostic plot} (b), masks the physical ion contribution $\left(m_{e}/m_{i}\right)\nabla_{\|}p_{1,i\|}^{n+1}$.
The residual imbalance between the two large terms can potentially produce spurious fields that dominate the true physical dynamics, leading to severe inaccuracies in the simulation results \cite{ChenY2009, ChenH2021, ChenH2023}. Overcoming the cancellation problem requires both small grid sizes and small timesteps.}

\revB{To ensure a more accurate numerical cancellation of the leading-order terms in the parallel Ohm's law, previous studies propose to introduce discrete particle and finite grid-size effects into the electric field term $E_{1\|}^{n+1}$ via the following approximation \cite{ChenY2011}}
\begin{equation}
 E_{1\|}^{n+1}\left(\mathbf{x}_{g}\right)\approx \frac{m_e}{m_i N_p} \sum_j E_{1 \|}^{n+1}\left(\mathbf{x}_{e j}^{n+1}\right) \left(v_{e j \|}^{n+1}\right)^{2} S\left(\mathbf{x}_g-\mathbf{x}_{ej}^{n+1}\right),
 \label{eq:e_para approximation}
\end{equation}
where $E_{1\|}^{n+1} \left(\mathbf{x}_{g}\right)$ and $E_{1\|}^{n+1} \left(\mathbf{x}_{ej}^{n+1}\right)$ denote the parallel electric field at the grid point and at the electron position, respectively.
\revB{Substituting Eq.~(\ref{eq:e_para approximation}) into the parallel Ohm's law (\ref{eq:parallel Ohm's law}), we obtain}
\begin{equation}
\begin{aligned}
  &\frac{m_e}{m_i N_p} \sum_j E_{1 \|}^{n+1}\left(\mathbf{x}_{e j}^{n+1}\right) \left(v_{e j \|}^{n+1}\right)^{2} S\left(\mathbf{x}_g-\mathbf{x}_{ej}^{n+1}\right)+\frac{m_e}{m_i}E_{1 \|}^{n+1}+\frac{m_e}{m_i} \frac{1}{\beta_e} \mathbf{b} \cdot \nabla \times \nabla \times \mathbf{E}_1^{n+1}-\Delta t\left(\nabla \times \mathbf{E}_1^{n+1}\right) \cdot \left(\nabla p_{0,e\perp}-\frac{m_{e}}{m_{i}}\nabla p_{0,i\perp}\right)\\
  &=-\nabla_{\|} p_{1,e\|}^{n+1}+\frac{m_e}{m_i} \nabla_{\|} p_{1,i\|}^{n+1}-\mathbf{B}_1^n \cdot \left(\nabla p_{0,e\perp}-\frac{m_{e}}{m_{i}}\nabla p_{0,i\perp}\right).
\end{aligned}
\label{eq:numerical parallel Ohm's law}
\end{equation}
\revB{In practice, Eq.~(\ref{eq:numerical parallel Ohm's law}) is solved by the iterative method,} 
\begin{equation}
  \begin{aligned}
    &\left(1+\frac{m_e}{m_i}\right)E_{1 \|}^{k+1}+\frac{m_e}{m_i} \frac{1}{\beta_e} \mathbf{b} \cdot \nabla \times \nabla \times \mathbf{E}_1^{k+1}-\Delta t\left(\nabla \times \mathbf{E}_1^{k+1}\right) \cdot \left(\nabla p_{0,e\perp}-\frac{m_{e}}{m_{i}}\nabla p_{0,i\perp}\right)\\
    &=-\nabla_{\|} p_{1,e\|}^{n+1}+\frac{m_e}{m_i} \nabla_{\|} p_{1,i\|}^{n+1}-\mathbf{B}_1^n \cdot \left(\nabla p_{0,e\perp}-\frac{m_{e}}{m_{i}}\nabla p_{0,i\perp}\right)+\left[E_{1 \|}^{k}-\frac{m_e}{m_i N_p} \sum_j E_{1 \|}^{k}\left(\mathbf{x}_{e j}^{n+1}\right) \left(v_{e j \|}^{n+1}\right)^{2} S\left(\mathbf{x}_g-\mathbf{x}_{ej}^{n+1}\right)\right],
  \end{aligned}
  \label{eq:iterative parallel Ohm's law}
\end{equation}
\revB{where the iterative process is similar to Eqs.~(\ref{eq:iterative implicit ampere law}) and (\ref{eq:iterative implicit perpendicular Ohm's law}).
In the following simulations, the explicit discretization scheme uses Eq.~(\ref{eq:iterative parallel Ohm's law}), instead of Eq.~(\ref{eq:parallel Ohm's law}).
}

\section{Simulation examples}
\label{sec:simulation examples}
Our testing of the two schemes encompasses both high- and low-frequency physics.
Simulations of perpendicular and parallel waves evaluate their performance in high-frequency regimes, whereas simulations of low-frequency IAW and ITG are designed to compare their ability to overcome the cancellation problem. 
\revA{The mass ratio used in simulations is $m_{i}/m_{e}=1836$.}
                                                   
\subsection{Perpendicular and parallel waves}

\begin{figure}[h]
  \centering
  \includegraphics[width=0.5\textwidth]{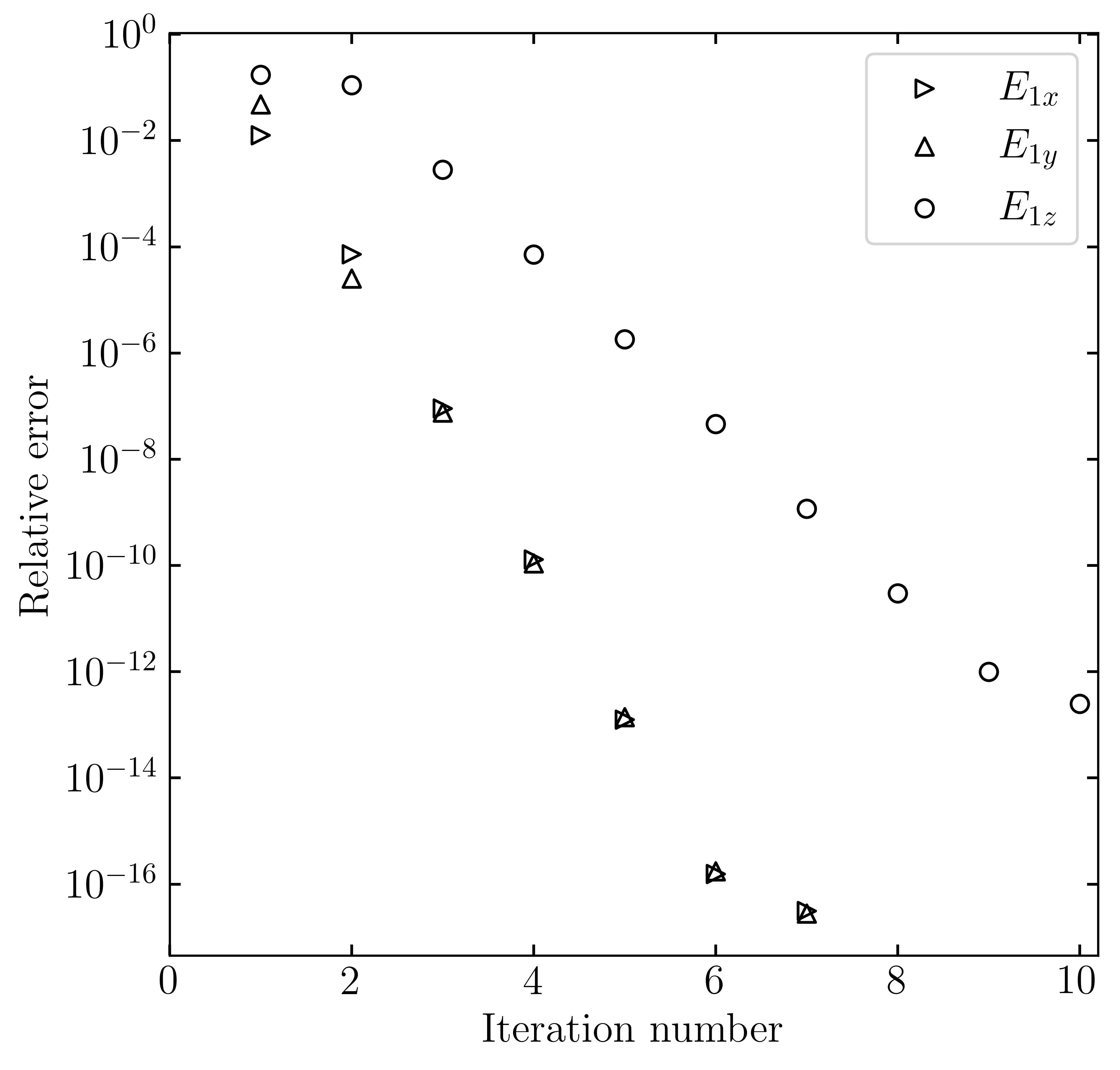}
  \caption{\label{fig: iterative_solver} Convergence test for the iterative solver of the implicit parallel Ampere's law (\ref{eq:iterative implicit ampere law}) and perpendicular Ohm's law (\ref{eq:iterative implicit perpendicular Ohm's law}). 
  The relative error at iteration $k$ is defined as $\|\mathbf{E}_{1d}^{k+1}-\mathbf{E}_{1d}^{k}\|_{2}/\|\mathbf{E}_{1d}^{k}\|_{2}$, $(d=x,~y,~z)$.
  The simulation uses a grid of $n_{x}=n_{y}=n_{z}=16$ with $N_{p}=16$ particles per grid cell, a timestep $\Omega_{ci}\Delta t=0.05$, and the wave parameters $\beta_e=0.05,T_{i}/T_{e}=1,k_{x}\rho_{s}=0.1,k_{y}\rho_{s}=k_{z}\rho_{s}=0$. }
\end{figure}

\begin{figure*}[t]
  \centering
  \subfigure[Perpendicular wave]
  {
    \begin{minipage}[b]{.45\textwidth}
      \centering
      \includegraphics[width=\textwidth]{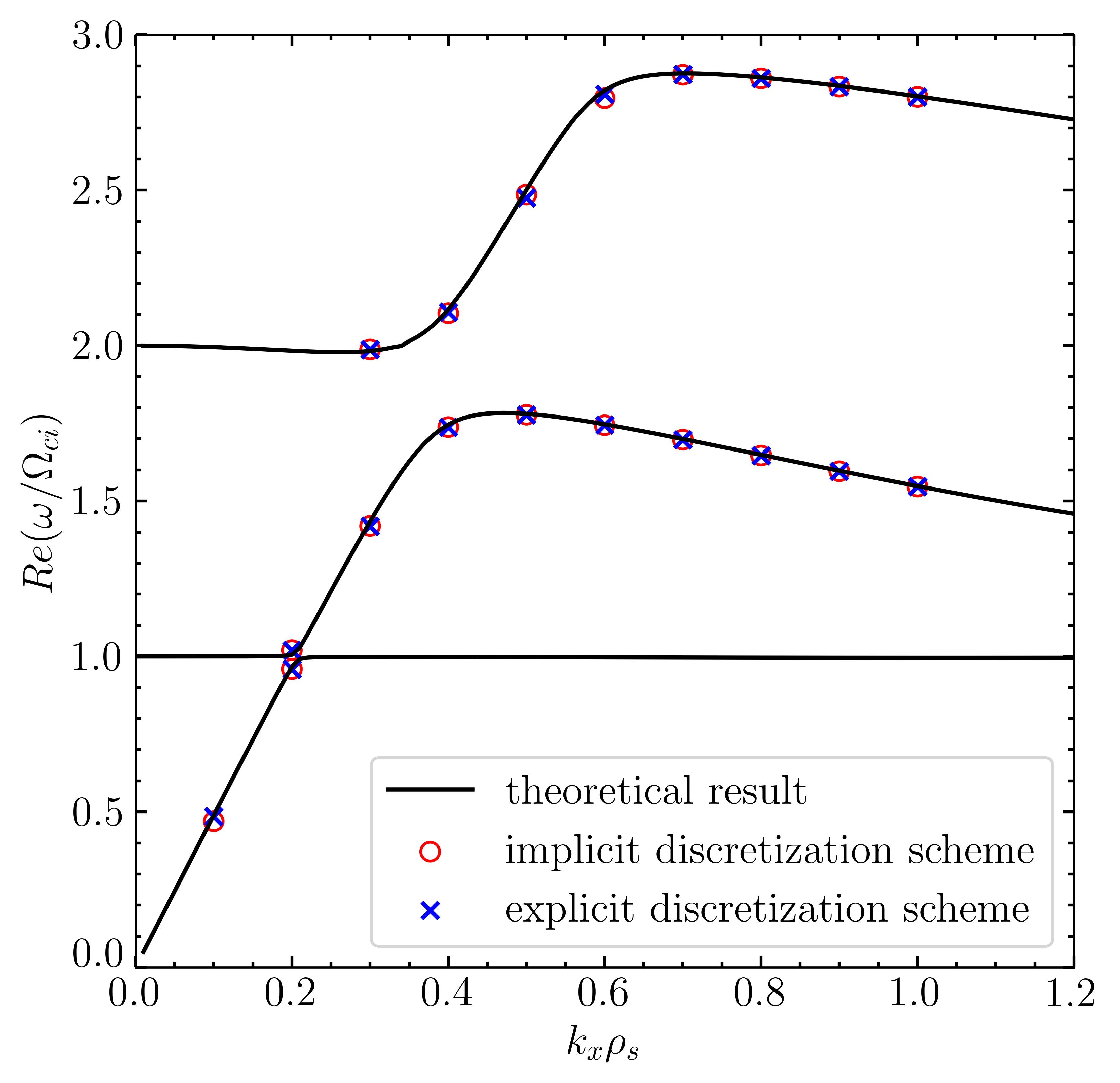}
    \end{minipage}
  }
  \subfigure[Parallel wave]
  {
    \begin{minipage}[b]{.45\textwidth}
      \centering
      \includegraphics[width=\textwidth]{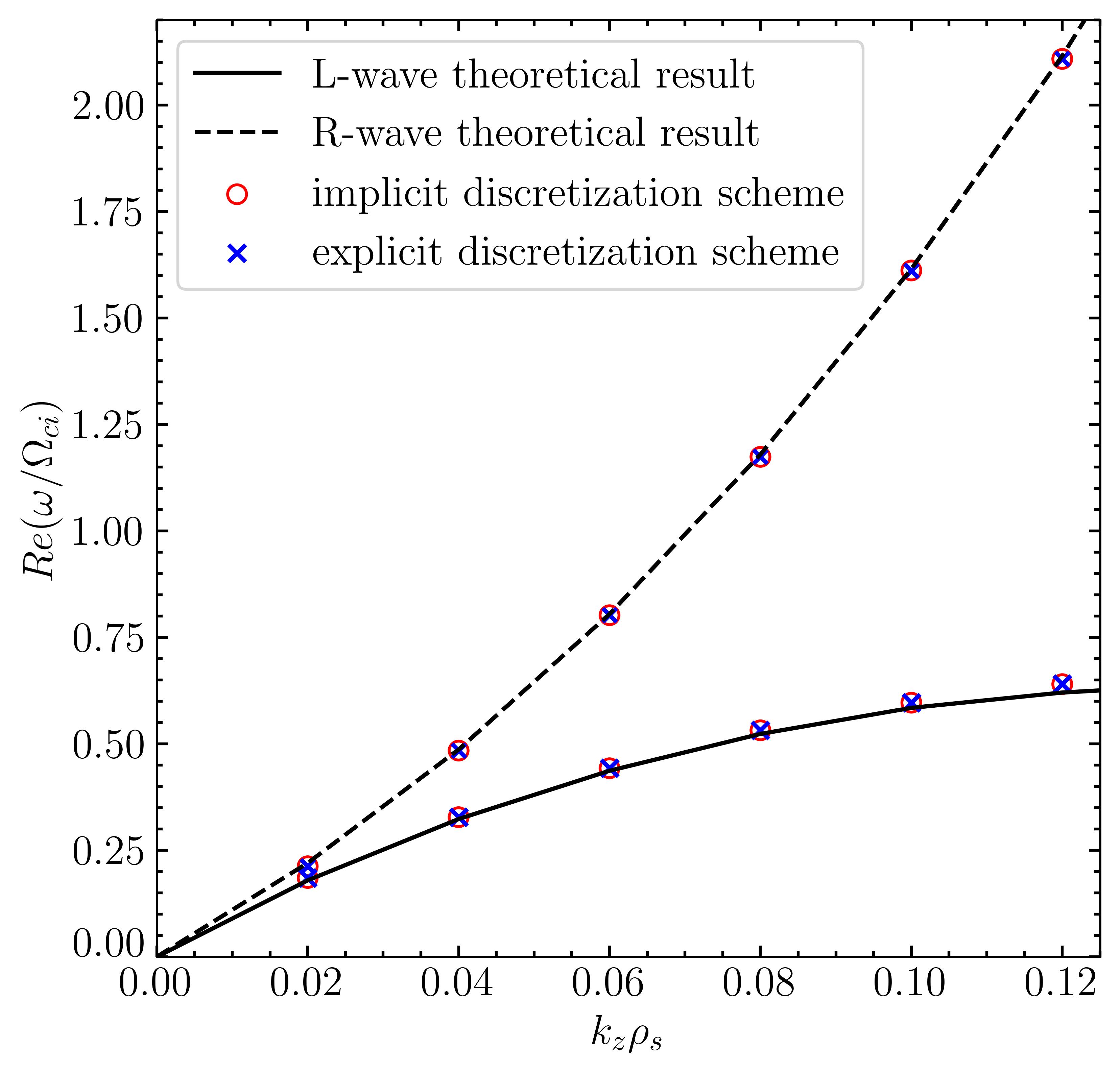}
    \end{minipage}
  }
  \caption{\label{fig:high-frequency summary}  Simulation results of perpendicular waves (a) and parallel waves (b). For perpendicular waves, the number of grids $n_{x}=n_{y}=n_{z}=32$, the particle number in one grid $N_{p}=64$, timestep $\Omega_{ci}\Delta t=0.05$, and the wave parameters $\beta_e=0.05,T_{i}/T_{e}=1,k_{y}\rho_{s}=k_{z}\rho_{s}=0$.
  For parallel waves, the simulation parameters are $\beta_e=0.01,T_{i}/T_{e}=1,k_{x}\rho_{s}=k_{y}\rho_{s}=0, n_{x}=n_{y}=2,n_{z}=64, N_{p}=256, \Omega_{ci}\Delta t=0.05$.}
\end{figure*}

We first test the two schemes with perpendicular waves.
\revA{Since the implicit parallel Ampere's law (\ref{eq:iterative implicit ampere law}) and perpendicular Ohm's law (\ref{eq:iterative implicit perpendicular Ohm's law}) are solved iteratively, figure \ref{fig: iterative_solver} presents a convergence test of the iterative solver.
For the parameters used in Fig.~\ref{fig: iterative_solver} ($N_{p}=16$ particles per grid cell and $n_{x}=16$ grids in one wavelength), the relative iterative error falls below the tolerance $(10^{-4})$ within a few iterations. }
The simulation results for perpendicular waves in uniform plasmas with $\beta=0.05,\mathbf{k}=k_{x}\mathbf{\hat{x}}$ are summarized in Fig.~\ref{fig:high-frequency summary} (a), where the theoretical results are calculated from the dispersion relation via the generalized argument principle code ZPL \cite{ChenH2022,ChenH2021b,LiZ2025}. 
Both schemes recover the correct real frequencies and capture the characteristic mode conversion from compressional Alfvén waves to ion Bernstein waves. 
Imaginary frequency components are omitted, as the theoretical damping rates for these modes are zero.

% \begin{figure}[t]
%   \centering
%   \includegraphics[width=0.5\textwidth]{caw_summary_real_frequency.jpg}
%   \caption{\label{fig:caw summary} A summary of perpendicular wave simulation results. Here the number of grids $n_{x}=n_{y}=n_{z}=32$, the particle number in one grid $N_{p}=64$, timestep $\Omega_{ci}\Delta t=0.05$. The wave parameters are $\beta_e=0.05,k_{y}\rho_{s}=k_{z}\rho_{s}=0$.}
% \end{figure}

We further evaluate the two schemes using left-handed (L) and right-handed (R) circularly polarized waves, which converge to shear Alfvén waves in the low-frequency regime $(\omega\ll\Omega_{ci})$.
As the frequency approaches the ion cyclotron frequency $(\omega\approx\Omega_{ci})$, the L-wave undergoes strong cyclotron damping \cite{LiZ2025}.
As summarized in Fig.~\ref{fig:high-frequency summary} (b), both schemes correctly yield the real frequencies of the waves across these regimes. 
Nevertheless, accurately capturing the cyclotron damping rate of the L-wave via PIC simulation remains challenging. 
This is because the L- and R-waves are coupled in the simulation, and the damped L-wave signal is readily obscured by undamped R-wave signal.
This difficulty represents an inherent limitation of the PIC methodology itself, rather than a shortcoming of either specific scheme.

% \begin{figure}[t]
%   \centering
%   \includegraphics[width=0.5\textwidth]{parallel_wave_real_frequency.jpg}
%   \caption{\label{fig:saw summary} A summary of parallel wave simulation results. Here $\beta_e=0.01,k_{x}\rho_{s}=k_{y}\rho_{s}=0, n_{x}=n_{y}=2,n_{z}=64, N_{p}=256, \Omega_{ci}\Delta t=0.05$.}
% \end{figure}

\subsection{Ion acoustic wave}
 
\revAB{The IAW, characterized by $\omega\sim k_{\|}v_{ti}\ll k_{\|}v_{te}$, serves as an ideal test case for comparing the ability of the two schemes to overcome the cancellation problem \cite{ChenY2009,ChenH2021,ChenH2023}.
The theoretical IAW dispersion relation is given by \cite{ChenH2021b}}
\begin{equation}
  \begin{aligned}
    \sum_{s=i,~e}\frac{1}{T_{s}}\left[1+\frac{\omega}{k_{z}v_{ts}}Z\left(\frac{\omega}{k_{z}v_{ts}}\right)\right]=0,
  \end{aligned}
  \label{eq:theoretical dispersion relation}
\end{equation}
\revAB{where the normalized thermal velocities are $v_{te}=\sqrt{2m_{i}/m_{e}}, v_{ti}=\sqrt{2T_{i}/T_{e}}$, and $Z\left(x\right)$ is the plasma dispersion function.}

\revAB{Before presenting the simulation results, we first provide an analytical argument to demonstrate why the implicit parallel Ampere's law mitigates the cancellation problem.
For simplicity and thereby clarity,  we make the following assumptions.}

\revAB{(1) The perturbed field is electrostatic, where $\mathbf{E}_{1}=E_{1z}\hat{\mathbf{z}}$ aligns with the z direction (the equilibrium magnetic field direction) and spatial variation depends only on the z coordinate.}

\revAB{(2) Only the linear particle motion in the z direction is considered,}
\begin{equation}
  \frac{z_{s}^{n+1}-z_{s}^{n}}{\Delta t}=v_{sz}^{n}, ~\frac{v_{sz}^{n+1}-v_{sz}^{n}}{\Delta t}=0, ~(s=i,~e),
  \label{eq:simplified particle motion}
\end{equation}
\revAB{where the parallel velocity $v_{sz}^{n}$ is constant and its superscript $n$ is dropped in the following derivation. }

\revAB{For the explicit discretization scheme, the particle weight equations are}
\begin{equation}
  \frac{w_{i}^{n+1}-w_{i}^{n}}{\Delta t}=\frac{T_{e}}{T_{i}}v_{iz}E_{1z}^{n}\left(z_{i}^{n}\right), ~\frac{w_{e}^{n+1}-w_{e}^{n}}{\Delta t}=-v_{ez}E_{1z}^{n}\left(z_{e}^{n}\right),
\end{equation}
\revAB{thus the particle weights at time step $t^{n+1}$ can be expressed as} 
\begin{equation}
  w_{i}^{n+1}=w_{i}^{0}+\Delta t\frac{T_{e}}{T_{i}}v_{iz}\sum_{m=0}^{n}E_{1z}^{m}\left(z_{i}^{m}\right), ~w_{e}^{n+1}=w_{e}^{0}-\Delta t v_{ez}\sum_{m=0}^{n}E_{1z}^{m}\left(z_{e}^{m}\right).
  \label{eq:IAW explicit particle weight}
\end{equation}

\revAB{Substituting above equations into Eq.~(\ref{eq:numerical parallel Ohm's law}), we obtain the discretized form of the parallel Ohm's law}
\begin{equation}
  \begin{aligned}
    &\frac{1}{N_{p}}\sum_{j}\frac{m_{e}}{m_{i}}v_{ejz}^{2}E_{1z}^{n+1}\left(z_{ej}^{n+1}\right)S\left(z_{g}-z_{ej}^{n+1}\right)+\frac{m_{e}}{m_{i}}E_{1z}^{n+1}\left(z_{g}\right)\\
    &=-\frac{\partial}{\partial z_{g}}\left\{\frac{1}{N_{p}}\sum_{j}\frac{m_{e}}{m_{i}}v_{ejz}^{2}\left[w_{ej}^{0}-\Delta t v_{ejz}\sum_{m=0}^{n}E_{1z}^{m}\left(z_{ej}^{m}\right)\right]S\left(z_{g}-z_{ej}^{n+1}\right)\right\}\\
    &+\frac{m_{e}}{m_{i}}\frac{\partial}{\partial z_{g}}\left\{\frac{1}{N_{p}}\sum_{j}v_{ijz}^{2}\left[w_{ij}^{0}+\frac{T_{e}}{T_{i}}\Delta t v_{ijz}\sum_{m=0}^{n}E_{1z}^{m}\left(z_{ij}^{m}\right)\right]S\left(z_{g}-z_{ij}^{n+1}\right)\right\}.
  \end{aligned}
  \label{eq:discretized parallel Ohm's law}
\end{equation}

\revAB{For the implicit discretization scheme, the electron weight equation is }
  \begin{equation}
  \begin{aligned}
    \frac{w_{e}^{*}-w_{e}^{n}}{\Delta t}=0, ~\frac{w_{e}^{n+1}-w_{e}^{*}}{\Delta t}=-v_{ez}E_{1z}^{n+1}\left(z_{e}^{n+1}\right),
  \end{aligned}
\end{equation}
\revAB{so the intermediate electron weight $w_{e}^{*}$ advanced from $w_{e}^{n}$ takes the form }
\begin{equation}
  w_{e}^{*}=w_{e}^{0}-\Delta t v_{ez}\sum_{m=1}^{n}E_{1z}^{m}\left(z_{e}^{m}\right).
\end{equation}
\revAB{The ion weight expression is the same as Eq.~(\ref{eq:IAW explicit particle weight}). Substituting the particle weight expressions into Eq.~(\ref{eq:implicit parallel Ampere's law}), we obtain the discretized form of the implicit parallel Ampere's law}
\begin{equation}
  \begin{aligned}
    &\frac{\Delta t}{N_{p}}\sum_{j}v_{ejz}^{2}E_{1z}^{n+1}\left(z_{ej}^{n+1}\right)S\left(z_{g}-z_{ej}^{n+1}\right)\\
    &=\frac{1}{N_{p}}\sum_{j}v_{ejz}\left[w_{ej}^{0}-\Delta t v_{ejz}\sum_{m=1}^{n}E_{1z}^{m}\left(z_{ej}^{m}\right)\right]S\left(z_{g}-z_{ej}^{n+1}\right)\\
    &-\frac{1}{N_{p}}\sum_{j}v_{ijz}\left[w_{ij}^{0}+\Delta t\frac{T_{e}}{T_{i}}v_{ijz}\sum_{m=0}^{n}E_{1z}^{m}\left(z_{ij}^{m}\right)\right]S\left(z_{g}-z_{ij}^{n+1}\right).
  \end{aligned}
  \label{eq:discretized parallel Ampere's law}
\end{equation}

\revAB{
In IAW cases, Eq.~(\ref{eq:discretized parallel Ampere's law}) does not suffer from the numerical cancellation problem.
The only theoretical cancellation involved is that between the ion parallel current and the electron parallel current, which yields the IAW dispersion relation.
In contrast, for Eq.~(\ref{eq:discretized parallel Ohm's law}), the leading-order terms $E_{1z}^{n+1}$ and $-{\partial p_{1,ez}^{n+1}}/{\partial z_{g}}$ are expected to nearly cancel to recover the accurate IAW mode.
Intuitively, the higher-order electron velocity moment and spatial gradient in $-{\partial p_{1,ez}^{n+1}}/{\partial z_{g}}$ make the cancellation problem in the parallel Ohm's law more severe, which requires finer grids and more particles to resolve.
Furthermore, we can next show that even in the limit of infinite particle number and grid number, the implicit parallel Ampere's law yields a more accurate IAW frequency than the parallel Ohm's law under the same timestep $\Delta t$.} 

\revAB{For further simplification, we use the following assumptions}

\revAB{(3) The electric field is filtered to a single $k_{z}$ mode at each time step, i.e., $E_{1z}^{n}(z_{g})=\tilde{E}_{1z}^{n}exp\left(ik_{z}z_{g}\right)+c.c.$, where the wavenumber $k_{z}$ is fixed and the complex amplitude $\tilde{E}_{1z}^{n}$ depends on the time step $t^{n}$.}

\revAB{(4) The initial electron and ion weights are loaded as $w_{ej}^{0}=\tilde{w}_{e}^{0}exp(ik_{z}z_{ej}^{0})+c.c.$ and $w_{ij}^{0}=\tilde{w}_{i}^{0}exp(ik_{z}z_{ij}^{0})+c.c.$, respectively.}

\revAB{By inserting $\delta\left(z-z_{ej}^{n+1}\right)\delta \left(v_{z}-v_{ejz}\right)$ and $\delta\left(z-z_{ij}^{n+1}\right)\delta \left(v_{z}-v_{ijz}\right)$ into Eq.~(\ref{eq:discretized parallel Ampere's law}), the discretized form of the implicit parallel Ampere's law becomes }
\begin{equation}
   \begin{aligned}
      &\frac{\Delta t}{\Delta z}\tilde{E}_{1z}^{n+1}e^{ik_{z}z_{g}}\sum_{p=-\infty}^{+\infty}\iint_{\mathbf{R}^{2}} v_{z}^{2}e^{ik_{z}p\Delta z}S\left(z_{g}+p\Delta z-z\right)S\left(z_{g}-z\right)\frac{\Delta z}{N_{p}}\sum_{j}\delta \left(z-z_{ej}^{n+1}\right)\delta\left(v_{z}-v_{ejz}\right)dz dv_{z}\\
      &=\frac{1}{\Delta z}\iint_{\mathbf{R}^{2}}v_{z}\left[\tilde{w}_{e}^{0}e^{ik_{z}\left(z-(n+1)v_{z}\Delta t\right)}-\Delta t v_{z}\sum_{m=1}^{n}\sum_{p=-\infty}^{+\infty}\tilde{E}_{1z}^{m}e^{ik_{z}\left(z_{g}+p\Delta z\right)}S\left(z_{g}+p\Delta z-z+(n+1-m)v_{z}\Delta t\right)\right]\\
      &S\left(z_{g}-z\right) \frac{\Delta z}{N_{p}}\sum_{j}\delta (z-z_{ej}^{n+1})\delta \left(v_{z}-v_{ejz}\right)dz dv_{z}\\
      &-\frac{1}{\Delta z}\iint_{\mathbf{R}^{2}}v_{z}\left[\tilde{w}_{i}^{0}e^{ik_{z}\left(z-(n+1)v_{z}\Delta t\right)}+\Delta t \frac{T_{e}}{T_{i}}v_{z}\sum_{m=0}^{n}\sum_{p=-\infty}^{+\infty}\tilde{E}_{1z}^{m}e^{ik_{z}\left(z_{g}+p\Delta z\right)}S\left(z_{g}+p\Delta z-z+(n+1-m)v_{z}\Delta t\right)\right]\\
      &S\left(z_{g}-z\right) \frac{\Delta z}{N_{p}}\sum_{j}\delta (z-z_{ij}^{n+1})\delta \left(v_{z}-v_{ijz}\right)dz dv_{z}.
   \end{aligned}
   \label{eq:interemdiate discretized form 1}
\end{equation}
\revAB{Then, we replace the summation terms over $j$ with the normalized equilibrium distribution functions under the assumption:} 

\revAB{(5) The particle number in each grid is large enough that the following equations hold }
\begin{equation}
  \begin{aligned}
    &\frac{\Delta z}{N_{p}}\sum_{j}\delta\left(z_{g}-z_{ej}^{n+1}\right)\delta\left(v_{z}-v_{ejz}\right)\approx f_{e0}=\frac{exp\left[-m_{e}v_{z}^{2}/\left(2m_{i}\right)\right]}{\sqrt{2\pi m_{i}/m_{e}}},\\
    &\frac{\Delta z}{N_{p}}\sum_{j}\delta\left(z_{g}-z_{ij}^{n+1}\right)\delta\left(v_{z}-v_{ijz}\right)\approx f_{i0}=\frac{exp\left[-T_{e}v_{z}^{2}/\left(2T_{i}\right)\right]}{\sqrt{2\pi T_{i}/T_{e}}}.\\
  \end{aligned}
  \label{eq:normalized distribution function}
\end{equation}

\revAB{In Eq.~(\ref{eq:interemdiate discretized form 1}), the integral of the shape functions over z can be defined as $I(\xi)=\int_{-\infty}^{+\infty}S(Z+\xi)S(Z)dZ$ and we notice that}
\begin{equation}
   \begin{aligned}
      &\tilde{I}(k)=\iint_{\mathbf{R}^{2}}S\left(Z+\xi\right)S\left(Z\right)e^{-ik\xi}d\xi dZ=|\tilde{S}\left(k\right)|^{2},\\
      &I(\xi)=\frac{1}{2\pi}\int_{-\infty}^{+\infty}|\tilde{S}\left(k\right)|^{2}e^{ik\xi}dk,
   \end{aligned}
\end{equation}
\revAB{with $\tilde{S}\left(k\right)=\int_{-\infty}^{+\infty} S(Z)e^{-ikZ} dZ$. Therefore, the double summation terms of $\tilde{E}_{1z}^{m}$ in Eq.~(\ref{eq:interemdiate discretized form 1}) can be simplified as}
\begin{equation}
   \begin{aligned}
   &-\frac{\Delta t}{\Delta z}\sum_{m=1}^{n}\tilde{E}_{1z}^{m}e^{ik_{z}z_{g}}\iint_{\mathbf{R}^{2}}v_{z}^{2}\sum_{p=-\infty}^{+\infty}e^{ik_{z}p\Delta z}S\left(z_{g}+p\Delta z-z+(n+1-m)v_{z}\Delta t\right)S\left(z_{g}-z\right) f_{e0}dz dv_{z}\\
   &=-\frac{\Delta t}{\Delta z}\sum_{m=1}^{n}\tilde{E}_{1z}^{m}e^{ik_{z}z_{g}}\int_{-\infty}^{+\infty}dv_{z} \left\{v_{z}^{2}f_{e0}\left[\sum_{p=-\infty}^{+\infty} \frac{e^{ik_{z} p\Delta z}}{2\pi}\int_{-\infty}^{+\infty}|\tilde{S}\left(k\right)|^{2}e^{ik p\Delta z}e^{ikv_z\left(n+1-m\right)\Delta t}dk\right]\right\}\\
   &=-\frac{\Delta t}{\Delta z}\sum_{m=1}^{n}\tilde{E}_{1z}^{m}e^{ik_{z}z_{g}}\int_{-\infty}^{+\infty}dv_{z} \left\{v_{z}^{2}f_{e0}\left[\frac{1}{\Delta z}\int_{-\infty}^{+\infty}|\tilde{S}\left(k\right)|^{2}e^{ikv_{z}\left(n+1-m\right)\Delta t}\sum_{p=-\infty}^{+\infty} \delta \left(k+k_{z}-\frac{2\pi p}{\Delta z}\right)dk\right]\right\}\\
   &=-\frac{\Delta t}{\Delta z^{2}}\sum_{m=1}^{n}\sum_{p=-\infty}^{+\infty}\tilde{E}_{1z}^{m}e^{ik_{z}z_{g}}\int_{-\infty}^{+\infty} v_{z}^{2}f_{e0}e^{-ik_{p}v_{z}(n+1-m)\Delta t}|\tilde{S}(k_{p})|^{2}dv_{z},
   \end{aligned}
\end{equation}
\revAB{with $k_{p}=k_{z}-2p\pi/\Delta z$. Here the poisson summation formula is used}
\begin{equation}
   \sum_{p=-\infty}^{+\infty}e^{i\left(k+k_{z}\right)p\Delta z}=\frac{2\pi}{\Delta z}\sum_{p=-\infty}^{+\infty}\delta \left(k+k_{z}-\frac{2\pi p}{\Delta z}\right).
\end{equation}

\revAB{After simplifying the integrals over $z$, the discretized form of the implicit parallel Ampere's law can be expressed as}
\begin{equation}
   \begin{aligned}
      &\Delta t\frac{2+cos(k_{z}\Delta z)}{3}\tilde{E}_{1z}^{n+1}e^{ik_{z}z_{g}}\int_{-\infty}^{+\infty}v_z^{2} f_{e0} dv_{z}\\
      &=\frac{\tilde{S}(k_{z})}{\Delta z}\tilde{w}_{e}^{0}e^{ik_{z}z_{g}}\int_{-\infty}^{+\infty} v_{z}e^{-ik_{z}v_{z}(n+1)\Delta t} f_{e0} dv_{z}-\Delta t\sum_{m=1}^{n}\sum_{p=-\infty}^{+\infty}\frac{|\tilde{S}(k_{p})|^{2}}{\Delta z^{2}}\tilde{E}_{1z}^{m}e^{ik_{z}z_{g}}\int_{-\infty}^{+\infty} v_{z}^{2}f_{e0}e^{-ik_{p}v_{z}(n+1-m)\Delta t}dv_{z}\\
      &-\frac{\tilde{S}(k_{z})}{\Delta z}\tilde{w}_{i}^{0}e^{ik_{z}z_{g}}\int_{-\infty}^{+\infty} v_{z}e^{-ik_{z}v_{z}(n+1)\Delta t} f_{i0} dv_{z}-\Delta t\sum_{m=0}^{n}\sum_{p=-\infty}^{+\infty}\frac{|\tilde{S}(k_{p})|^{2}}{\Delta z^{2}}\frac{T_{e}}{T_{i}}\tilde{E}_{1z}^{m}e^{ik_{z}z_{g}}\int_{-\infty}^{+\infty} v_{z}^{2}f_{i0}e^{-ik_{p}v_{z}(n+1-m)\Delta t}dv_{z}.
   \end{aligned}
\end{equation}

\revAB{We multiply $\sum_{n=0}^{+\infty}e^{i\omega \left(n+1\right)\Delta t}$ with the implicit parallel Ampere's law and notice that}
\begin{equation}
  \begin{aligned}
    &\sum_{n=0}^{+\infty}\sum_{m=1}^{n}\tilde{E}_{1z}^{m}e^{i(\omega-k_{p}v_{z})\left(n+1\right)\Delta t}e^{ik_{p}v_{z}m\Delta t}=\sum_{m=1}^{+\infty}\tilde{E}_{1z}^{m}e^{ik_{p}v_{z}m\Delta t}\sum_{n=m}^{+\infty}e^{i(\omega-k_{p}v_{z})\left(n+1\right)\Delta t}\\
    &=\sum_{m=1}^{+\infty}\tilde{E}_{1z}^{m}e^{ik_{p}v_{z}m\Delta t}\frac{e^{i(\omega-k_{p}v_{z})\left(m+1\right)\Delta t}}{1-e^{i(\omega-k_{p}v_{z})\Delta t}}=\left[\hat{E}_{1z}\left(k_{z},\omega\right)-\tilde{E}_{1z}^{0}\left(k_{z}\right)\right]\frac{e^{i(\omega-k_{p}v_{z})\Delta t}}{1-e^{i(\omega-k_{p}v_{z})\Delta t}},
  \end{aligned}
\end{equation}
\revAB{where $\hat{E}_{1z}\left(k_{z},\omega\right)$ denotes the electric field after a discrete-time Fourier transformation, defined as $\hat{E}_{1z}\left(k_{z},\omega\right)=\sum_{m=0}^{+\infty}\tilde{E}_{1z}^{m}e^{i\omega m\Delta t}$.}

\revAB{Therefore, the discretized form of the implicit parallel Ampere's law becomes}
\begin{equation}
   \begin{aligned}
      &\left[\Delta t\frac{2+cos(k_{z}\Delta z)}{3}\int v_z^{2} f_{e0} dv_{z}+\Delta t\sum_{p=-\infty}^{+\infty}\frac{|\tilde{S}(k_{p})|^{2}}{\Delta z^{2}}\int_{-\infty}^{+\infty} v_{z}^{2}\left(f_{e0}+\frac{T_{e}}{T_{i}}f_{i0}\right)\frac{e^{i\left(\omega-k_{p}v_{z}\right)\Delta t}}{1-e^{i\left(\omega-k_{p}v_{z}\right)\Delta t}}dv_{z}\right]\hat{E}_{1z}\left(k_{z},\omega\right)\\
      &=\Delta t\tilde{E}_{1z}^{0}\left(k_{z}\right)\frac{2+cos(k_{z}\Delta z)}{3}\int v_z^{2} f_{e0} dv_{z}
      +\Delta t\tilde{E}_{1z}^{0}\left(k_{z}\right)\sum_{p=-\infty}^{+\infty}\frac{|\tilde{S}(k_{p})|^{2}}{\Delta z^{2}}\int_{-\infty}^{+\infty} v_{z}^{2}f_{e0}\frac{e^{i\left(\omega-k_{p}v_{z}\right)\Delta t}}{1-e^{i\left(\omega-k_{p}v_{z}\right)\Delta t}}dv_{z}\\
      &+\frac{\tilde{S}\left(k_{z}\right)}{\Delta z}\int v_{z} \left(\tilde{w}_{e}^{0}f_{e0}-\tilde{w}_{i}^{0}f_{i0}\right) \frac{e^{i\left(\omega-k_{z}v_{z}\right)\Delta t}}{1-e^{i\left(\omega-k_{z}v_{z}\right)\Delta t}} dv_{z}.
   \end{aligned}
\end{equation}
\revAB{
The spectrum peak appears at the position where the coefficient of $\hat{E}_{1z}\left(k_{z},\omega\right)$ is zero, which gives the numerical dispersion relation of IAW modified by the finite grid size and the finite timestep
}
\begin{equation}
  \begin{aligned}
    \Delta t\frac{m_{i}}{m_{e}} \frac{2+cos(k_{z}\Delta z)}{3}-\Delta t\sum_{s=i,~e}\sum_{p=-\infty}^{+\infty}\frac{m_{i}}{m_{s}}\frac{|\tilde{S}(k_{p})|^{2}}{\Delta z^{2}}\left\{\frac{1}{2}+\frac{2i}{\Delta t}\sum_{q=-\infty}^{+\infty}\frac{\omega_{q}}{k_{p}^{2}v_{ts}^{2}}\left[1+\frac{\omega_{q}}{|k_{p}|v_{ts}}Z\left(\frac{\omega_{q}}{|k_{p}|v_{ts}}\right)\right]\right\}=0,
  \end{aligned}
  \label{eq:implicit dispersion relation}
\end{equation}
\revAB{with $\omega_{q}=\omega-2q\pi/\Delta t$ and $k_{p}=k_{z}-2p\pi/\Delta z$. Here we have used the expansion}
\begin{equation}
  \begin{aligned}
    \frac{1}{1-e^{i\left(\omega-k_{p}v_{z}\right)\Delta t}}=\frac{1}{2}+\frac{i}{\Delta t}\sum_{q=-\infty}^{+\infty}\frac{1}{\omega-k_{p}v_{z}-2q\pi/\Delta t},
  \end{aligned}
\end{equation}
\revAB{which converges in the sense of the principal value under the symmetric limit on $q$.}

\revAB{Following a similar derivation, we obtain the numerical dispersion relation of IAW from the parallel Ohm's law}
\begin{equation}
  \begin{aligned}
    \frac{2+cos(k_{z}\Delta z)}{3}\frac{m_{i}}{m_{e}}+1-\frac{2sin(k_{z}\Delta z)}{\Delta z} \sum_{s=i,~e}\sum_{p=-\infty}^{+\infty} \sum_{q=-\infty}^{+\infty} \frac{|\tilde{S}(k_{p})|^{2}}{k_{p}\Delta z^{2}}\frac{m_{i}}{m_{s}}\left[\frac{\omega_{q}^{2}}{k_{p}^{2}v_{ts}^{2}}+\frac{1}{2}+\frac{\omega_{q}^{3}}{|k_{p}^{3}|v_{ts}^{3}}Z\left(\frac{\omega_{q}}{|k_{p}|v_{ts}}\right)\right]=0.
  \end{aligned}
  \label{eq:explicit dispersion relation}
\end{equation}
\revAB{In the derivation of Eq.~(\ref{eq:explicit dispersion relation}), $\partial/\partial z_{g}$ is discretized via central differencing. The shape function $S(z)$ is }
\begin{equation}
S(z) =
\begin{cases} 
1 - {|z|}/{\Delta z} & \text{for } |z| \leq \Delta z, \\
0 & \text{for } |z| > \Delta z,
\end{cases}
\end{equation}
\revAB{with $\tilde{S}(k)=\Delta z [sin(k\Delta z/2)/(k\Delta z/2)]^{2}$.}

\begin{figure*}[t]
  \centering
  \subfigure[Real frequency]
  {
    \begin{minipage}[b]{.45\textwidth}
      \centering
      \includegraphics[width=\textwidth]{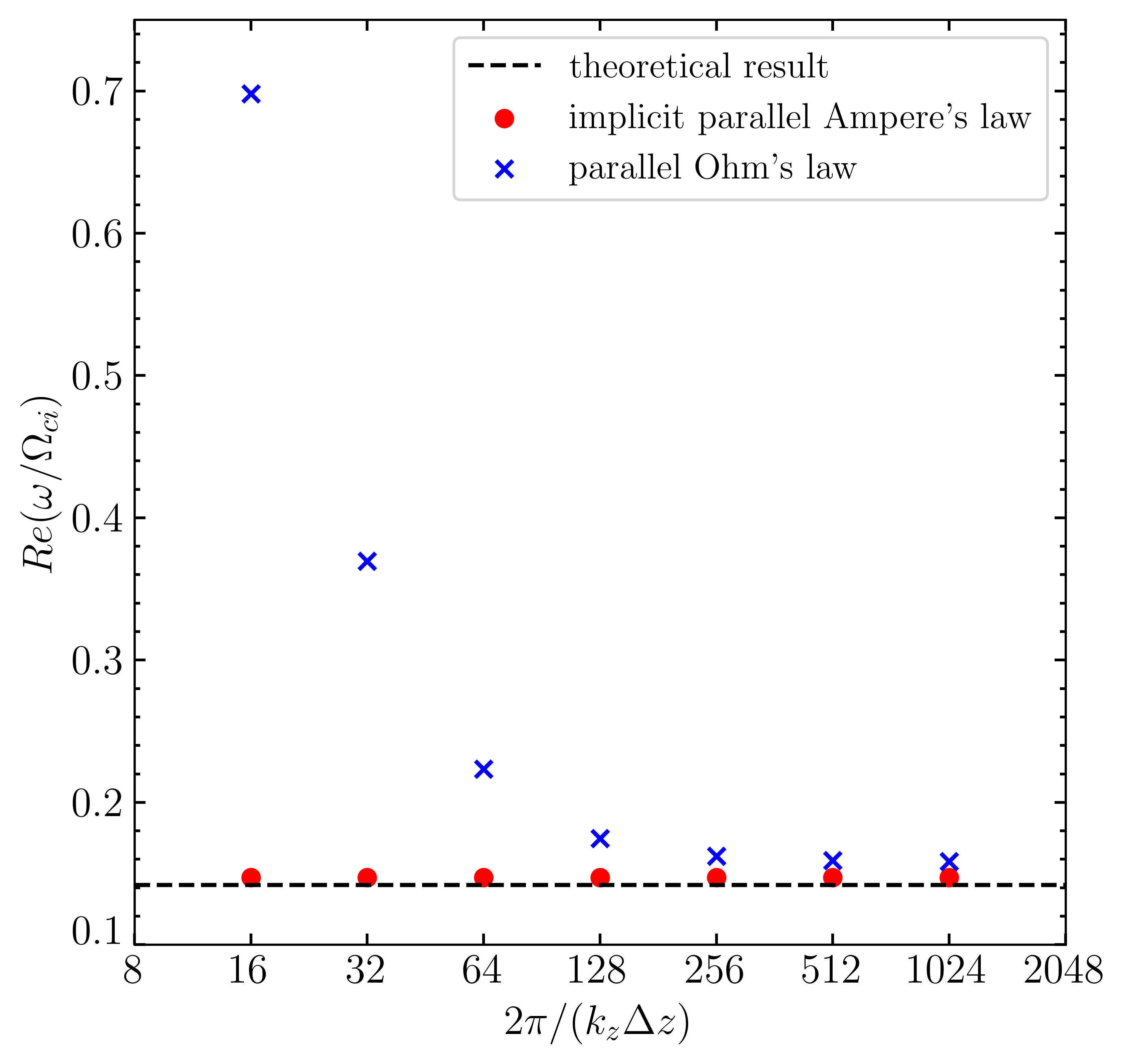}
    \end{minipage}
  }
  \subfigure[Damping rate]
  {
    \begin{minipage}[b]{.47\textwidth}
      \centering
      \includegraphics[width=\textwidth]{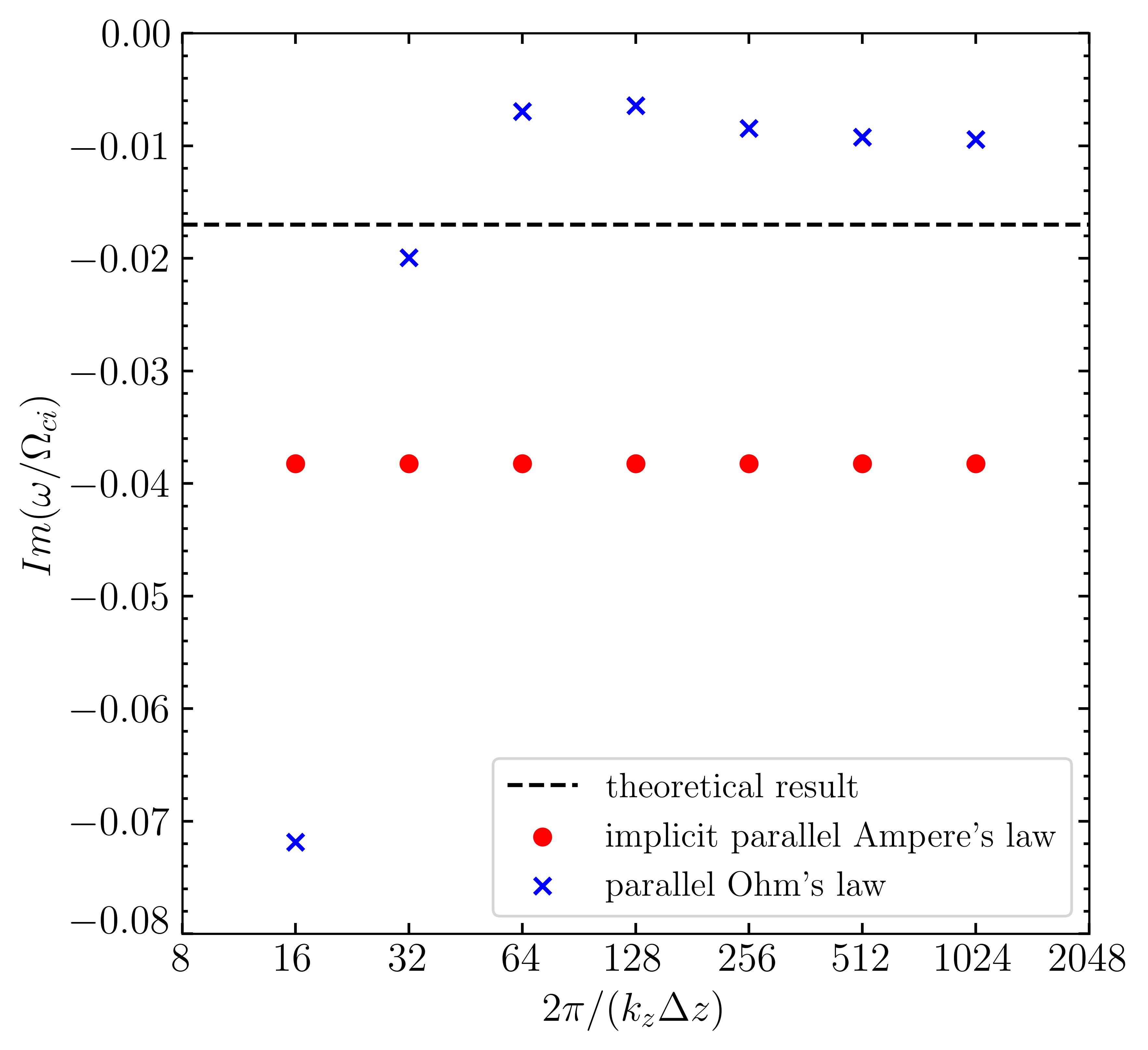}
    \end{minipage}
  }
  \caption{\label{fig:cancellation grid} Effect of the finite grid size on the numerical dispersion relations of the IAW. The dashed line shows the theoretical results from the IAW dispersion relation (\ref{eq:theoretical dispersion relation}). The red circles and blue crosses represent the numerical dispersion relations Eqs.~(\ref{eq:implicit dispersion relation}) and (\ref{eq:explicit dispersion relation}),  obtained using the implicit parallel Ampere's law and the parallel Ohm's law, respectively. IAW parameters are $T_{i}/T_{e}=0.25, k_{z}\rho_{s}=0.1, \Omega_{ci}\Delta t=0.01$.}
\end{figure*}

\revAB{Figure~\ref{fig:cancellation grid} demonstrates the eigenvalue results obtained from the theoretical and numerical dispersion relations, Eqs.~(\ref{eq:theoretical dispersion relation}), (\ref{eq:implicit dispersion relation}) and (\ref{eq:explicit dispersion relation}).
For the implicit parallel Ampere's law, Eq.~(\ref{eq:implicit dispersion relation}), the eigenvalues converge when the number of grid points per wavelength exceeds 16.
In contrast, the parallel Ohm's law, Eq.~(\ref{eq:explicit dispersion relation}), requires more than 256 grid points per wavelength to achieve convergence.
This confirms that the cancellation problem in the parallel Ohm's law requires substantially finer grid resolution.
}

\begin{figure*}[t]
  \centering
  \subfigure[Real frequency]
  {
    \begin{minipage}[b]{.45\textwidth}
      \centering
      \includegraphics[width=\textwidth]{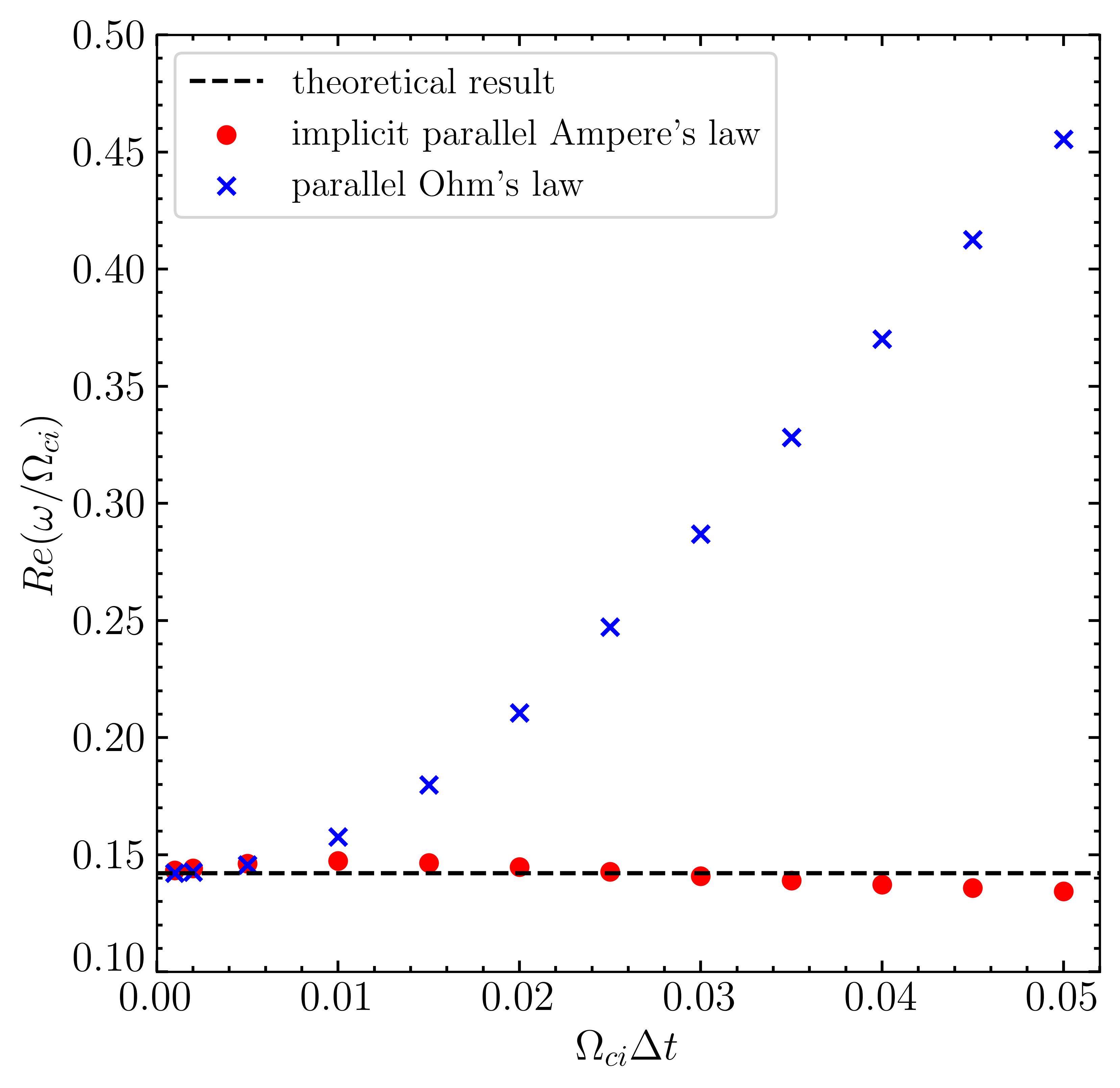}
    \end{minipage}
  }
  \subfigure[Damping rate]
  {
    \begin{minipage}[b]{.46\textwidth}
      \centering
      \includegraphics[width=\textwidth]{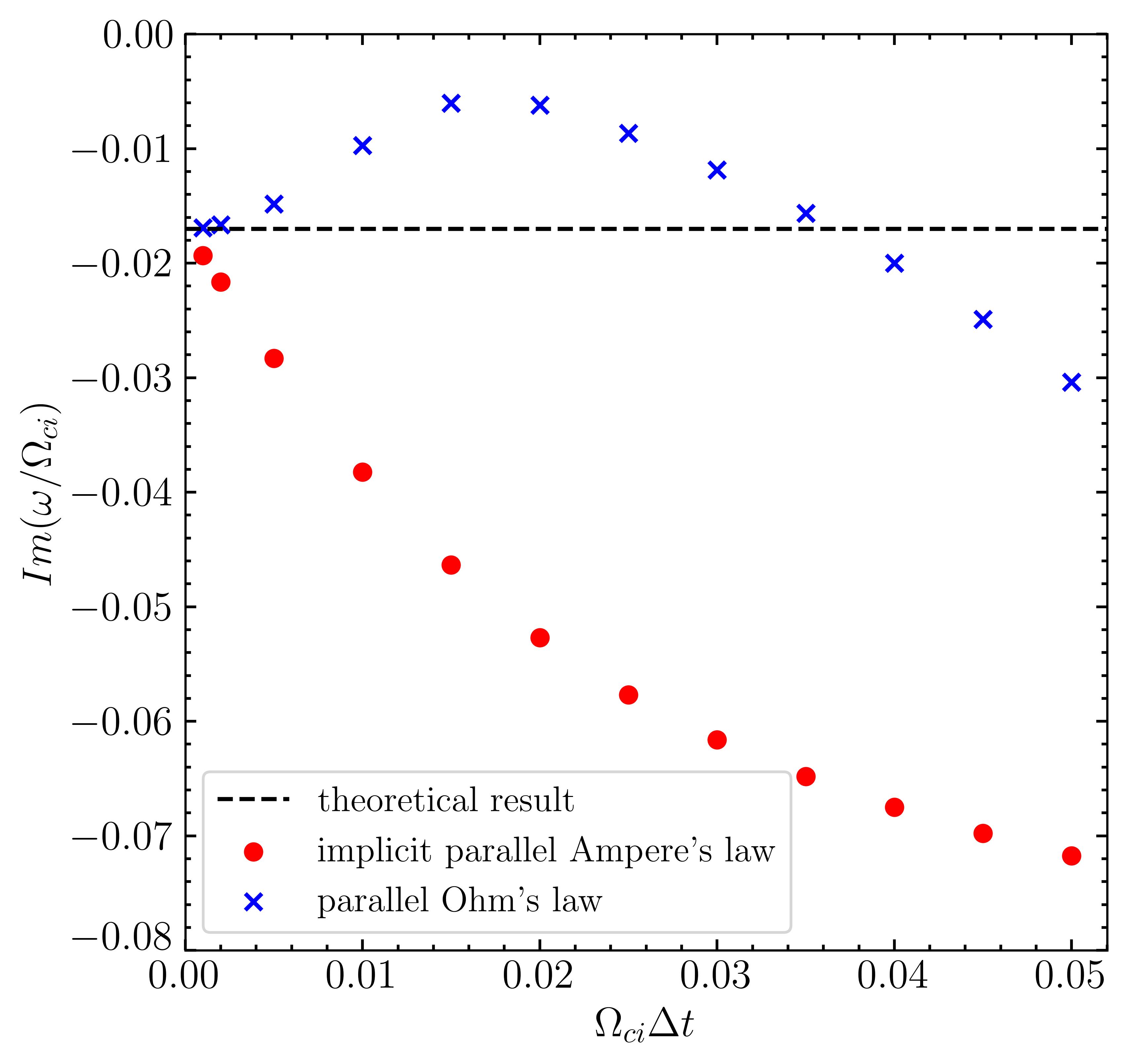}
    \end{minipage}
  }
  \caption{\label{fig:cancellation} Effect of the finite timestep on the numerical IAW dispersion relations in the limit of infinite grid resolution. The dashed line shows the theoretical dispersion relation (\ref{eq:theoretical dispersion relation}). The red circles and blue crosses correspond to the numerical dispersion relations obtained from the implicit parallel Ampere's law (\ref{eq:implicit dispersion relation 2}) and the parallel Ohm's law (\ref{eq:explicit dispersion relation 2}),  respectively. IAW parameters are $T_{i}/T_{e}=0.25, k_{z}\rho_{s}=0.1$.}
\end{figure*}

\revAB{More importantly, the cancellation problem in the parallel Ohm's law demands not only finer spatial grids but also smaller timesteps.
Taking the limit of infinite grid resolution $(\Delta z\rightarrow 0)$ in Eq.~(\ref{eq:implicit dispersion relation}) yields the numerical dispersion relation of the implicit parallel Ampere's law, modified solely by the finite timestep effect,}
\begin{equation}
  \begin{aligned}
    &\frac{\Delta t}{2}\frac{m_{i}}{m_{e}}-\frac{\Delta t}{2}-2 i\sum_{s=i,~e}\sum_{q=-\infty}^{+\infty}\frac{m_{i}}{m_{s}}\frac{\omega-2q\pi/\Delta t}{k_{z}^{2}v_{ts}^{2}}\left[1+\frac{\omega-2q\pi/\Delta t}{k_{z}v_{ts}}Z\left(\frac{\omega-2q\pi/\Delta t}{k_{z}v_{ts}}\right)\right]=0.
  \end{aligned}
  \label{eq:implicit dispersion relation 2}
\end{equation}
\revAB{Similarly, under infinite grid resolution, the numerical dispersion relation of the parallel Ohm's law reduces to}
\begin{equation}
  \begin{aligned}
    \frac{m_i}{m_{e}}+1-2\sum_{s=i,~e}\sum_{q=-\infty}^{+\infty}\frac{m_{i}}{m_{s}}\left[\left(\frac{\omega-2q\pi/\Delta t}{k_{z}v_{ts}}\right)^{2}+\frac{1}{2}+\left(\frac{\omega-2q\pi/\Delta t}{k_{z}v_{ts}}\right)^{3}Z\left(\frac{\omega-2q\pi/\Delta t}{k_{z}v_{ts}}\right)\right]=0.
  \end{aligned}
  \label{eq:explicit dispersion relation 2}
\end{equation}
\revAB{
The eigenvalue results obtained from the theoretical and numerical dispersion relations, Eqs.~(\ref{eq:theoretical dispersion relation}), (\ref{eq:implicit dispersion relation 2}) and (\ref{eq:explicit dispersion relation 2}), are presented in Fig.~\ref{fig:cancellation}.
The real frequencies from the implicit parallel Ampere's law, Eq.~(\ref{eq:implicit dispersion relation 2}), remain in good agreement with the theoretical values across the entire range of $\Omega_{ci}\Delta t \in (0,0.05]$.
In contrast, those from the parallel Ohm's law, Eq.~(\ref{eq:explicit dispersion relation 2}), exhibit significant deviations for $\Omega_{ci}\Delta t>0.01$, even though $\Omega_{ci}\Delta t<0.05$ is already sufficiently small to accurately resolve particle motions.
Agreement with the theoretical result is recovered only when $\Omega_{ci}\Delta t$ is reduced to $0.005$ or smaller.
Consequently, unless the timestep $\Omega_{ci}\Delta t$ is chosen sufficiently small, the parallel Ohm's law fails to reproduce accurate IAW frequency results, even in the limit of infinite grid resolution and particle number.
This strong sensitivity of the parallel Ohm's law to $\Delta t$ highlights the greater difficulty in overcoming the cancellation problem compared with the implicit parallel Ampere's law.}

\revAB{
Regarding the damping rates, both numerical dispersion relations deviate from theoretical results unless $\Omega_{ci}\Delta t$ is smaller than $0.002$, as shown by Fig.~\ref{fig:cancellation} (b).
The implicit parallel Ampere's law tends to overestimate the damping rate due to the numerical stabilizing effect inherent to the implicit time-stepping scheme, which is a common phenomenon not limited to the IAW.
As can be seen in section \ref{sec:second order}, this numerical damping can be effectively reduced by using a second-order pushing scheme.
}

\begin{figure*}[t]
  \centering
  \subfigure[Real frequency]
  {
    \begin{minipage}[b]{.45\textwidth}
      \centering
      \includegraphics[width=\textwidth]{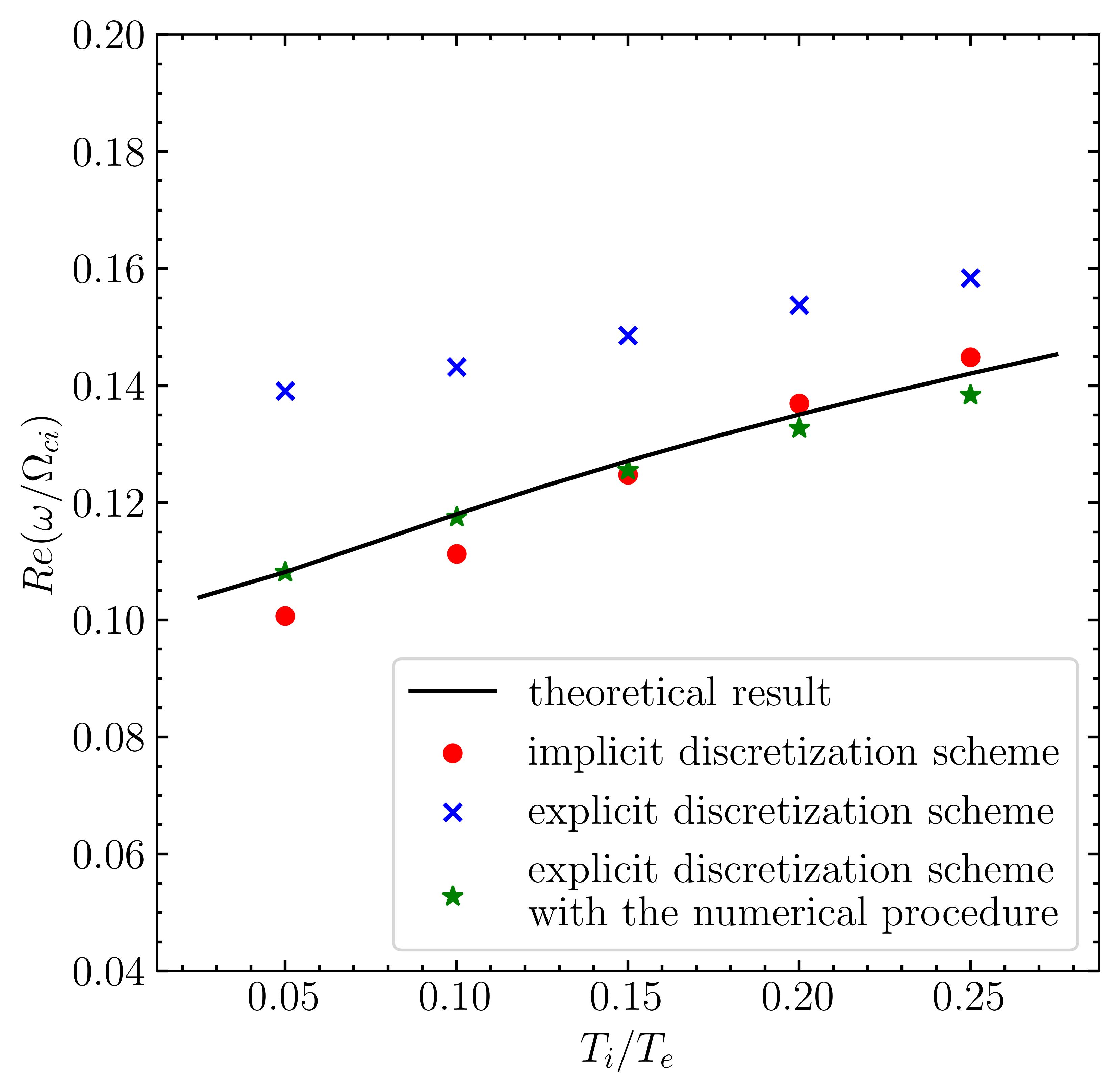}
    \end{minipage}
  }
  \subfigure[Damping rate]
  {
    \begin{minipage}[b]{.45\textwidth}
      \centering
      \includegraphics[width=\textwidth]{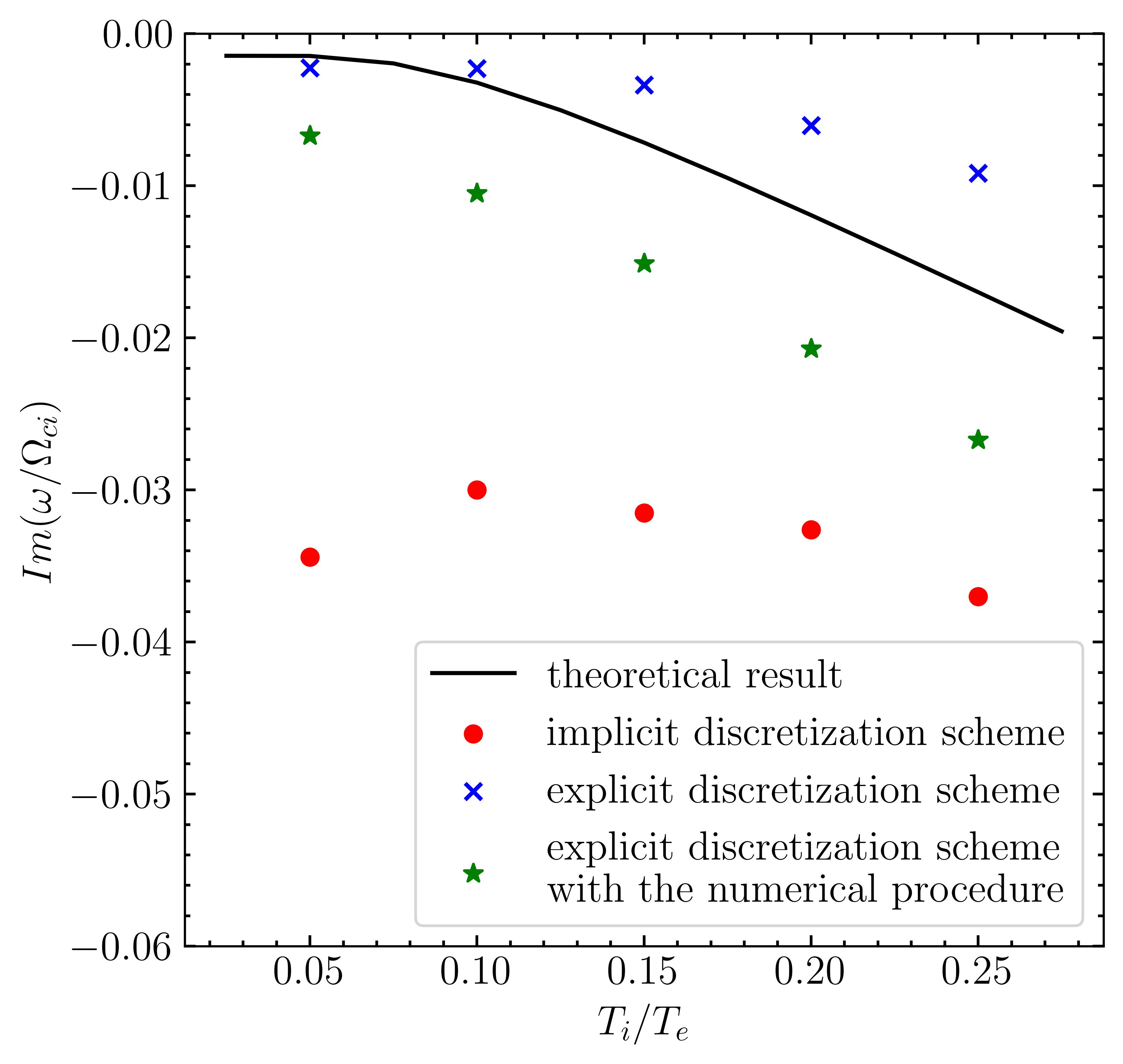}
    \end{minipage}
  }
  \caption{\label{fig:IAW summary 1} Simulation results of IAW for different schemes. Simulation parameters are $n_{x}=n_{y}=2, n_{z}=128, N_{p}=256, \Omega_{ci}\Delta t=0.01$. The wave parameters are $\beta_e=0.01,k_{x}\rho_{s}=k_{y}\rho_{s}=0, k_{z}\rho_{s}=0.1$.}
\end{figure*}

\revB{Building on the analytical IAW argument, we now turn to numerical simulations to evaluate how different schemes perform in practice.}
\revAB{Figure \ref{fig:IAW summary 1} compares the IAW simulation results for three schemes: the implicit discretization scheme, the explicit discretization scheme, and the explicit discretization scheme using the numerical procedure that replaces $p_{1, e\|}$ with $p_{1, e\|}+T_{e}(n_{1, i}-n_{1, e})$.}
\revB{This procedure, proposed in previous work \cite{ChenY2009,ChengJ2013}, is designed to enforce the IAW dispersion relation (quasi-neutrality condition) at leading order, thereby mitigating the cancellation problem.
Although effective, this procedure introduces high-frequency oscillations into the simulation \cite{ChenY2009}.
Consequently, it requires additional numerical operations—such as adjusting particle weight in the end of each timestep to enforce charge neutrality—to suppress the high-frequency noise \cite{ChenY2009}.}

\revB{As shown in Fig.~\ref{fig:IAW summary 1} (a), the explicit discretization scheme with the numerical procedure yields the most accurate real frequencies, closely matching the theoretical dispersion relation.
However, this technique lacks robustness, as it does not perform well when applied to other benchmark problems like the ITG case (see section \ref{subsec:itg}).
Without the numerical procedure, the explicit discretization scheme produces real frequencies that are noticeably less accurate than those obtained with the implicit discretization scheme.
This result confirms our analytical argument that the implicit parallel Ampere's law is more effective at mitigating the cancellation problem.}

\revB{For the damping rates, Fig.~\ref{fig:IAW summary 1} (b) shows that none of the three schemes accurately reproduces the theoretical results under the current simulation parameters $(\Omega_{ci}\Delta t=0.01, n_{z}=128,N_{p}=256)$.
While this timestep is sufficient to resolve particle motions, the implicit discretization scheme overestimates the damping rates due to significant numerical damping introduced by its implicit weight-pushing scheme.
Accurate damping rates for both the implicit and explicit discretization schemes are only achieved when the timestep is reduced to $\Omega_{ci}\Delta t=0.001$.
This strong dependence on timestep motivates the second-order time-stepping scheme for pushing electron and ion weights, which is presented in section \ref{sec:second order}.}

\subsection{Ion temperature gradient driven instability}
\label{subsec:itg}

The ITG also suffers from the cancellation problem \cite{Cummings1994, ChenY2001}. 
To benchmark against simulation results, we first present our analytical model.
We assume the equilibrium density and temperature are $n_{i}=n_{e}=n_{0}(1+\kappa_{n}x), T_{i}=T_{0}(1+\kappa_{ti}x),T_{e}=T_{0}(1+\kappa_{te}x)$ where $\kappa_{n}$, $\kappa_{ti}$ and $\kappa_{te}$ are constant number.
Under low $\beta$ assumption, the equilibrium magnetic field is uniform $\mathbf{B}_{0}=B_{0}\mathbf{\hat{z}}$.
As a function of constant motion, the ion equilibrium distribution function is assumed as 
\begin{equation}
f_{i 0}=\frac{n_0\left(1+\kappa_n \eta\right)}{\left(2\pi T_{0}/m_{i}\right)^{\frac{3}{2}}\left(1+\kappa_{ti} \eta\right)^{\frac{3}{2}}} e^{-\frac{m_i v^2}{2 T_{0}\left(1+\kappa_{ti} \eta\right)}},
\end{equation}
where $\eta=x+m_{i}v_{y}/qB_{0}$.
The linearized Vlasov equation becomes 
\begin{equation}
\begin{aligned}
&\frac{d f_{i1}}{dt}=\frac{n_0 q_i\left(1+\kappa_n \eta\right)}{\left(2\pi /m_{i}\right)^{\frac{3}{2}}T_{0}^{\frac{5}{2}}\left(1+\kappa_{ti} \eta\right)^{\frac{5}{2}}} e^{-\frac{m_i v^2}{2 T_{0}\left(1+\kappa_{ti} \eta\right)}} \left(v_x E_{1 x}+v_y E_{1 y}+v_z E_{1 z}\right)\\
& -\frac{n_0\left(1+\kappa_n \eta\right)}{B_0\left({2 \pi T_{0}}/{m_i}\right)^{\frac{3}{2}} \left(1+\kappa_{ti} \eta\right)^{\frac{3}{2}}} e^{-\frac{m_i v^2}{2 T_{0}\left(1+\kappa_{ti} \eta\right)}}\left(E_{1 y}+v_z B_{1 x}-v_x B_{1 z}\right)\left\{\frac{\kappa_n}{1+\kappa_n \eta}+\left[\frac{m_i v^2}{2 T_i(\eta)}-\frac{3}{2}\right] \frac{\kappa_{ti}}{1+\kappa_{ti} \eta}\right\}.\\
\end{aligned}
\label{eq:consistent treatment}
\end{equation}
Then Eq.~(\ref{eq:consistent treatment}) is Fourier transformed with \revA{$\partial /\partial x\rightarrow ik_{x}$} and $x\rightarrow i{\partial}/{\partial k_{x}}$.
Solving the linearized Vlasov equation and substituting $f_{i1}$ into the field equations, the final dispersion relation comprises a complex set of differential equations in $k_{x}$, which is difficult to solve.

\begin{figure*}[t]
  \centering
  \subfigure[Theoretical eigenvalues]
  {
    \begin{minipage}[b]{.45\textwidth}
      \centering
      \includegraphics[width=\textwidth]{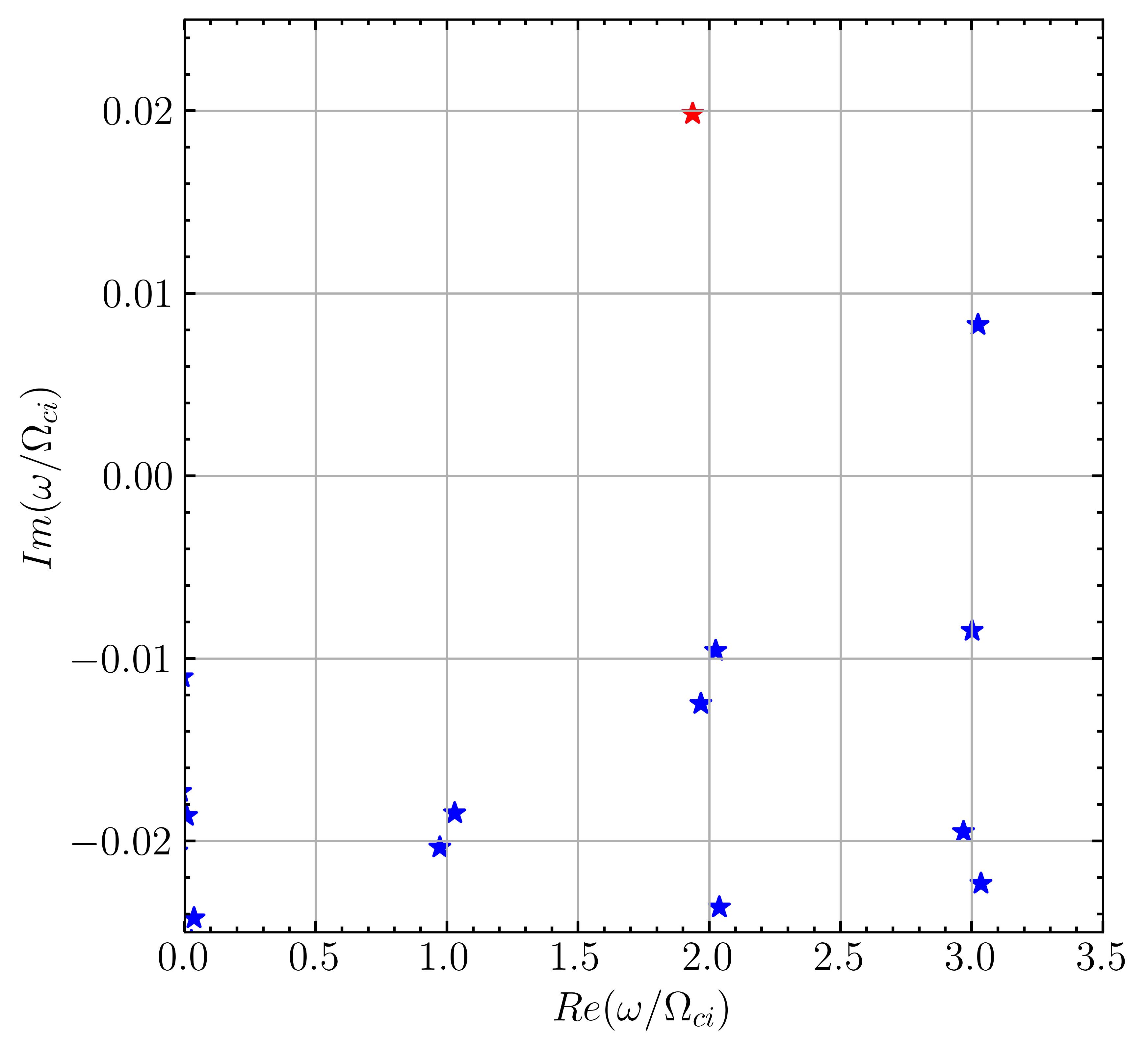}
    \end{minipage}
  }
  \subfigure[Simulation result]
  {
    \begin{minipage}[b]{.45\textwidth}
      \centering
      \includegraphics[width=\textwidth]{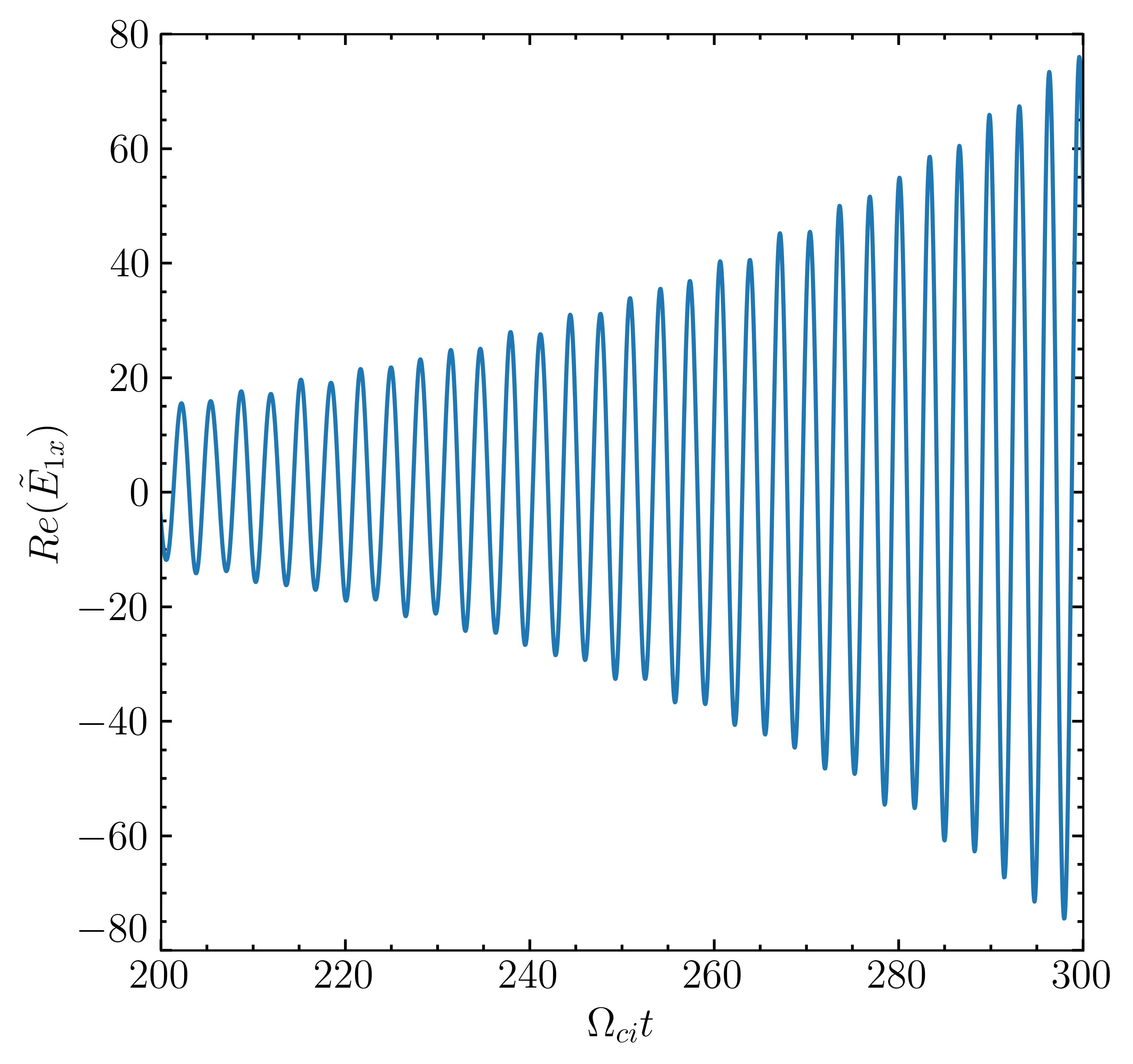}
    \end{minipage}
  }
  \caption{\label{fig:ITG 1} (a) The theoretical eigenvalues of the dispersion relation derived from Eq.~(\ref{eq:inconsistent treatment}). (b) The \revA{linear} simulation results of implicit discretization scheme when Eq.~(\ref{eq:inconsistent treatment}) is used as the ion weight pushing equation. Simulation parameters are $n_{x}=n_{y}=32, n_{z}=64, N_{p}=128, \Omega_{ci}\Delta t=0.05$. The plasma parameters are $\beta_e=0.001,\kappa_{n}\rho_{s}=0,\kappa_{ti}\rho_{s}=-0.3,\kappa_{te}\rho_{s}=0,k_{x}\rho_{s}=0.2,k_{y}\rho_{s}=0.4, k_{z}\rho_{s}=0.01$. \revA{In Fig.~\ref{fig:ITG 1} (b), $Re(\tilde {E}_{1x})$ denotes the real part of $E_{1x}$ after discrete spatial Fourier transformation, evaluated at the wavenumber $k_{x}\rho_{s}=0.2,k_{y}\rho_{s}=0.4, k_{z}\rho_{s}=0.01$. 
  The simulated mode in Fig.~\ref{fig:ITG 1} (b) exhibits a complex frequency $\omega/\Omega_{ci}=1.936+0.0173i$, which corresponds to the eigenvalue $\omega/\Omega_{ci}=1.936+0.0199i$ highlighted by the red star symbol in Fig.~\ref{fig:ITG 1} (a).} }
\end{figure*}

Since the purpose of ITG simulations here is to compare the two schemes' capability in addressing the cancellation problem, we employ a simplified theoretical treatment to derive an algebraic dispersion relation.
Specifically, in Eq.~(\ref{eq:consistent treatment}), we make the local assumption which sets $x=0$ and neglects $O(\kappa_{n}^{2}),O(\kappa_{n}\kappa_{ti}),O(\kappa_{ti}^{2})$ terms
\begin{equation}
  \begin{aligned}
    \frac{d f_{i1}}{dt}\approx &\frac{q_i}{T_{0}} f_M\left\{1+\left[\kappa_n+\left(\frac{m_i v^2}{2 T_{0}}-\frac{5}{2}\right) \kappa_{ti}\right] \frac{m_{i}v_{y}}{qB_{0}}\right\}\left(v_x E_{1 x}+v_y E_{1 y}+v_z E_{1 z}\right) \\
    & -\frac{f_M}{B_0}\left(E_{1 y}+v_z B_{1 x}-v_x B_{1 z}\right)\left[\kappa_n+\left(\frac{m_i v^2}{2 T_{0}}-\frac{3}{2}\right) \kappa_{ti}\right].\\
  \end{aligned}
  \label{eq:inconsistent treatment}
\end{equation}
Here all fractional terms with $m_{i}v_{y}/qB_{0}$ in the denominator have been Taylor expanded and $f_{M}$ denotes the uniform Maxwell distribution.
However, both the dispersion relation and PIC simulation based on Eq.~(\ref{eq:inconsistent treatment}) show that there may exist high-frequency drift cyclotron instabilities, as can be seen in Fig.~\ref{fig:ITG 1}.

\begin{figure*}[t]
  \centering
  \subfigure[Theoretical eigenvalues]
  {
    \begin{minipage}[b]{.45\textwidth}
      \centering
      \includegraphics[width=\textwidth]{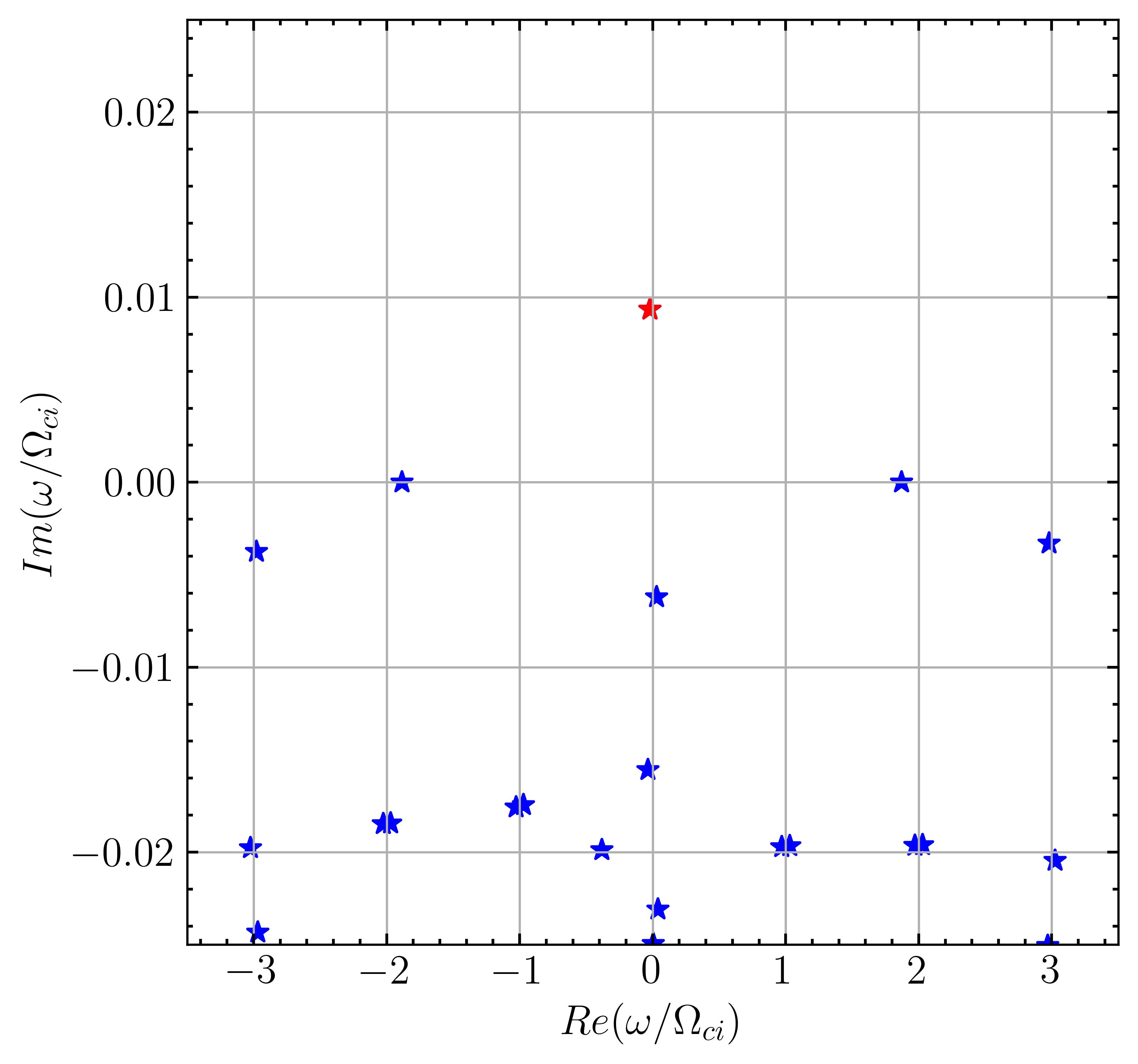}
    \end{minipage}
  }
  \subfigure[Simulation result]
  {
    \begin{minipage}[b]{.45\textwidth}
      \centering
      \includegraphics[width=\textwidth]{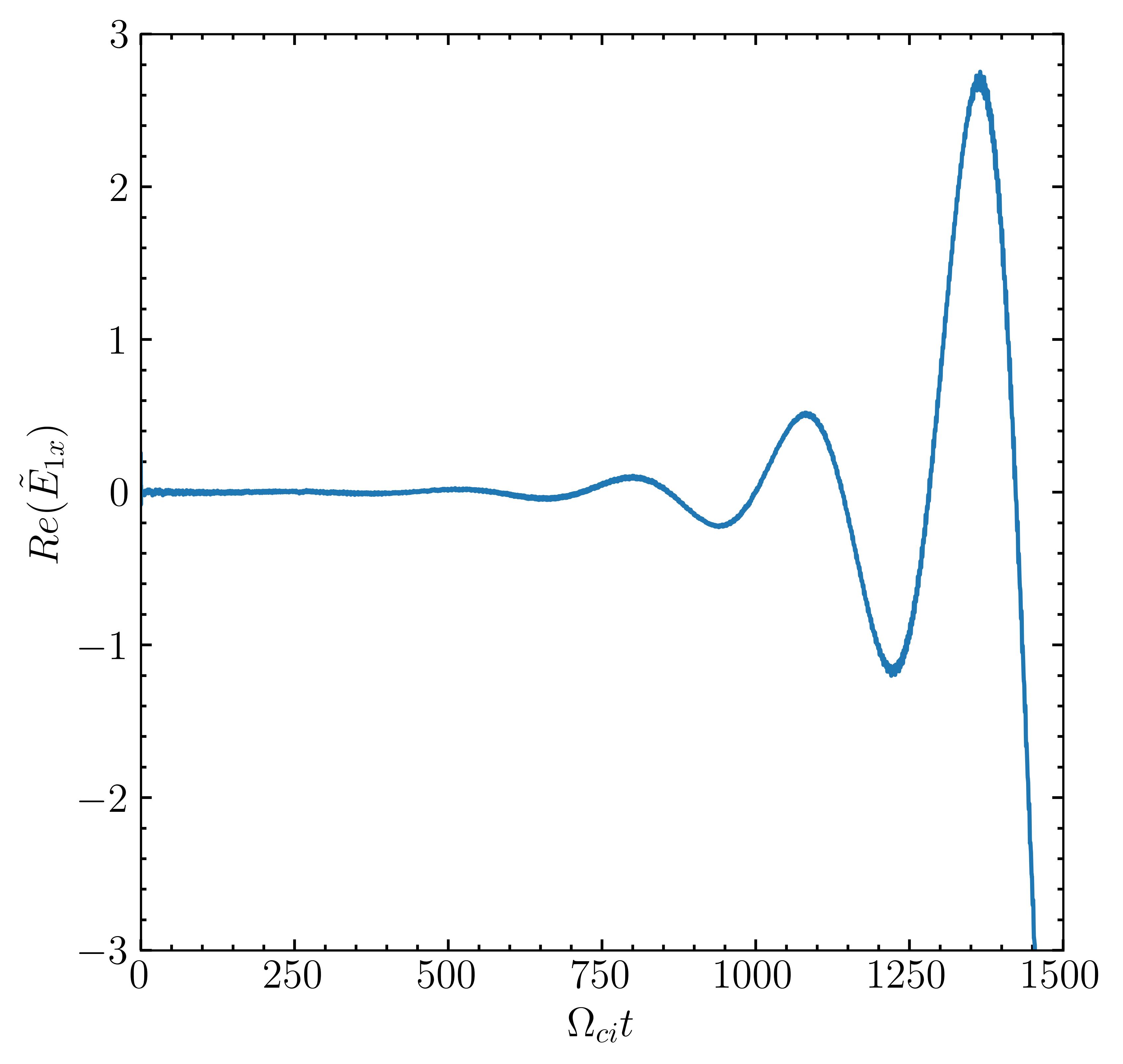}
    \end{minipage}
  }
  \caption{\label{fig:ITG 2} (a) Theoretical eigenvalues of the dispersion relation
  (b) \revA{Linear} simulation results obtained with the implicit discretization scheme
  both under the Boussinesq assumption. Figure \ref{fig:ITG 2} is presented for direct comparison with Figure \ref{fig:ITG 1}, both employing the same simulation parameters.
  \revA{The complex frequency of the simulated mode shown in Fig.~\ref{fig:ITG 2} (b) is $\omega/\Omega_{ci}=-0.0222+0.00588i$, which corresponds to the eigenvalue $\omega/\Omega_{ci}=-0.0222+0.00934i$ indicated by the red star symbol in Fig.~\ref{fig:ITG 2} (a).}}
\end{figure*}

To suppress drift cyclotron instabilities, we introduce a further simplification by setting $\eta=0$ in Eq.~(\ref{eq:consistent treatment}), leading to the following form
\begin{equation}
  \begin{aligned}
    \frac{d f_{i1}}{dt}\approx &\frac{q_i}{T_{0}} f_M\left(v_x E_{1 x}+v_y E_{1 y}+v_z E_{1 z}\right) -\frac{f_M}{B_0}\left(E_{1 y}+v_z B_{1 x}-v_x B_{1 z}\right)\left[\kappa_n+\left(\frac{m_i v^2}{2 T_{0}}-\frac{3}{2}\right) \kappa_{ti}\right].\\
  \end{aligned}
  \label{eq:Boussinesq treatment}
\end{equation}
This simplification, referred to as the Boussinesq assumption \cite{Raeth2024, Raeth2024p}, has been incorporated into the ion weight pushing equations (\ref{eq:ion weight pushing}) presented in section \ref{subsec:model I}.
Physically, this assumption corresponds to neglecting the constant of ion gyromotion $(\eta=0)$ while retaining only one dominant nonuniformity term in the linearized Vlasov equation. 
Figure \ref{fig:ITG 2} shows the eigenvalues of the dispersion relation and PIC simulation results under this assumption with identical parameters, confirming that the model correctly captures the ITG mode while avoiding high-frequency instabilities.

\begin{figure*}[t]
  \centering
  \subfigure[Real frequency]
  {
    \begin{minipage}[b]{.45\textwidth}
      \centering
      \includegraphics[width=\textwidth]{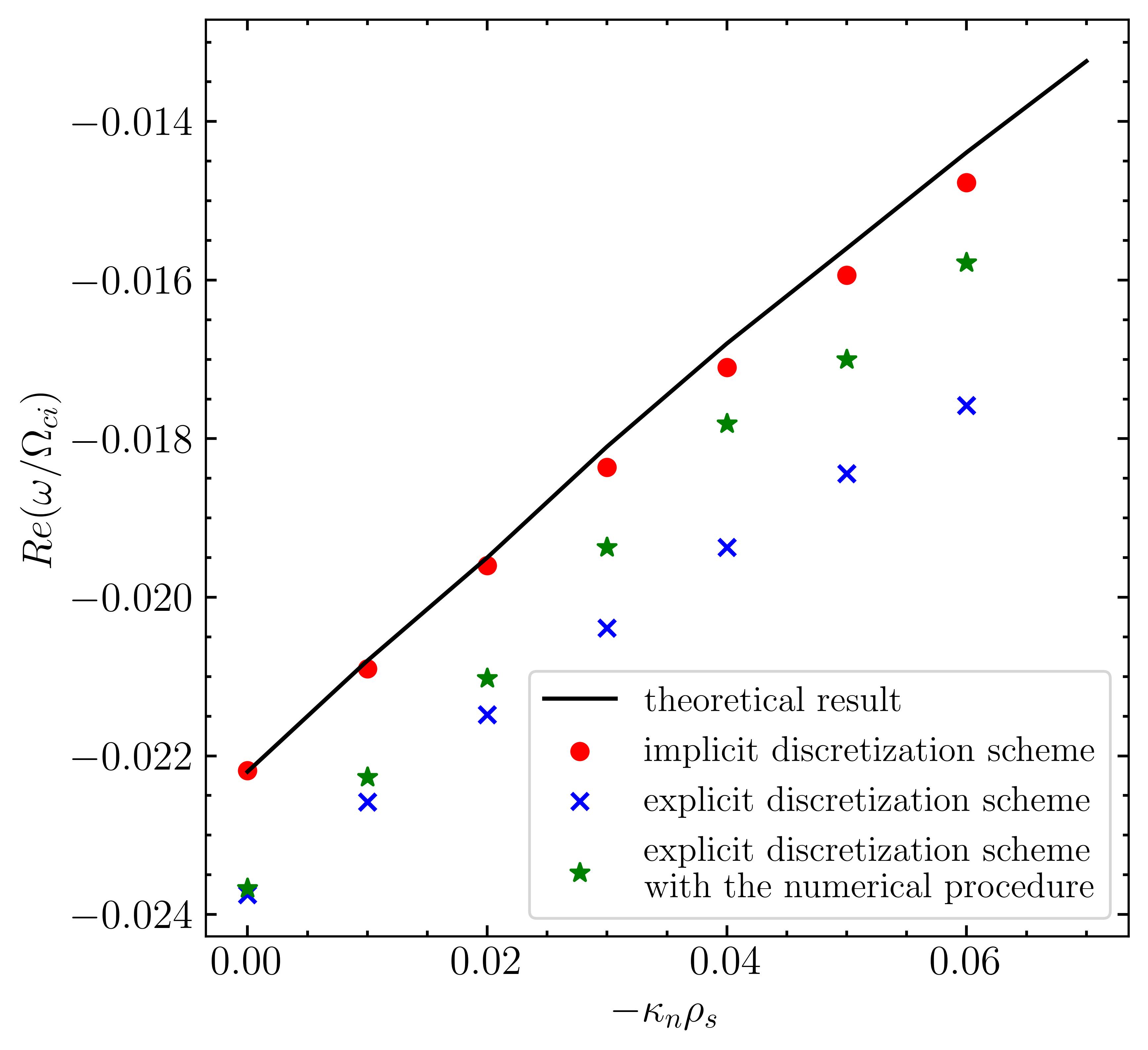}
    \end{minipage}
  }
  \subfigure[Growth rate]
  {
    \begin{minipage}[b]{.45\textwidth}
      \centering
      \includegraphics[width=\textwidth]{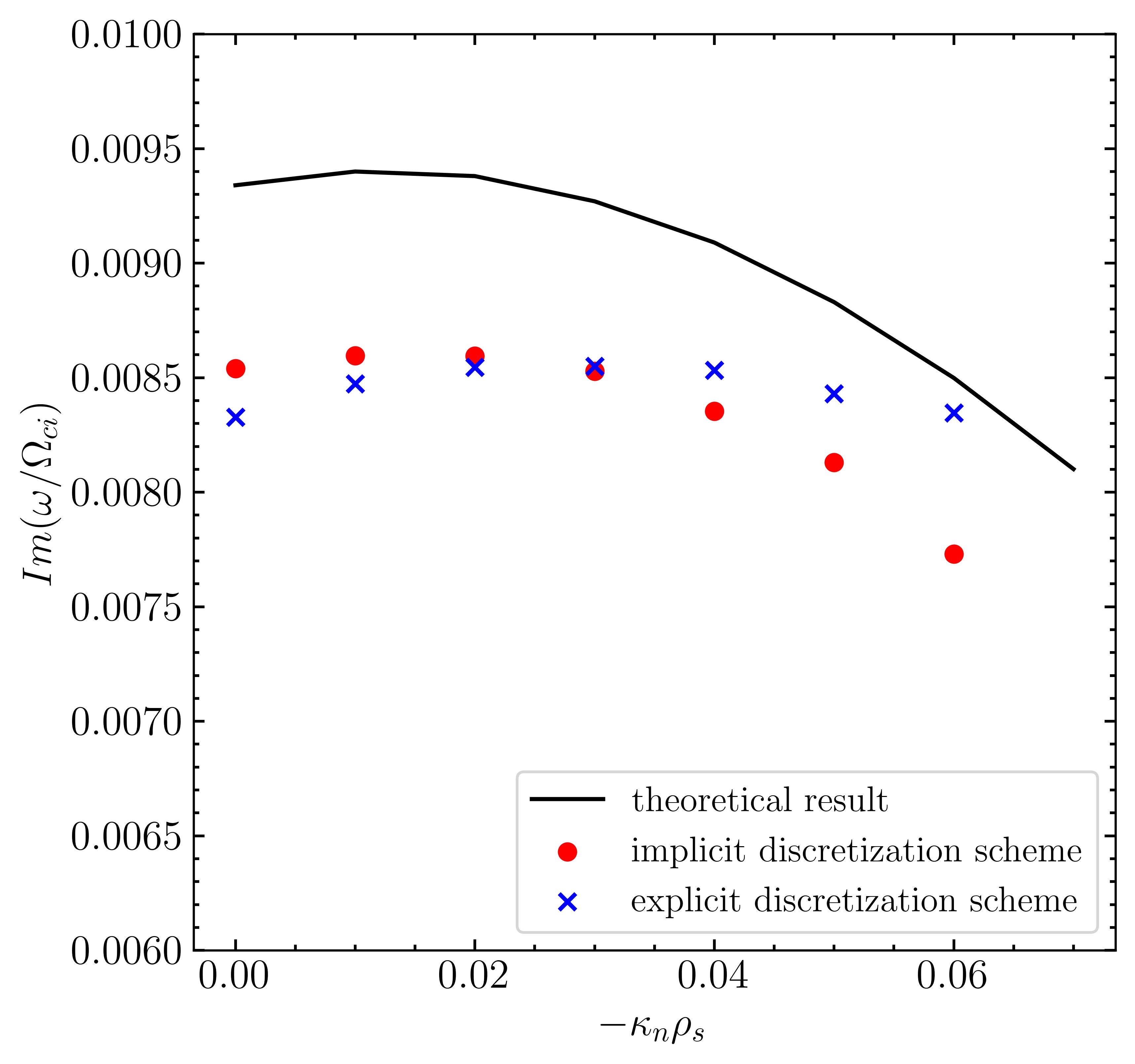}
    \end{minipage}
  }
  \caption{\label{fig:ITG 3} Simulation results of ion temperature gradient driven instabilities (ITG) for different schemes under the Boussinesq assumption. Here simulation parameters are $n_{x}=n_{y}=32,n_{z}=64,N_{p}=128,\Omega_{ci}\Delta t=0.01$.The plasma parameters are $\beta_e=0.001,\kappa_{ti}\rho_{s}=-0.3,\kappa_{te}\rho_{s}=0,k_{x}\rho_{s}=0.2,k_{y}\rho_{s}=0.4, k_{z}\rho_{s}=0.01$. The explicit discretization scheme with the numerical procedure yields damped ITG modes, whose damping rates are not shown in figure (b).}
\end{figure*}

Figure \ref{fig:ITG 3} summarizes the PIC simulation results under the Boussinesq assumption.
The implicit discretization scheme exhibits the closest agreement with the theoretical predictions in real frequencies, confirming its better ability to mitigate the cancellation problem compared to the explicit discretization scheme.
Regarding growth rates, both schemes exhibit a systematic underestimation relative to theory (Fig.~\ref{fig:ITG 3}(b)), due to the numerical damping from the implicit ion weight pushing equations (\ref{eq:ion weight pushing}).
This makes it difficult to compare their performance in predicting the ITG growth rate.
Crucially, the explicit discretization scheme with the numerical procedure produces an artificially damped ITG mode, underscoring that this expedient approach lacks universal applicability.

The ITG simulations use $\Omega_{ci}\Delta t=0.01$, a timestep sufficient to resolve ion gyromotion and to reduce numerical damping. 
Nevertheless, a $10\%$ error remains in the growth rates computed by the implicit discretization scheme. 
Achieving a better accuracy requires further reduction of the timestep, which motivates the development of a second-order scheme for particle pushing.

\section{Second-order time-stepping scheme}
\label{sec:second order}

\subsection{Numerical scheme}
\revB{We develop a second-order time-stepping scheme based on a semi-implicit formulation \cite{ChengJ2013}, where the particle weights are advanced using}
\begin{equation}
  w^{n+1}=w^{n}+\frac{\Delta t}{2}\left(\frac{d w^{n}}{dt}+\frac{d w^{n+1}}{dt}\right).
\end{equation}
\revB{Accordingly, the discrete forms of the ion and electron weight pushing equations are given by}
\begin{equation}
  \begin{aligned}
      \frac{w_{i}^{*}-w_{i}^{n}}{\Delta t/2}=&\frac{T_{e}}{T_{i}}\mathbf{v}_{i}^{n}\cdot\mathbf{E}_{1}^{n}\left(\mathbf{x}_{i}^{n}\right)-\left[E_{1y}^{n}\left(\mathbf{x}_{i}^{n}\right)+v_{iz}^{n}B_{1x}^{n}\left(\mathbf{x}_{i}^{n}\right)-v_{ix}^{n}B_{1z}^{n}\left(\mathbf{x}_{i}^{n}\right)\right]\kappa_{i}^{n},\\
      \frac{w_{i}^{n+1}-w_{i}^{*}}{\Delta t/2}=&\frac{T_{e}}{T_{i}}\mathbf{v}_{i}^{n+1}\cdot\mathbf{E}_{1}^{n+1}\left(\mathbf{x}_{i}^{n+1}\right)-\left[E_{1y}^{n+1}\left(\mathbf{x}_{i}^{n+1}\right)+v_{iz}^{n+1}B_{1x}^{n+1}\left(\mathbf{x}_{i}^{n+1}\right)-v_{ix}^{n+1}B_{1z}^{n+1}\left(\mathbf{x}_{i}^{n+1}\right)\right]\kappa_{i}^{n+1}.\\
  \end{aligned}
  \label{eq:second order ion weight pushing}
\end{equation}
\begin{equation}
  \begin{aligned}
  \frac{w_e^*-w_e^n}{\Delta t/2}=&-\mu_{e} \mathbf{b}\cdot\left(\nabla\times\mathbf{E}_{1}^{n}\right)\left(\mathbf{x}_{e}^{n}\right)-v_{e z}^{n} E_{1z}^{n}\left(\mathbf{x}_{e}^{n}\right)-\left[E_{1 y}^{n}\left(\mathbf{x}_{e}^{n}\right)+v_{ez}^n B_{1 x}^n\left(\mathbf{x}_{e}^{n}\right)\right]\kappa_{e}^{n},\\
  \frac{w_e^{n+1}-w_e^*}{\Delta t/2}=&-\mu_{e} \mathbf{b}\cdot\left(\nabla\times\mathbf{E}_{1}^{n+1}\right)\left(\mathbf{x}_{e}^{n+1}\right)-v_{e z}^{n+1} E_{1z}^{n+1}\left(\mathbf{x}_{e}^{n+1}\right)-\left[E_{1 y}^{n+1}\left(\mathbf{x}_{e}^{n+1}\right)+v_{ez}^{n+1} B_{1 x}^{n+1}\left(\mathbf{x}_{e}^{n+1}\right)\right]\kappa_{e}^{n+1}.
  \end{aligned}
  \label{eq:second order electron weight pushing}
\end{equation}
\revB{Here $\kappa_{i}^{n+1}$ and $\kappa_{e}^{n+1}$ are defined as}
\begin{equation}
  \begin{aligned}
    &\kappa_{i}^{n+1}=\frac{\partial ln n_{i}}{\partial x}+\left[\frac{T_{e}\left(v_{i}^{n+1}\right)^{2}}{2T_{i}}-\frac{3}{2}\right]\frac{\partial ln T_{i}}{\partial x},\\
    &\kappa_{e}^{n+1}=\frac{\partial ln n_{e}}{\partial x}+\left[\frac{m_{e}\left(v_{ez}^{n+1}\right)^{2}}{2m_{i}}+\mu_{e}B_{0}-\frac{3}{2}\right] \frac{\partial \ln T_e}{\partial x}.
  \end{aligned}
\end{equation}

\revB{In this paper, the second-order scheme refers to the semi-implicit formulations used to advance the particle weights, as given by Eqs.~(\ref{eq:second order ion weight pushing}) and (\ref{eq:second order electron weight pushing}).
The electric field equations—namely, the implicit parallel Ampere's law and perpendicular Ohm's law—do not themselves involve time derivatives.
Their compatibility with the second-order time-stepping scheme is achieved by consistently splitting the implicit parts of the particle weights.
Under Eqs.~(\ref{eq:second order ion weight pushing}) and (\ref{eq:second order electron weight pushing}), the particle current and pressure at time step $t^{n+1}$ can be split into intermediate and implicit parts as follows,}
\begin{equation}
  \mathbf{J}_{i}^{n+1}=\mathbf{J}_{i}^{*}+\mathbf{J}_{i}^{\text{Imp}}, \quad {J}_{ez}^{n+1}=J_{ez}^{*}+J_{ez}^{\text{Imp}}, \quad p_{1,e\perp}^{n+1}=p_{1,e\perp}^{*}+p_{1,e\perp}^{\text{Imp}},
\end{equation}
\revB{where the superscripts $*$ and $\text{Imp}$ denote the intermediate and implicit parts, respectively.
The detailed expressions of the above terms are}
\begin{equation}
  \begin{aligned}
    &\mathbf{J}_{i}^{*}\left(\mathbf{x}_{g}\right)=\frac{1}{N_{p}}\sum_{j}\mathbf{v}_{ij}^{n+1}w_{ij}^{*}S\left(\mathbf{x}_{g}-\mathbf{x}_{ij}^{n+1}\right),\\
    &\mathbf{J}_{i}^{\text{Imp}}\left(\mathbf{x}_{g}\right)=\frac{\Delta t}{2N_{p}}\sum_{j}\mathbf{v}_{ij}^{n+1}\left\{\frac{T_{e}}{T_{i}}\mathbf{v}_{ij}^{n+1}\cdot\mathbf{E}_{1}^{n+1}\left(\mathbf{x}_{ij}^{n+1}\right)-\left[E_{1y}^{n+1}\left(\mathbf{x}_{ij}^{n+1}\right)+v_{ijz}^{n+1}B_{1x}^{n+1}\left(\mathbf{x}_{ij}^{n+1}\right)-v_{ijx}^{n+1}B_{1z}^{n+1}\left(\mathbf{x}_{ij}^{n+1}\right)\right]\kappa_{ij}^{n+1}\right\}S\left(\mathbf{x}_{g}-\mathbf{x}_{ij}^{n+1}\right),\\
  \end{aligned}
\end{equation}

\begin{equation}
  \begin{aligned}
    J_{ez}^{*}\left(\mathbf{x}_{g}\right)=&-\frac{1}{N_{p}}\sum_{j}{v}_{ejz}^{n+1}w_{ej}^{*}S\left(\mathbf{x}_{g}-\mathbf{x}_{ej}^{n+1}\right),\\
    J_{ez}^{\text{Imp}}(\mathbf{x}_{g})=&\frac{\Delta t}{2 N_{p}}\sum_{j}v_{ejz}^{n+1}S\left(\mathbf{x}_{g}-\mathbf{x}_{ej}^{n+1}\right)\left\{\mu_{ej} \mathbf{b}\cdot\left(\nabla\times\mathbf{E}_{1}^{n+1}\right)\left(\mathbf{x}_{ej}^{n+1}\right)\right.\\
    &\left.+v_{ejz}^{n+1}E_{1z}^{n+1}\left(\mathbf{x}_{ej}^{n+1}\right)+\left[E_{1y}^{n+1}\left(\mathbf{x}_{ej}^{n+1}\right)+v_{ejz}^{n+1}B_{1x}^{n+1}\left(\mathbf{x}_{ej}^{n+1}\right)\right]\kappa_{ej}^{n+1}\right\},
  \end{aligned}
\end{equation}

\begin{equation}
  \begin{aligned}
  p_{1,e\perp}^{*}(\mathbf{x}_{g})=&\frac{1}{N_{p}}\sum_{j} \left(\mu_{ej}B_{0}\right)w_{ej}^{*}S\left(\mathbf{x}_{g}-\mathbf{x}_{ej}^{n+1}\right),\\
  p_{1,e\perp}^{\text{Imp}}(\mathbf{x}_{g})=&-\frac{\Delta t}{2 N_{p}}\sum_{j}\left(\mu_{ej}B_{0}\right) S\left(\mathbf{x}_{g}-\mathbf{x}_{ej}^{n+1}\right)\left\{\mu_{ej} \mathbf{b}\cdot\left(\nabla\times\mathbf{E}_{1}^{n+1}\right)\left(\mathbf{x}_{ej}^{n+1}\right)\right.\\
  &\left.+v_{ejz}^{n+1}E_{1z}^{n+1}\left(\mathbf{x}_{ej}^{n+1}\right)+\left[E_{1y}^{n+1}\left(\mathbf{x}_{ej}^{n+1}\right)+v_{ejz}^{n+1}B_{1x}^{n+1}\left(\mathbf{x}_{ej}^{n+1}\right)\right]\kappa_{ej}^{n+1}\right\}.
  \end{aligned}
\end{equation}

\revB{Therefore, formally, the perpendicular Ohm's law can be expressed as }
\begin{equation}
  \begin{aligned}
  &\beta_{e}\mathbf{E}_{1 \perp}^{n+1}-\frac{\Delta t}{2} \mathbf{b} \times\left(\nabla \times \nabla \times \mathbf{E}_1^{n+1}\right)+\beta_{e}\mathbf{J}_{i\perp}^{\text{Imp}}\times\mathbf{b}+\beta_{e}\nabla_{\perp}p_{1,e\perp}^{\text{Imp}}=-\mathbf{b} \times\left(\nabla \times \mathbf{B}_1^{*}\right)-\beta_{e}\nabla_{\perp} p_{1,e\perp }^{*}-\beta_{e}\mathbf{J}_{i \perp}^* \times \mathbf{b}.
  \label{eq:second order perpendicular ohm law}
  \end{aligned}
\end{equation}

\revB{The implicit parallel Ampere's law becomes}
\begin{equation}
  \begin{aligned}
    &\frac{\Delta t}{2}\mathbf{b}\cdot\nabla\times\left(\nabla\times\mathbf{E}_{1}^{n+1}\right)+\beta_{e}\left(J_{ez}^{\text{Imp}}+J_{iz}^{\text{Imp}}\right)={\mathbf{b}\cdot\nabla\times\mathbf{B}_{1}^{*}}-\beta_{e}\left(J_{ez}^{*}+J_{iz}^{*}\right).
  \end{aligned}
  \label{eq:second order ampere law}
\end{equation}
\revB{Here, to ensure a consistent order of accuracy in time, the Faraday's law is also discretized using the second-order scheme}
\begin{equation}
  \mathbf{B}_{1}^{n+1}=\mathbf{B}_{1}^{*}-\frac{\Delta t}{2}\left(\nabla\times\mathbf{E}_{1}^{n+1}\right),
  \label{eq:second order faraday law}
\end{equation}
\revB{with $\mathbf{B}_{1}^{*}=\mathbf{B}_{1}^{n}-\left(\nabla\times\mathbf{E}_{1}^{n}\right){\Delta t}/{2}$.}

\revB{In practice, Eq.~(\ref{eq:second order perpendicular ohm law}) and (\ref{eq:second order ampere law}) are solved using an iterative method analogous to Eq.~(\ref{eq:iterative implicit ampere law}). 
By following a derivation similar to the one leading from Eq.~(\ref{eq:implicit parallel Ampere's law}) to Eq.~(\ref{eq:iterative implicit ampere law}), the corresponding iterative forms of the field equations are obtained.}

\revB{1. The iterative perpendicular Ohm's law}
\begin{equation}
  \begin{aligned}
    &\beta_{e}\mathbf{E}_{1\perp}^{k+1}-\frac{\Delta t}{2}\mathbf{b}\times\left(\nabla \times\nabla\times \mathbf{E}_{1}^{k+1}\right)-\frac{\Delta t^{2}\beta_{e}}{4}\frac{T_{i}}{T_{e}}\frac{\partial ln n_{i} T_{i}}{\partial x}\left(\mathbf{\hat{x}}\times\mathbf{b}\right)\mathbf{b}\cdot\left(\nabla \times \mathbf{E}_{1}^{k+1}\right)\\
    &+\frac{\Delta t\beta_{e}}{2}\mathbf{E}_{1\perp}^{k+1}\times\mathbf{b}-\frac{\Delta t\beta_{e}}{B_{0}}\nabla_{\perp}\left[\mathbf{b}\cdot\left(\nabla\times\mathbf{E}_{1}^{k+1}\right)\right]-\frac{\Delta t\beta_{e}}{2}\frac{\partial ln n_{e} T_{e}}{\partial x}\nabla_{\perp}E_{1y}^{k+1}\\
    &=-\mathbf{b}\times\left(\nabla \times \mathbf{B}_{1}^{*}\right)-\beta_{e}\nabla_{\perp} p_{1,e\perp}^{*}-\beta_{e}\mathbf{J}_{i\perp}^{*}\times\mathbf{B}_{0}-\frac{\Delta t\beta_{e}}{2N_{p}}\left(\mathbf{\hat{x}}\times\mathbf{b}\right)\sum_{j}\left(v_{ijx}^{n+1}\right)^{2}B_{1z}^{*}\left(\mathbf{x}_{ij}^{n+1}\right)\kappa_{ij}^{n+1}S\left(\mathbf{x}_{g}-\mathbf{x}_{ij}^{n+1}\right)\\
    &-\frac{\left(\mathbf{\hat{x}}\times\mathbf{b}\right)\Delta t^{2}\beta_{e}}{4}\left[\frac{T_{i}}{T_{e}}\frac{\partial ln n_{i} T_{i}}{\partial x}\mathbf{b}\cdot\left(\nabla \times \mathbf{E}_{1}^{k}\right)-\frac{1}{N_{p}}\sum_{j}\left(v_{ijx}^{n+1}\right)^{2}\mathbf{b}\cdot\left(\nabla\times\mathbf{E}_{1}^{k}\right)\left(\mathbf{x}_{ij}^{n+1}\right)\kappa_{ij}^{n+1}S\left(\mathbf{x}_{g}-\mathbf{x}_{ij}^{n+1}\right)\right]\\
    &+\frac{\Delta t\beta_{e}}{2}\left\{\mathbf{E}_{1\perp}^{k}-\frac{1}{N_{p}}\frac{T_{e}}{T_{i}}\sum_{j}\left[\mathbf{v}_{ij\perp}^{n+1}\cdot\mathbf{E}_{1\perp}^{k}\left(\mathbf{x}_{ij}^{n+1}\right)\right]\mathbf{v}_{ij\perp}^{n+1}S\left(\mathbf{x}_{g}-\mathbf{x}_{ij}^{n+1}\right)\right\}\times\mathbf{b}\\
    &-\Delta t\beta_{e}\left\{\frac{1}{B_{0}}\nabla_{\perp}\left[\mathbf{b}\cdot\left(\nabla\times\mathbf{E}_{1}^{k}\right)\right]-\frac{1}{2N_{p}}\nabla_{\perp}\sum_{j}\left(\mu_{ej}^{2}B_{0}\right)\mathbf{b}\cdot\left(\nabla\times\mathbf{E}_{1}^{k}\right)\left(\mathbf{x}_{ej}^{n+1}\right)S\left(\mathbf{x}_{g}-\mathbf{x}_{ej}^{n+1}\right)\right\}\\
    &-\frac{\Delta t\beta_{e}}{2}\left[\frac{\partial ln n_{e} T_{e}}{\partial x}\nabla_{\perp}E_{1y}^{k}-\frac{1}{N_{p}}\nabla_{\perp}\sum_{j}\left(\mu_{ej}B_{0}\right)E_{1y}^{k}\left(\mathbf{x}_{ej}^{n+1}\right)\kappa_{ej}^{n+1}S\left(\mathbf{x}_{g}-\mathbf{x}_{ej}^{n+1}\right)\right].
  \end{aligned}
  \label{eq: second order iterative perpendicular ohm law}
\end{equation}

\revB{2. The iterative form of the implicit parallel Ampere's law}
\begin{equation}
  \begin{aligned}
    &\frac{\Delta t}{2}\mathbf{b}\cdot\nabla\times\left(\nabla\times\mathbf{E}_{1}^{k+1}\right)+\frac{\Delta t\beta_{e}}{2}\left(\frac{m_{i}}{m_{e}}+1\right)E_{1z}^{k+1}-\frac{\Delta t^{2}\beta_{e}}{4}\mathbf{\hat{x}}\cdot\left(\nabla\times\mathbf{E}_{1}^{k+1}\right)\left(\frac{m_{i}}{m_{e}}\frac{\partial ln n_{e} T_{e}}{\partial x}-\frac{T_{i}}{T_{e}}\frac{\partial ln n_{i} T_{i}}{\partial x}\right)\\
    &=\mathbf{b}\cdot\nabla\times\mathbf{B}_{1}^{*}-\beta_{e}\left(J_{ez}^{*}+J_{iz}^{*}\right)-\frac{\Delta t \beta_{e}}{2 N_{p}}\left[\sum_{j}\left(v_{ejz}^{n+1}\right)^{2}B_{1x}^{*}\left(\mathbf{x}_{ej}^{n+1}\right)\kappa_{ej}^{n+1}S\left(\mathbf{x}_{g}-\mathbf{x}_{ej}^{n+1}\right)-\sum_{j}\left(v_{ijz}^{n+1}\right)^{2}B_{1x}^{*}\left(\mathbf{x}_{ij}^{n+1}\right)\kappa_{ij}^{n+1}S\left(\mathbf{x}_{g}-\mathbf{x}_{ij}^{n+1}\right)\right]\\
    &+\frac{\Delta t \beta_{e}}{2}\left[\frac{m_{i}}{m_{e}}E_{1z}^{k}-\frac{1}{N_{p}}\sum_{j}\left(v_{ejz}^{n+1}\right)^{2}E_{1z}^{k}\left(\mathbf{x}_{ej}^{n+1}\right)S\left(\mathbf{x}_{g}-\mathbf{x}_{ej}^{n+1}\right)\right]+\frac{\Delta t \beta_{e}}{2}\left[E_{1z}^{k}-\frac{1}{N_{p}}\frac{T_{e}}{T_{i}}\sum_{j}\left(v_{ijz}^{n+1}\right)^{2}E_{1z}^{k}\left(\mathbf{x}_{ij}^{n+1}\right)S\left(\mathbf{x}_{g}-\mathbf{x}_{ij}^{n+1}\right)\right]\\
    &-\frac{\Delta t^{2} \beta_{e}}{4}\left[\frac{m_{i}}{m_{e}}\frac{\partial ln n_{e}T_{e}}{\partial x}\mathbf{\hat{x}}\cdot\left(\nabla\times\mathbf{E}_{1}^{k}\right)-\frac{1}{N_{p}}\sum_{j}\left(v_{ejz}^{n+1}\right)^{2}\mathbf{\hat{x}}\cdot\left(\nabla\times\mathbf{E}_{1}^{k}\right)\left(\mathbf{x}_{ej}^{n+1}\right)\kappa_{ej}^{n+1}S\left(\mathbf{x}_{g}-\mathbf{x}_{ej}^{n+1}\right)\right]\\
    &+\frac{\Delta t^{2} \beta_{e}}{4}\left[\frac{T_{i}}{T_{e}}\frac{\partial ln n_{i} T_{i}}{\partial x}\mathbf{\hat{x}}\cdot\left(\nabla\times\mathbf{E}_{1}^{k}\right)-\frac{1}{N_{p}}\sum_{j}\left(v_{ijz}^{n+1}\right)^{2}\mathbf{\hat{x}}\cdot\left(\nabla\times\mathbf{E}_{1}^{k}\right)\left(\mathbf{x}_{ij}^{n+1}\right)\kappa_{ij}^{n+1}S\left(\mathbf{x}_{g}-\mathbf{x}_{ij}^{n+1}\right)\right].
  \end{aligned}
  \label{eq: second order iterative parallel ampere law}
\end{equation}
\revB{
Equations~(\ref{eq: second order iterative perpendicular ohm law}) and (\ref{eq: second order iterative parallel ampere law}) are employed to update the iterative electric field $\mathbf{E}_{1}^{k+1}$. 
On the right-hand side of these equations, the particle summation terms involving $\mathbf{E}_{1}^{k}$ are paired with their corresponding approximations in the limit of sufficient particles and grids to illustrate the iterative structure of the solver.
It should be noted that certain particle summation terms, such as $\sum_{j} v_{ijz}^{n+1}\left(\mathbf{v}_{ij\perp}^{n+1}\cdot\mathbf{E}_{1\perp}^{k}\right)S\left(\mathbf{x}_{g}-\mathbf{x}_{ij}^{n+1}\right)$, have been omitted in the written form of Eqs.~(\ref{eq: second order iterative perpendicular ohm law}) and (\ref{eq: second order iterative parallel ampere law}) because their corresponding approximations evaluate to zero analytically.
Nevertheless, in the actual implementation of the FIDES code, these terms are inherently retained, as the particle moments are accumulated monolithically rather than evaluated on a term-by-term basis.
}

\revB{In summary, the structure of this second-order semi-implicit scheme follows the same flow as that depicted in Fig.~\ref{fig: flowchart} for the implicit discretization scheme in section 2.2.
The only difference lies in the specific expressions used for the particle weights and field equations, which are now replaced by their second-order counterparts.
Specifically, the particle weights are advanced using Eqs.~(\ref{eq:second order ion weight pushing}) and (\ref{eq:second order electron weight pushing}).
The electric field is computed self-consistently from the perpendicular Ohm's law  (\ref{eq: second order iterative perpendicular ohm law}) and the implicit parallel Ampere's law (\ref{eq: second order iterative parallel ampere law}), after which the magnetic field is updated using Faraday's law in its second-order form (\ref{eq:second order faraday law}).}

\subsection{Odd-even decoupling problem}

\begin{figure*}[t]
  \centering
  \subfigure[Odd-even decoupling]
  {
    \begin{minipage}[b]{.45\textwidth}
      \centering
      \includegraphics[width=\textwidth]{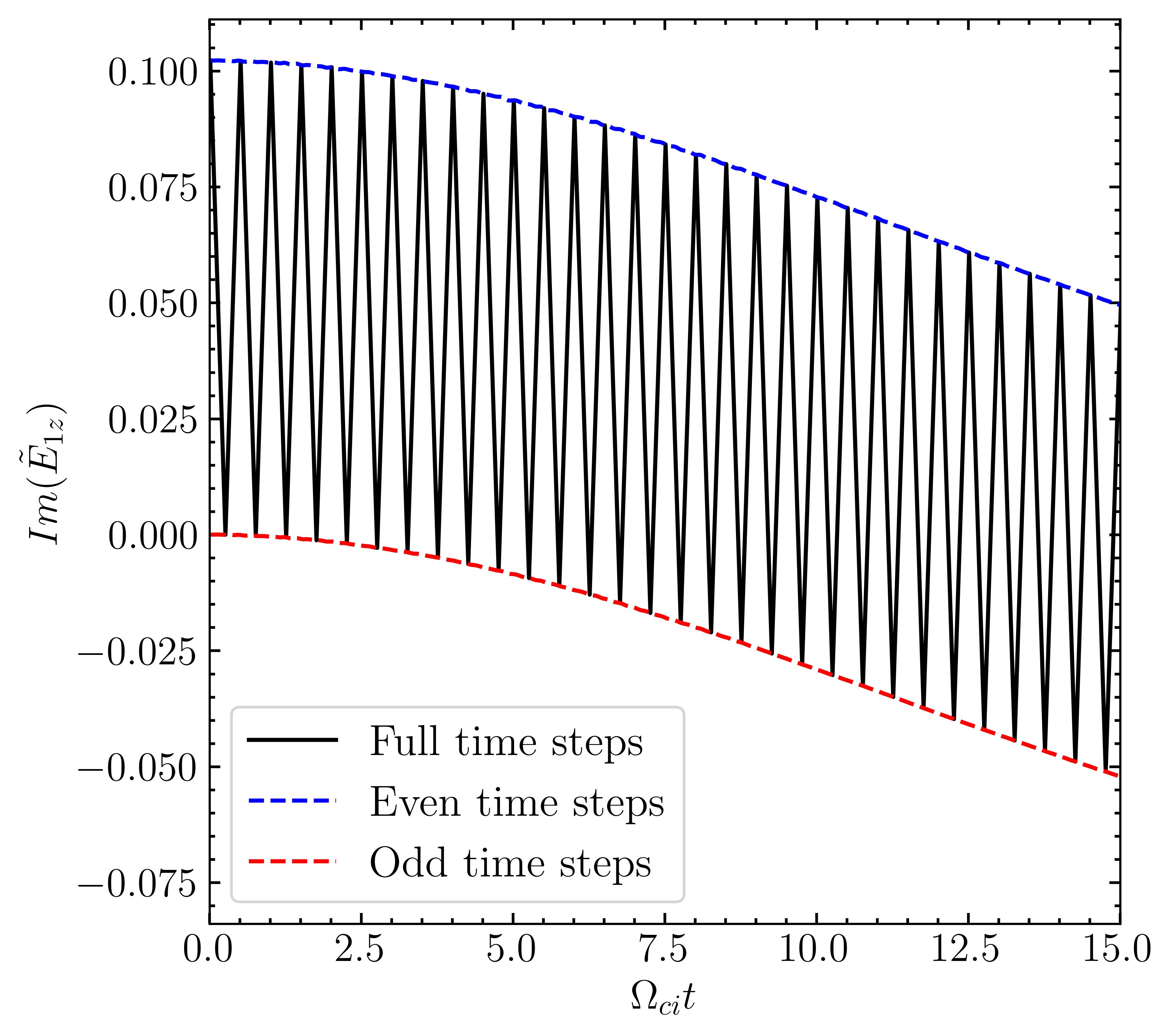}
    \end{minipage}
  }
  \subfigure[After optimization]
  {
    \begin{minipage}[b]{.45\textwidth}
      \centering
      \includegraphics[width=\textwidth]{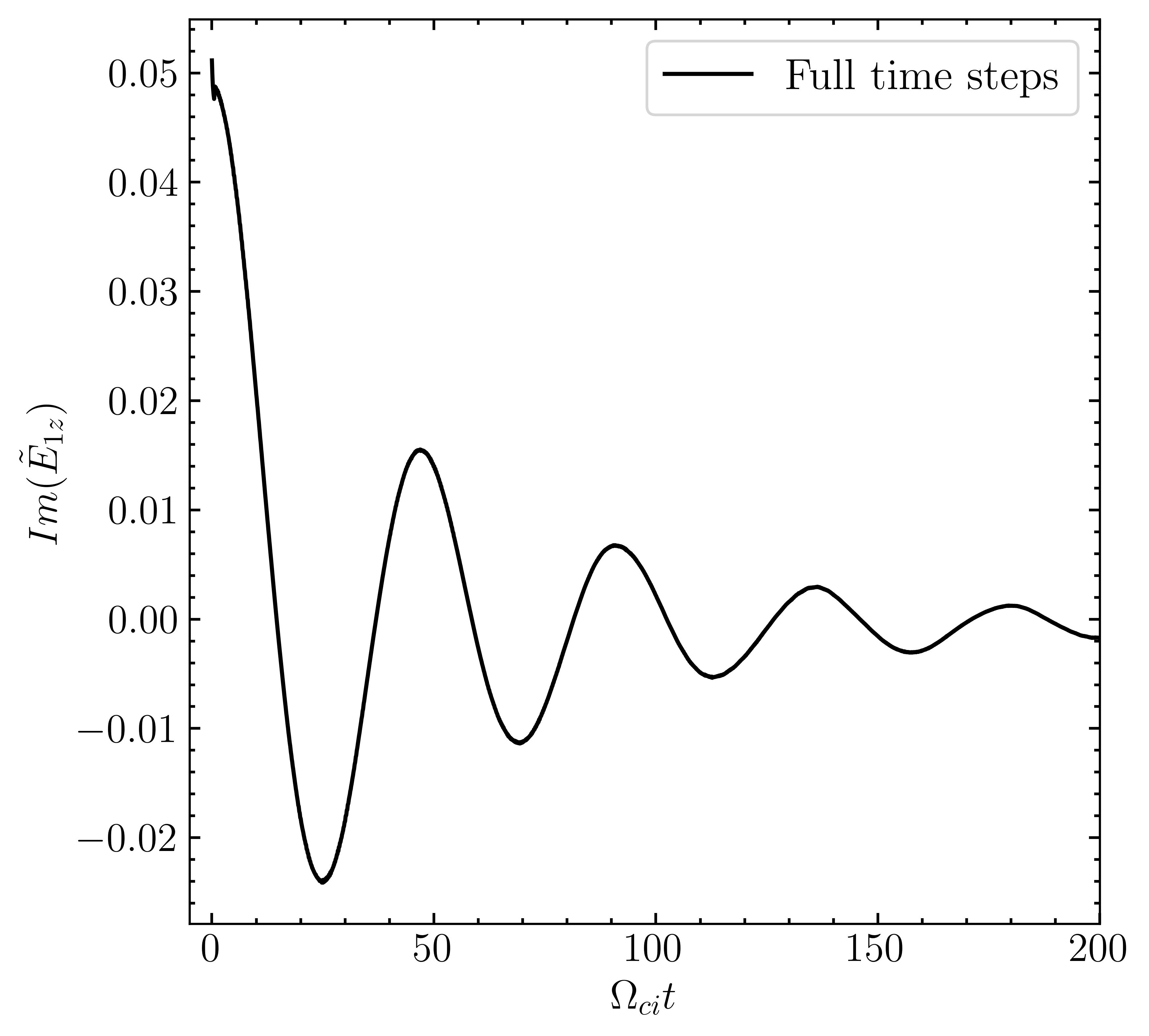}
    \end{minipage}
  }
  \caption{\label{fig:odd-even decoupling} (a) The odd-even decoupling problem in the IAW simulation for the second-order scheme. (b) The same IAW simulation after applying the proposed three-point second-order scheme with $a=0.01$ to advance electron weights. The first-order implicit scheme is used for initialization during the first 50 time steps. Here simulation parameters are $n_{x}=n_{y}=2,n_{z}=128,N_{p}=256,\Omega_{ci}\Delta t=0.01$.The plasma parameters are $\beta_e=0.01,T_{i}/T_{e}=0.25,k_{x}\rho_{s}=k_{y}\rho_{s}=0, k_{z}\rho_{s}=0.1$. For clarity, the plotting interval in (a) is $25\Omega_{ci}\Delta t$. }
\end{figure*}

The IAW simulations reveal a problem with the second-order scheme.
As shown in Fig.~\ref{fig:odd-even decoupling} (a), it seems that the solutions at odd and even timesteps become decoupled, producing a sawtooth-like oscillation in the $E_{z}-t$ diagram.
This phenomenon is the so-called odd-even decoupling in computational fluid dynamics \cite{Patankar1980}, though here it manifests in the time domain rather than in space.
The odd-even decoupling problem compromises the accuracy of simulation results and can trigger additional numerical instabilities.
We analyze its underlying mechanism below and propose targeted solutions.

\revB{To illustrate the odd-even decoupling problem, we make the following simplifications for the IAW case.}

\revB{(1) We restrict the discussion to a one-dimensional electrostatic configuration, where $\mathbf{E}_{1}=E_{1z}\hat{\mathbf{z}}$ aligns with the direction of the equilibrium magnetic field.}

\revB{(2) Since the electron current contributes the most part of particle current, only electron weights and motions are considered.
Specifically, }
\begin{equation}
  \begin{aligned}
  \frac{w_{e}^{*}-w_{e}^n}{\Delta t}=-\frac{1}{2} v_{ez}E_{1 z}^n\left(z_{e}^{n}\right),~\frac{w_{e}^{n+1}-w_{e}^{*}}{\Delta t}=-\frac{1}{2} v_{ez}E_{1 z}^{n+1}\left(z_{e}^{n+1}\right).
  \end{aligned}
\end{equation}
\revB{The linear electron motion equation in the z direction is the same as Eq.~(\ref{eq:simplified particle motion}). Therefore, we can get the intermediate electron weight $w_{e}^{*}$ advanced from $w_{e}^{n}$, }
\begin{equation}
  w_{e}^{*}=w_{e}^{0}-\frac{\Delta t}{2}v_{ez}E_{1z}^{0}\left(z_{e}^{0}\right)-\Delta t v_{ez}\sum_{m=1}^{n}E_{1z}^{m}\left(z_{e}^{m}\right).
\end{equation}

\revB{The governing field equation is the implicit parallel Ampere's law, which can be expressed as}
\begin{equation}
  \begin{aligned}
  &\frac{\Delta t}{2N_{p}}\sum_{j}\left(v_{ejz}^{2}\right)^{2} E_{1 z}^{n+1}\left(z_{ej}^{n+1}\right)S\left(z_{g}-z_{ej}^{n+1}\right)=\frac{1}{N_{p}}\sum_{j}v_{ejz}\left[w_{ej}^{0}-\frac{\Delta t}{2}v_{ejz}E_{1z}^{0}\left(z_{ej}^{0}\right)-\Delta t v_{ejz}\sum_{m=1}^{n}E_{1z}^{m}\left(z_{ej}^{m}\right)\right]S\left(z_{g}-z_{ej}^{n+1}\right).
  \end{aligned}
\end{equation}

\revB{For further derivation, we make the same assumptions (3)-(5) in section 3.2 and take the limit of infinite grid resolution, which leads to}
\begin{equation}
  \begin{aligned}
    \int \left(\frac{\Delta t}{2}\tilde{E}_{1z}^{0}v_{z}-\tilde{w}_{e}^{0}\right)v_{z}e^{-ik_{z}v_{z}\left(n+1\right)\Delta t}f_{e0}d v_{z}+\Delta t\sum_{m=1}^{n}\tilde{E}_{1z}^{m}\int v_{z}^{2}e^{-ik_{z}v_{z}\left(n+1-m\right)\Delta t}f_{e0}dv_{z}+\frac{\Delta t}{2}\tilde{E}_{1z}^{n+1}\int v_{z}^{2}f_{e0}d v_{z}=0,
  \end{aligned}
  \label{eq:field equation for odd-even decoupling}
\end{equation}
\revB{with $f_{e0}$ given by Eq.~(\ref{eq:normalized distribution function}).
The velocity integral in Eq.~(\ref{eq:field equation for odd-even decoupling}) can be calculated using the residue theorem,
}
\begin{equation}
  \begin{aligned}
    &\frac{1}{\sqrt{2\pi m_{i}/m_{e}}}\int_{-\infty}^{+\infty} v_{z}^{2} exp\left[-\frac{m_{e} v_{z}^{2}}{2 m_{i}}-ik_{z}v_{z}\left(n+1-m\right)\Delta t\right]d v_{z}\\
    &=\frac{1}{\sqrt{2\pi m_{i}/m_{e}}}exp\left[-k_{z}^{2}\left(n+1-m\right)^{2}\Delta t^{2}\frac{m_{i}}{2m_{e}}\right]\int_{-\infty-ik_{z}\left(n+1-m\right)\Delta t \frac{m_{i}}{m_{e}}}^{+\infty-ik_{z}\left(n+1-m\right)\Delta t \frac{m_{i}}{m_{e}}}v_{z}^{2}exp\left\{-\frac{m_{e}}{2m_{i}}\left[v_{z}+ik_{z}\left(n+1-m\right)\Delta t \frac{m_{i}}{m_{e}}\right]^{2}\right\}dv_{z}\\
    &=\frac{m_{i}}{m_{e}}\left[1-k_{z}^{2}\left(n+1-m\right)^{2}\Delta t^{2}\frac{m_{i}}{m_{e}}\right]exp\left[-k_{z}^{2}\left(n+1-m\right)^{2}\Delta t^{2}\frac{m_{i}}{2m_{e}}\right].\\
  \end{aligned}
\end{equation}
\revB{Therefore, we can obtain a recurrence relation of $E_{1z,k}^{n+1}$ from Eq.~(\ref{eq:field equation for odd-even decoupling}),}
\begin{equation}
  \begin{aligned}
  & i k_z(n+1) e^{-k_z^2(n+1)^2 \Delta t^2 \frac{m_i}{2 m_e}} \tilde{w}_{e}^0+\frac{1}{2} \tilde{E}_{1 z}^0 e^{-k_z^2(n+1)^2 \Delta t^2 \frac{m_i}{2 m_e}}\left[1-k_z^2(n+1)^2 \Delta t^2 \frac{m_i}{m_e}\right] \\
  & +\sum_{m=1}^n \tilde{E}_{1 z}^m e^{-k_z^2(n+1-m)^2 \Delta t^2 \frac{m_i}{2 m_e}}\left[1-k_z^2(n+1-m)^2 \Delta t^2 \frac{m_i}{m_e}\right]+\frac{1}{2} \tilde{E}_{1 z}^{n+1}=0.
  \end{aligned}
  \label{eq:recurrence relation}
\end{equation}

\revB{An explicit form of $\tilde{E}_{1z}^{n+1}$ is generally not available from the recurrence relation.
However, an intuitive understanding can be gained when the time step $n$ is moderate and $\Delta t$ is sufficiently small such that $k_{z}^{2}n^{2}\Delta t^{2}{m_{i}}/{m_{e}}\ll 1$ holds.
In this regime, the leading-order terms of the recurrence relation (\ref{eq:recurrence relation}) yield}
\begin{equation}
  \begin{aligned}
  \tilde{E}_{1 z}^{2 n}=\tilde{E}_{1 z}^0,\quad
  \tilde{E}_{1 z}^{2 n+1}=-\tilde{E}_{1 z}^0-2 i k_z \tilde{w}_{e}^0.
  \end{aligned}
\end{equation}
\revB{The leading-order expression of $\tilde{E}_{1 z}^{n}$ behaves as an alternating sequence determined by the parity of $n$, which accounts for the observed odd-even decoupling in the IAW simulations.
The time evolution of $\tilde{E}_{1z}^{n}$ in the order of $O(k_{z}^{2}\Delta t^{2}m_{i}/m_{e})$ is described by the higher-order terms of Eq.~(\ref{eq:recurrence relation}), which are not considered here.
}

To solve the odd-even decoupling problem, we generalize the preceding derivation.
If the electron weight pushing equation is
\begin{equation}
\frac{w_{e j}^{n+1}-w_{e j}^n}{\Delta t}=-v_{e j z}\left[a E_{1 z}^{n-1}\left(z_{e j}^{n-1}\right)+b E_{1 z}^n\left(z_{e j}^n\right)+c E_{1 z}^{n+1}\left(z_{e j}^{n+1}\right)\right],
\end{equation}
the leading-order terms of the implicit parallel Ampere's law for $k_{z}^{2}n^{2}\Delta t^{2}{m_{i}}/{m_{e}}\ll 1$ can be simplified as 
\begin{equation}
  ik_{z}\tilde{w}_{e}^{0}+a\tilde{E}_{1z}^{n-1}+b\tilde{E}_{1z}^{n}+c\tilde{E}_{1z}^{n+1}=0,
  \label{eq:generalized relation}
\end{equation}
which leads to the following two representative cases.

1. The first-order implicit scheme $(a=0,b=0,c=1)$. 
We can get \revB{$\tilde{E}_{1z}^{n}=-ik_{z}\tilde{w}_{e}^{0}$}, which can explain why the implicit discretization scheme doesn't encounter the odd-even decoupling problem.
Motivated by this, one may use the first-order implicit scheme during initialization to connect odd and even sequences.
Numerical experiments confirms that although the odd-even decoupling oscillation persists, its amplitude is significantly reduced.

2. The second-order scheme with three points $(a\in(0,0.25),b=0.5-2a,c=0.5+a)$.
The leading-order solution takes the form 
\begin{equation}
\tilde{E}_{1z}^{n}=
\begin{cases}\frac{1}{r_1-r_2}\left[\left(\tilde{E}_{1 z}^0+i k_z \tilde{w}_{e}^0\right)\left(r_1^{n+1}-r_2^{n+1}\right)-\frac{a}{0.5+a} i k_z \tilde{w}_{e}^0\left(r_1^n-r_2^n\right)\right]-i k_z \tilde{w}_{e}^0, & a \neq \frac{1}{16}, \\ {\left[\left(\tilde{E}_{1 z}^0+i k_z \tilde{w}_{e}^0\right)+n\left(\tilde{E}_{1 z}^0+\frac{4}{3} i k_z \tilde{w}_{e}^0\right)\right]\left(-\frac{1}{3}\right)^n-i k_z \tilde{w}_{e}^0,} & a=\frac{1}{16},\end{cases}
\end{equation}
where $r_{1,2}=\left(4a-1\pm\sqrt{1-16a}\right)/2(1+2a)$ are eigenvalues of Eq.~(\ref{eq:generalized relation}).
Since $|r_{1,2}|<1$, this scheme numerically damps odd-even decoupling oscillations.
In practice, choosing $a\ll 1$ can suppress the odd-even decoupling oscillations while preserving the accuracy of the underlying second-order scheme.

\begin{figure*}[t]
  \centering
  \subfigure[Odd-even decoupling]
  {
    \begin{minipage}[b]{.45\textwidth}
      \centering
      \includegraphics[width=\textwidth]{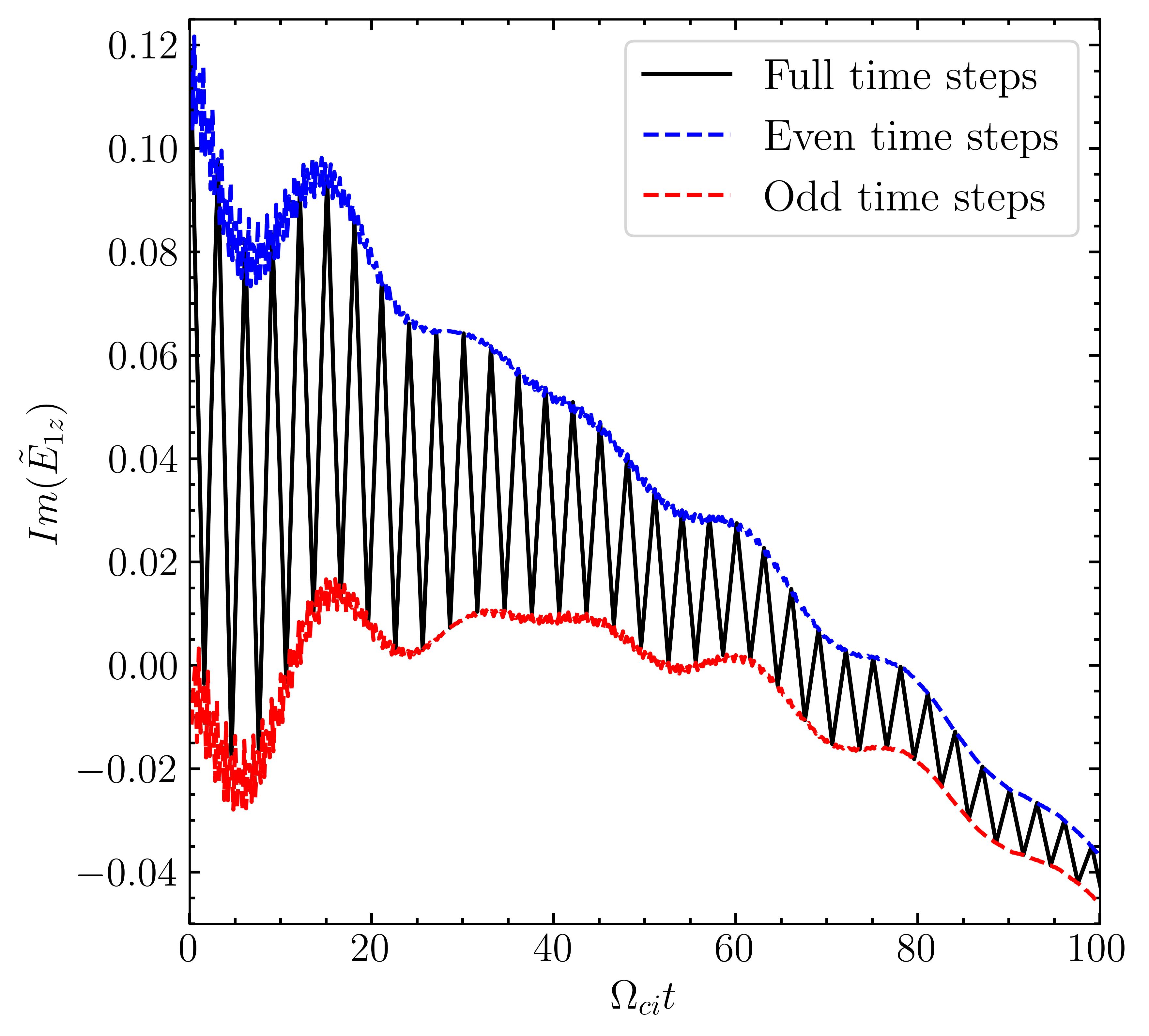}
    \end{minipage}
  }
  \subfigure[After optimization]
  {
    \begin{minipage}[b]{.45\textwidth}
      \centering
      \includegraphics[width=\textwidth]{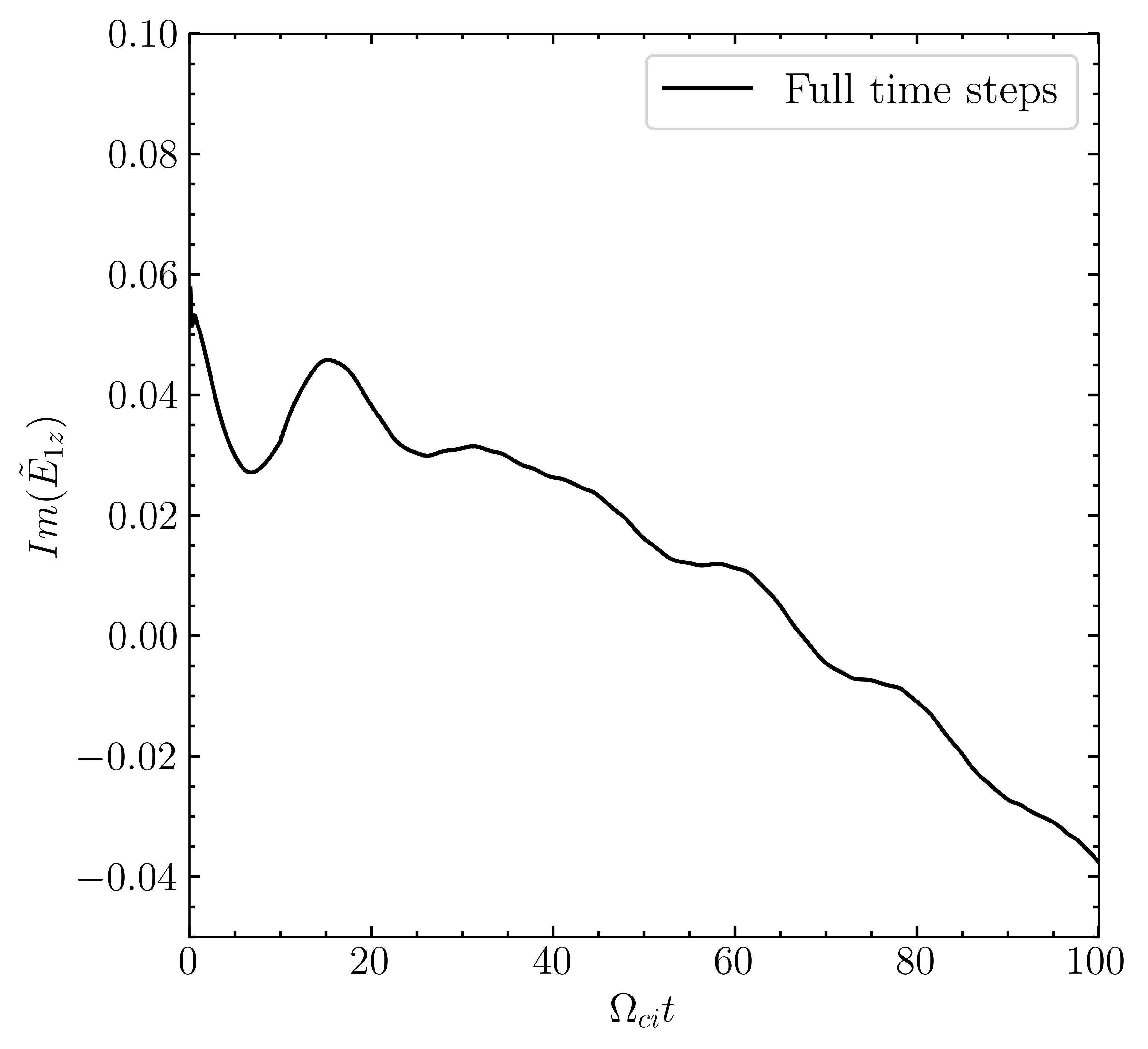}
    \end{minipage}
  }
  \caption{\label{fig:odd-even decoupling 2} (a) The odd-even decoupling problem observed in the ITG simulation when using the second-order scheme. (b) The same ITG simulation after applying the proposed solution, which uses the first-order implicit scheme for the first $50$ time steps and then employs the three-point second-order scheme with $a=0.01$ to advance electron weights in the running stage. Simulation parameters are $n_{x}=n_{y}=32,n_{z}=64,N_{p}=128,\Omega_{ci}\Delta t=0.1$. Plasma parameters are $\beta_e=0.001,T_{i}/T_{e}=1, \kappa_{n}\rho_{s}=0,\kappa_{ti}\rho_{s}=-0.3,\kappa_{te}\rho_{s}=0, k_{x}\rho_{s}=0.2,k_{y}\rho_{s}=0.4, k_{z}\rho_{s}=0.01$. For clarity, the plotting interval in (a) is $25\Omega_{ci}\Delta t$. }
\end{figure*}

\revAB{In the following simulations, the three-point scheme with the typical parameter $a=0.01$ is employed for advancing the electron weights.
Together with the first-order implicit scheme used during the initialization stage, this combination effectively resolves the odd-even decoupling problem, as demonstrated in Fig.~\ref{fig:odd-even decoupling} (b).}
\revB{Notably, the proposed method is not limited to the IAW simulations.
The key to overcoming the odd-even decoupling problem is to establish a connection between odd and even time steps.
We have verified that the method also eliminates this decoupling successfully in other scenarios, such as the ITG simulations shown in Fig.~\ref{fig:odd-even decoupling 2}.
}

\subsection{Numerical results}

\begin{figure*}[t]
  \centering
  \subfigure[Real frequency]
  {
    \begin{minipage}[b]{.45\textwidth}
      \centering
      \includegraphics[width=\textwidth]{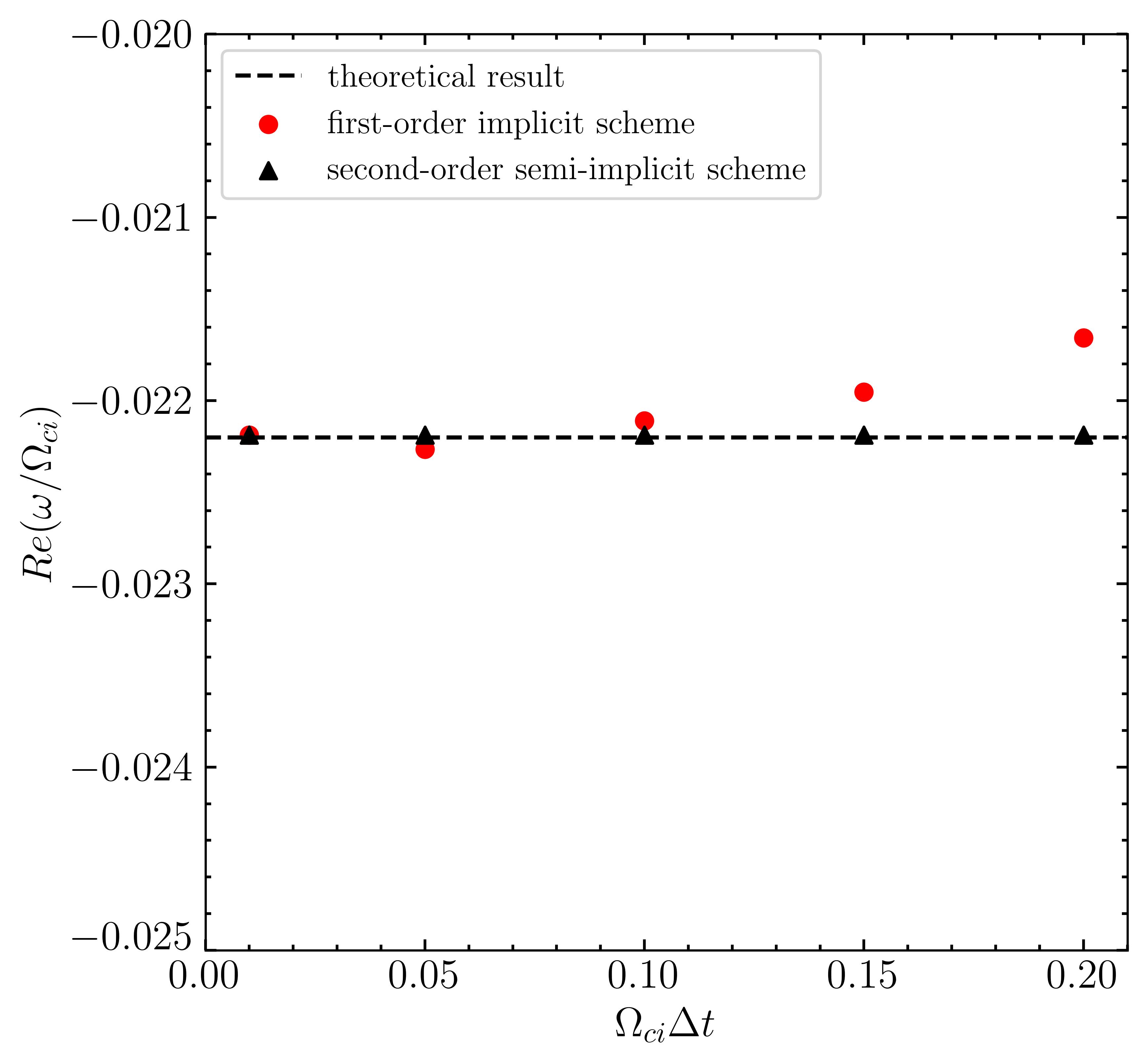}
    \end{minipage}
  }
  \subfigure[Growth rate]
  {
    \begin{minipage}[b]{.45\textwidth}
      \centering
      \includegraphics[width=\textwidth]{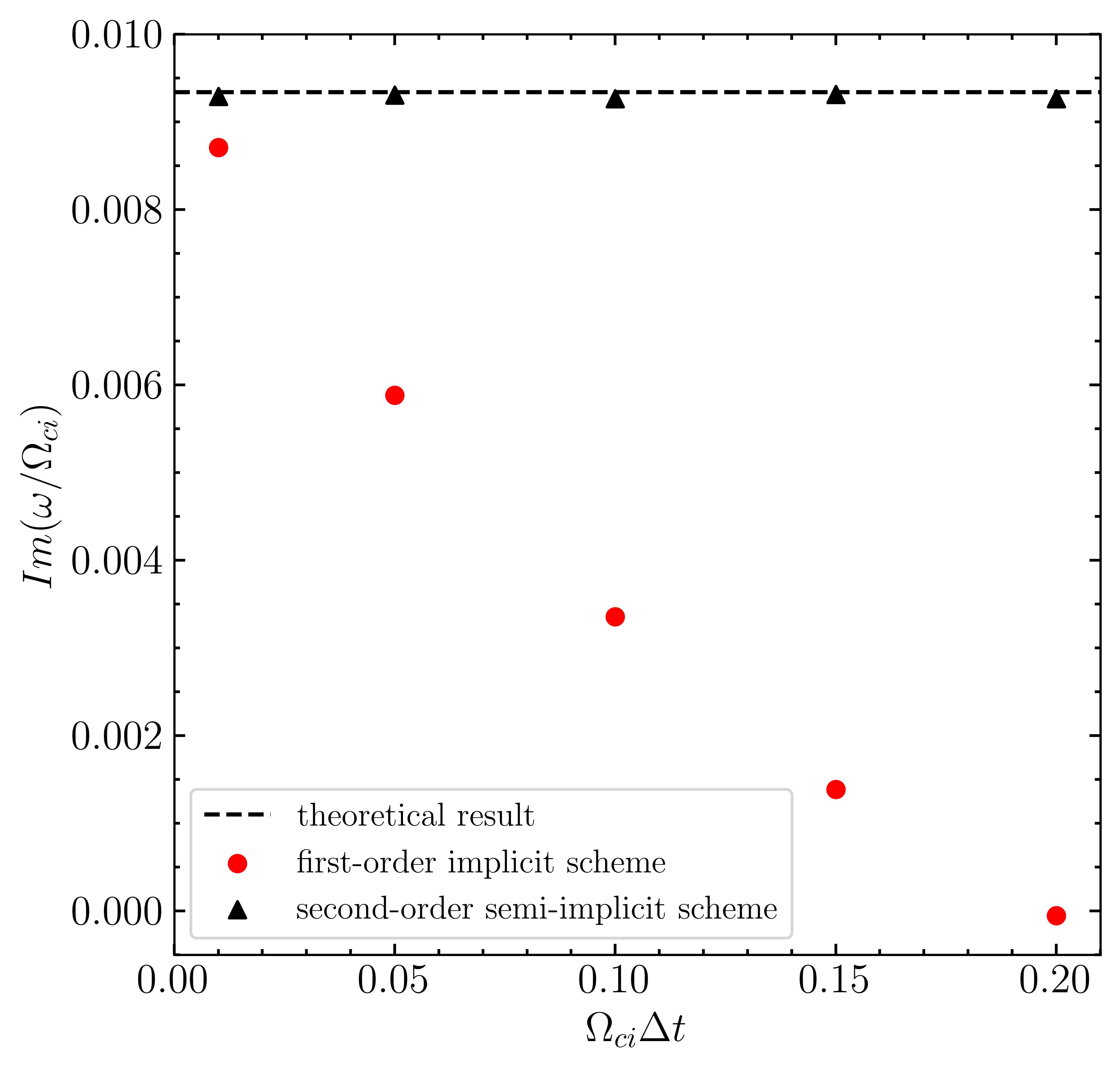}
    \end{minipage}
  }
  \caption{\label{fig: timestep refinement study} The timestep convergence study for ITG simulations between the first-order implicit and second-order semi-implicit schemes. Here simulation parameters are $n_{x}=n_{y}=32,n_{z}=64,N_{p}=128$. The plasma parameters are $\beta_e=0.001,\kappa_{n}\rho_{s}=0,\kappa_{ti}\rho_{s}=-0.3,\kappa_{te}\rho_{s}=0,k_{x}\rho_{s}=0.2,k_{y}\rho_{s}=0.4, k_{z}\rho_{s}=0.01$. }
\end{figure*}

\begin{figure*}[t]
  \centering
  \subfigure[Real frequency]
  {
    \begin{minipage}[b]{.45\textwidth}
      \centering
      \includegraphics[width=\textwidth]{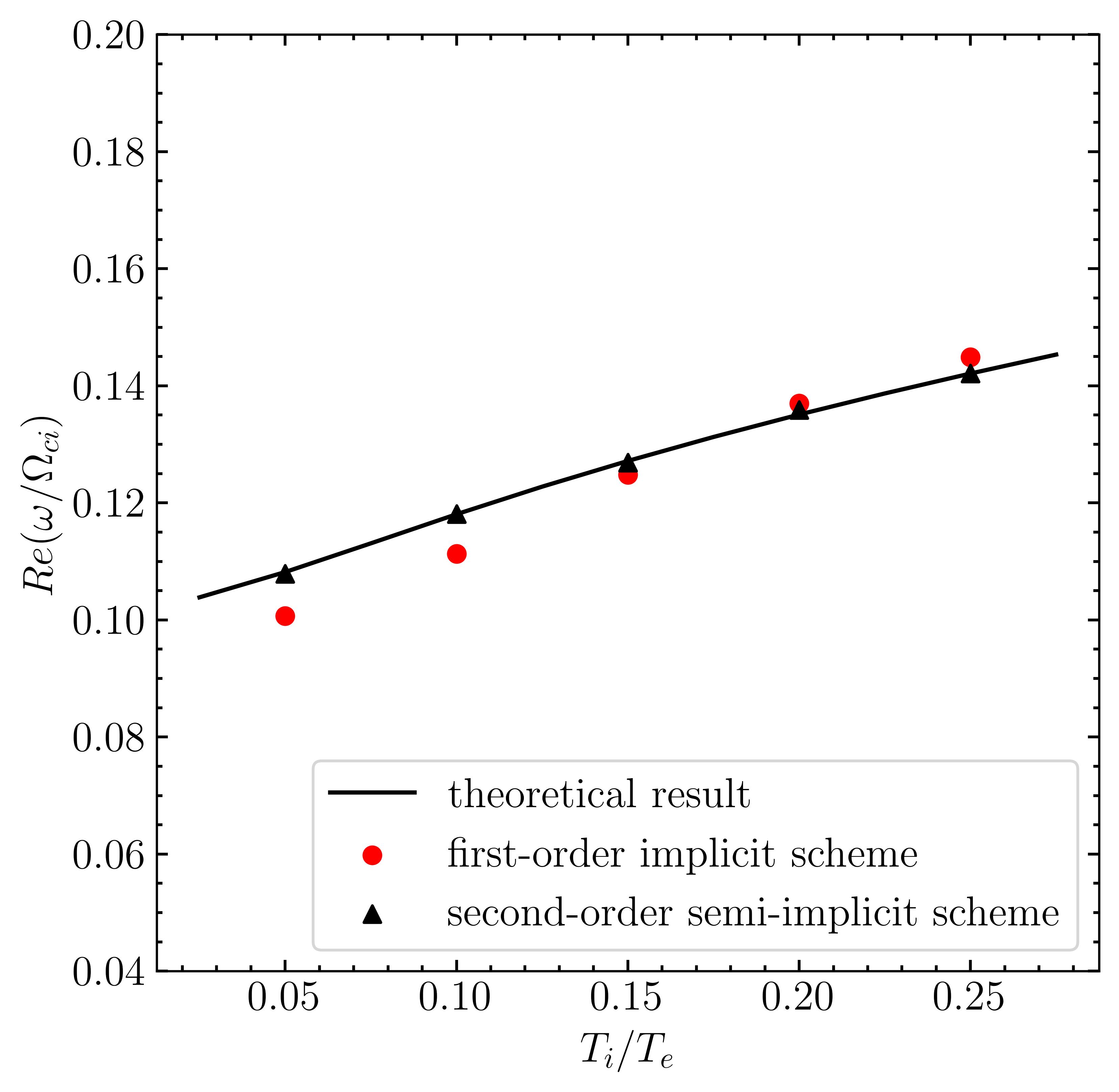}
    \end{minipage}
  }
  \subfigure[Damping rate]
  {
    \begin{minipage}[b]{.45\textwidth}
      \centering
      \includegraphics[width=\textwidth]{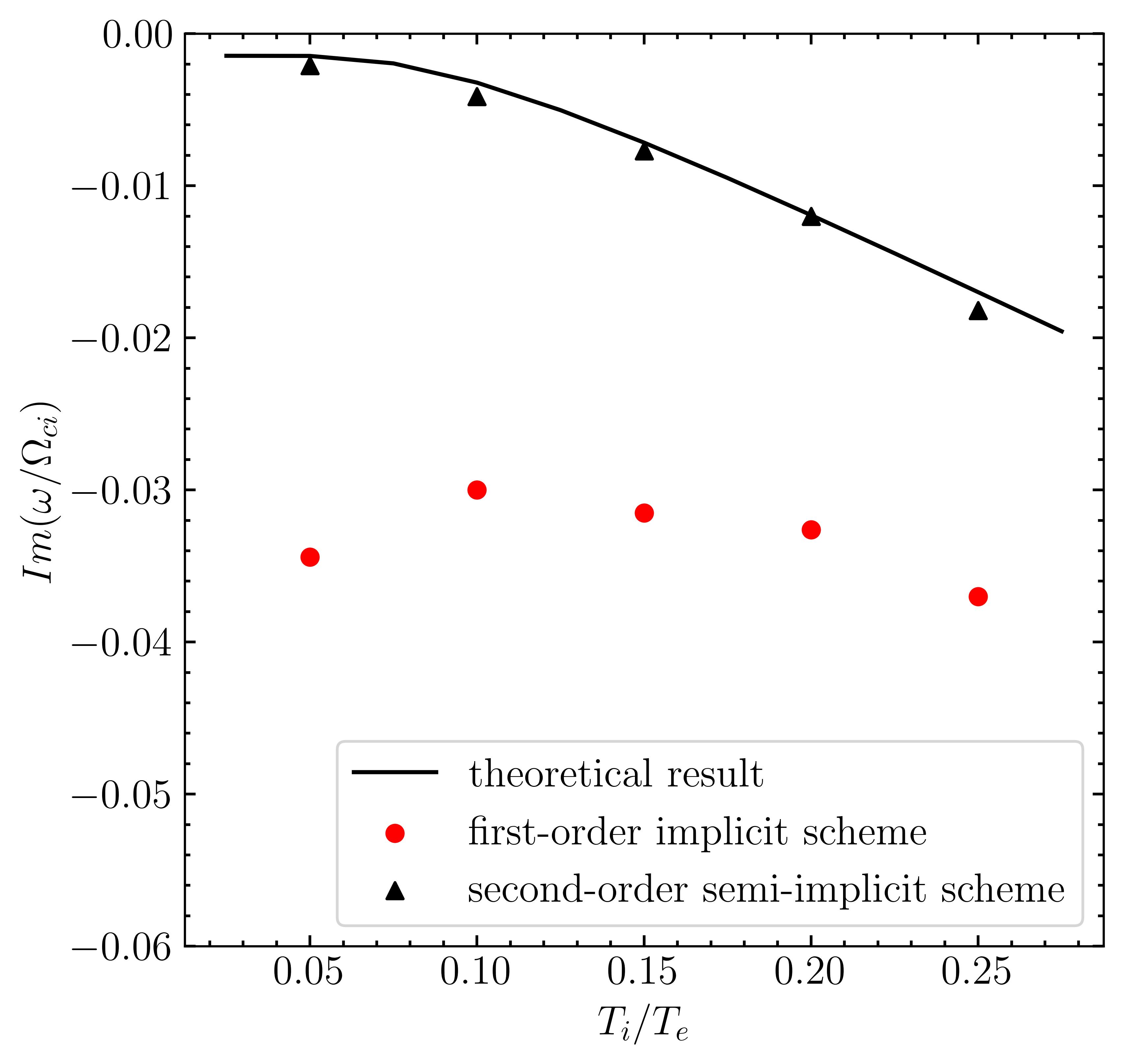}
    \end{minipage}
  }
  \caption{\label{fig:IAW summary 2} Comparison of IAW simulation results between the first-order ($\Omega_{ci}\Delta t=0.01$) and second-order ($\Omega_{ci}\Delta t=0.05$) schemes. All other parameters are consistent with those in Fig.~\ref{fig:IAW summary 1}.}
\end{figure*}

\begin{figure*}[t]
  \centering
  \subfigure[Real frequency]
  {
    \begin{minipage}[b]{.45\textwidth}
      \centering
      \includegraphics[width=\textwidth]{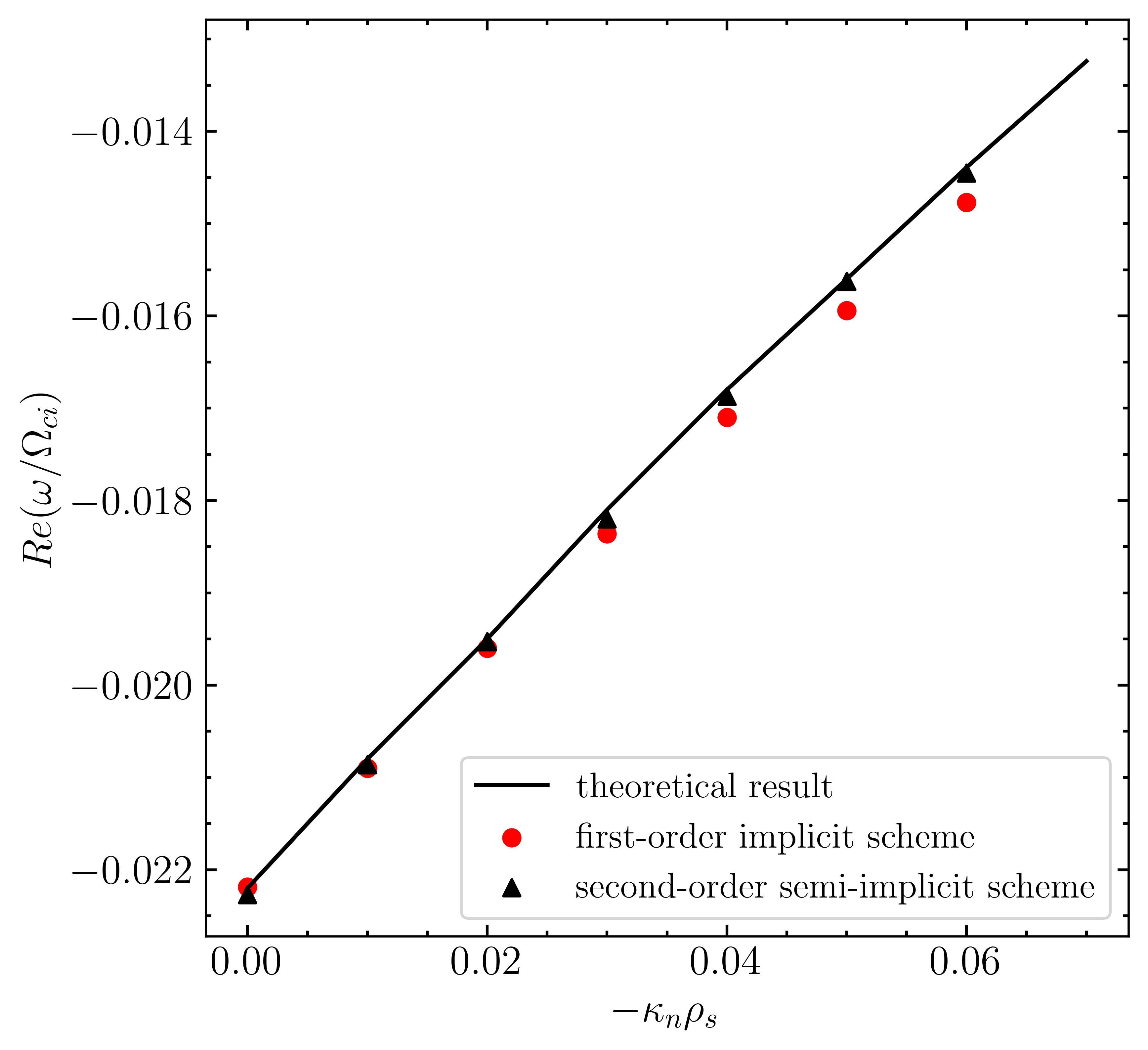}
    \end{minipage}
  }
  \subfigure[Growth rate]
  {
    \begin{minipage}[b]{.45\textwidth}
      \centering
      \includegraphics[width=\textwidth]{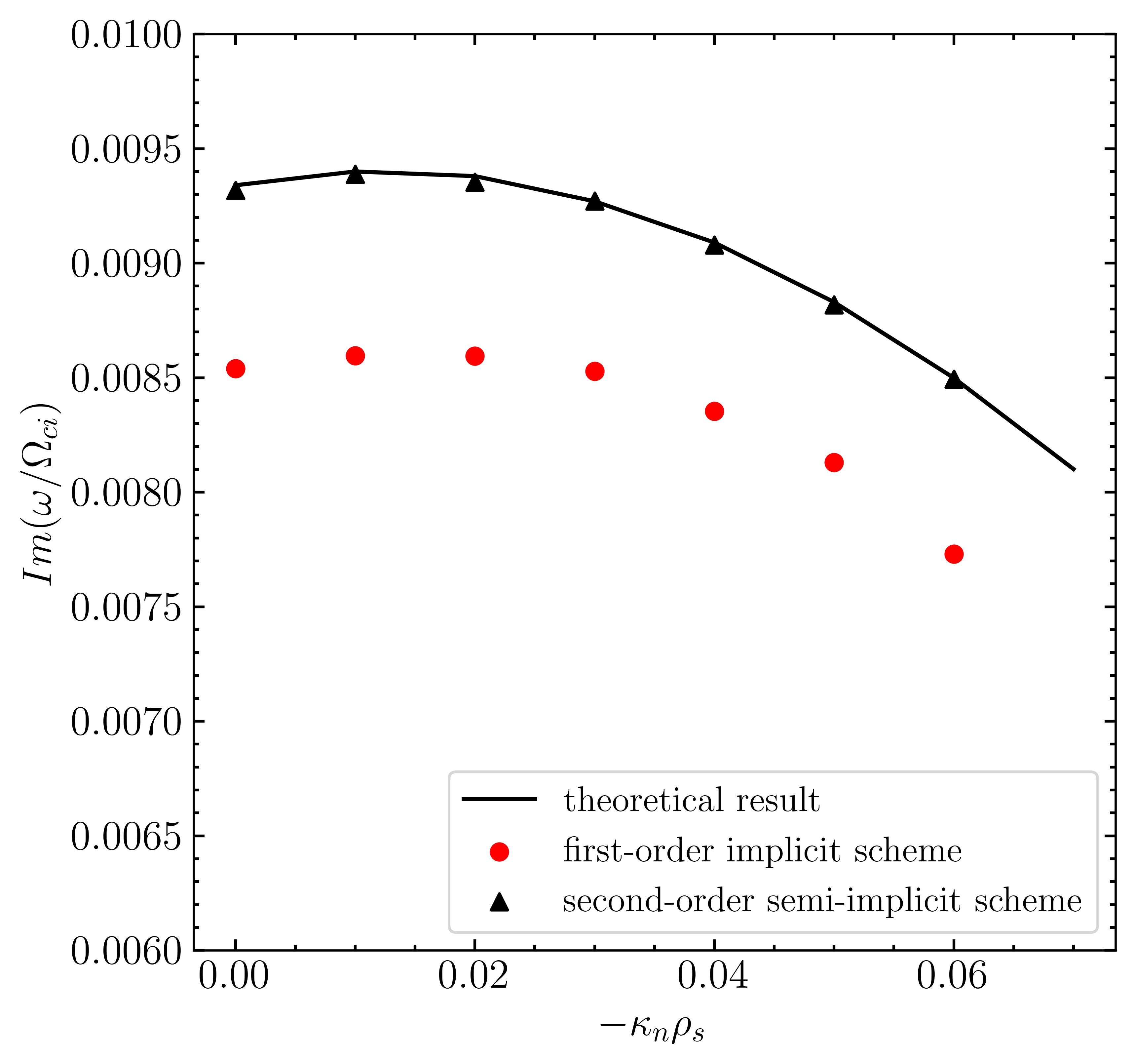}
    \end{minipage}
  }
  \caption{\label{fig:ITG 4} ITG simulation results for the first-order ($\Omega_{ci}\Delta t=0.01$) and second-order ($\Omega_{ci}\Delta t=0.1$) schemes, with other parameters as in Fig.~\ref{fig:ITG 3}. }
\end{figure*}

\revB{With the odd-even decoupling problem resolved, a clean comparison of the first-order implicit and second-order semi-implicit schemes becomes feasible. 
Figure~\ref{fig: timestep refinement study} presents a timestep refinement study of the ITG simulations for the two schemes.
The second-order scheme accurately reproduces the real frequency and growth rate over the entire range of tested timestep $(\Omega_{ci}\Delta t\leq 0.2)$.
In contrast, while the first-order implicit scheme yields accurate real frequencies, it requires a sufficiently small timestep $(\Omega_{ci}\Delta t\leq 0.01)$ to achieve good agreement for the growth rate.
These results confirm the expected order of accuracy and clearly demonstrate the advantage of the second-order semi-implicit scheme.}

\revB{The robustness of the second-order semi-implicit scheme is further illustrated in Figs.~\ref{fig:IAW summary 2} and \ref{fig:ITG 4} for different parameter regimes.
In both the IAW and ITG cases, the second-order scheme achieves more accurate damping rates (for IAW) and growth rates (for ITG) even with a larger timestep $\Delta t$ compared to the first-order implicit scheme. 
These results confirm that the second-order scheme effectively reduces numerical damping inherent in the implicit advancement.}

\subsection{Discussion}

\revA{In this work, three numerical schemes are presented.}

\revA{1. Explicit discretization scheme (conventional method). This scheme is based on the generalized perpendicular and parallel Ohm's law, given by Eqs.~(\ref{eq:iterative implicit perpendicular Ohm's law}) and (\ref{eq:iterative parallel Ohm's law}).}

\revA{2. First-order implicit discretization scheme. 
It advances electron weights via an implicit $E_{\|}$ scheme (\ref{eq:electron weight pushing}), and employs the implicit parallel Ampere's law (\ref{eq:iterative implicit ampere law}) and perpendicular Ohm's law (\ref{eq:iterative implicit perpendicular Ohm's law}) as electric field equations.} 

\revA{3. Second-order semi-implicit discretization scheme. 
This scheme advances particle weights with the second-order semi-implicit scheme, Eqs.~(\ref{eq:second order ion weight pushing}) and (\ref{eq:second order electron weight pushing}).
The field equations are the implicit parallel Ampere's law and perpendicular Ohm's law in the compatible second-order form, as illustrated in Eqs.~(\ref{eq: second order iterative perpendicular ohm law}) and (\ref{eq: second order iterative parallel ampere law}).} 

\revA{The first-order implicit discretization scheme is proposed to better mitigate the cancellation problem in the parallel Ohm's law. 
A key feature is that the implicit parallel Ampere's law employs the first-order velocity moment of the electron weights.
Analytical analysis and numerical experiments confirm that this scheme mitigates the cancellation problem.
Specifically, for both the IAW and ITG test cases, the real frequency is accurately obtained over a wide range of $\Delta t$.
However, the damping rate (for IAW) or growth rate (for ITG) remains inaccurate due to the numerical damping inherent in the implicit pushing.
}

\revA{To overcome this limitation, we develop the second-order semi-implicit scheme.
It retains the implicit parallel Ampere's law and perpendicular Ohm's law as field equations, but advances particle weights using a second-order semi-implicit scheme.
Numerical tests demonstrate that this scheme accurately captures both the real frequency and the growth/damping rate over a large range of $\Delta t$.
Compared with the conventional scheme,  the second-order semi-implicit scheme overcomes the cancellation problem in a comprehensive manner.
}

\revA{For simulations where the cancellation problem is severe, the second-order semi-implicit scheme achieves accurate results with the lowest computational cost among the three schemes.
However, this scheme is considerably more complex to implement than the other two.
}
\section{Conclusion}

In this paper, we have developed the full-kinetic ion drift-kinetic electron simulation (FIDES) code.
We present two numerical schemes and compare their performance in both high- and low-frequency cases.
In the implicit discretization scheme, the electric field is determined by the implicit parallel Ampere's law and implicit perpendicular Ohm's law.
The formulation of the implicit parallel Ampere's law requires an implicit $E_{\|}$ scheme for advancing electron weight.
To suppress unphysical high-frequency instabilities with perpendicular electric fields, an implicit $\mathbf{E}_{\perp}$ scheme is used for ion weights.
Benchmarks against perpendicular and parallel waves confirm that FIDES can correctly simulate high-frequency wave behavior.
The low-frequency IAW and ITG simulations demonstrate that the implicit parallel Ampere's law mitigates the cancellation problem more effectively than the conventional parallel Ohm's law.
The parallel Ohm's law, with its higher-order moment of electron weights and the spatial gradient $\nabla_{\|}$, complicates the numerical cancellation between $E_{1\|}$ and $-\nabla_{\|}\delta p_{e\|}$.
To achieve more accurate wave dynamics, we further develop a second-order scheme to reduce the numerical damping of the original implicit method.
However, this scheme introduces the odd-even decoupling problem in time domain.
To address this, we implement an integrated approach which uses the first-order implicit scheme in the initialization stage and advances particle weights via the second-order scheme with three points.
This strategy can effectively suppress the odd-even decoupling oscillations while preserving the accuracy of the second-order scheme.

\revAB{In closing, we emphasize that the scope of this paper is limited to the presentation of the new algorithm and the analysis of its linear characteristics. 
A comprehensive study for the nonlinear performance of the scheme will be reported in the future.}

\section*{Code availability}
\revA{The FIDES source code is archived at Zenodo with the DOI https://doi.org/10.5281/zenodo.19605309.
The repository is presently under restricted access.
Access will be granted immediately upon reasonable request directed to the corresponding author.}

\section*{Acknowledgements}
This work was supported by the National MCF Energy R \& D Program of China under Grant No.2024YFE03230300, National Natural Science Foundation of China under Grant No.12375213, 12125502 and 12335014, Natural Science Foundation of Sichuan Province under Grant No.2025ZNSFSC0061, China National Nuclear Corporation ‘Young Talents’ Project No.2024-QNYC-02 and the Innovation Program of Southwestern Institute of Physics (202301XWCX001).
This manuscript was first submitted to the Journal of Computational Physics on 21 December 2025.

% \appendix
% \section{Example Appendix Section}


%%       \cite{<label>} ==> [1]



% \bibliographystyle{elsarticle-num} 
% \bibliography{references}

\end{document}